\documentclass{aa}  

\usepackage{graphicx}
\usepackage{caption}
\usepackage{float}
\usepackage{txfonts}
\usepackage{booktabs}
\usepackage{caption}
\usepackage{adjustbox}
\usepackage{blindtext}
\usepackage{lipsum}
\usepackage{amsmath}
\usepackage{subfig}

\numberwithin{equation}{section}

\usepackage{appendix}

\usepackage{natbib}
\bibpunct{(}{)}{;}{a}{}{,}

\begin{document} 

    \title{First systematic high-precision survey of bright supernovae}

   \subtitle{I. Methodology for identifying early bumps}

   \author{E. Paraskeva
          \inst{1,2}
          \and
          A. Z. Bonanos\inst{1}
          \and
          A. Liakos\inst{1}
          \and
          Z. T. Spetsieri\inst{1,2}
          \and
          Justyn R. Maund\inst{3}
          }

   \institute{IAASARS, National Observatory of Athens,
              15236, Penteli, Greece\\
              \email{eparaskeva@astro.noa.gr}
         \and
             Department of Astrophysics, Astronomy \& Mechanics, Faculty of Physics, National and Kapodistrian University of Athens, 15784 Athens, Greece
        \and
            Department of Physics and Astronomy, University of Sheffield, Hicks Building, Hounsfield Road, Sheffield S3 7RH, UK }

   \date{Received February 5, 2020; accepted , 2020}

  \abstract{Rapid variability before and near the maximum brightness of supernovae has the potential
to provide a better understanding of nearly every aspect of supernovae, from the physics of the explosion up to their progenitors and the circumstellar environment. Thanks to modern time-domain optical
surveys, which are discovering supernovae in the early stage of their evolution, we have the unique
opportunity to capture their intraday behavior before maximum. We present high-cadence photometric monitoring (on the order of seconds-minutes) of the optical light curves of three Type Ia and two Type II SNe over several nights
before and near maximum light, using the fast imagers available on the 2.3~m Aristarchos telescope at Helmos Observatory and the 1.2~m telescope at Kryoneri Observatory in Greece.
We applied differential aperture photometry techniques using optimal apertures and we present reconstructed light curves after
implementing a seeing correction and the Trend Filtering Algorithm \citep[TFA,][]{2005MNRAS.356..557K}. TFA yielded the best results, achieving a typical precision between $0.01-0.04$~mag. We did not detect significant bumps with amplitudes greater than 0.05~mag in any of the SNe targets in the \textit{VR-, R-,} and \textit{I-} bands light curves obtained.  We measured the intraday slope for each light curve, which ranges between -0.37$-$0.36 mag day$^{-1}$ in broadband \textit{VR}, -0.19$-$0.31 mag day$^{-1}$ in \textit{R} band, and -0.13$-$0.10 mag day$^{-1}$ in \textit{I} band. We used SNe light curve fitting templates for SN 2018gv, SN 2018hgc and SN 2018hhn to photometrically classify the light curves and to calculate the time of maximum. We provide values for the maximum of SN 2018zd after applying a low-order polynomial fit and SN 2018hhn for the first time. We conclude that optimal aperture photometry in combination with TFA provides the highest-precision light curves for SNe that are relatively well separated from the centers of their host galaxies. This work aims to inspire the use of ground-based, high-cadence and high-precision photometry to study SNe with the purpose of revealing clues and properties of the explosion environment of both core-collapse and Type Ia supernovae, the explosion mechanisms, binary star interaction and  progenitor channels. We suggest monitoring early supernovae light curves in hotter (bluer) bands with a cadence of hours as a promising way of investigating the post-explosion photometric behavior of the progenitor stars.}


\maketitle

\section{Introduction}

    \begin{table*}[h!]
\tiny
\centering
\caption{Log of Observations}
\begin{adjustbox}{width=1\textwidth}
\begin{tabular}{@{}lccclcccc@{}}
\toprule
Name & \begin{tabular}[c]{@{}c@{}}RA\\ (J2000)\end{tabular} & \begin{tabular}[c]{@{}c@{}}DEC\\ (J2000)\end{tabular} & Type & Filter & \begin{tabular}[c]{@{}c@{}}Total number\\ of frames\end{tabular} & Telescope & Discovery ref. & Classification ref. \\ \midrule \midrule \vspace{0.1cm}
SN 2018gv & 08:05:34.61 & -11:26:16.30 & Ia & \textit{R, I} & 236, 236 & Kryoneri & \cite{2018ATel12063....1B} & \cite{2018ATel11175....1S, 2018ATel11177....1B} \\ \vspace{0.1cm}
SN 2018zd & 06:18:03.18 & +78:22:00.90 & IIL$^{a}$ & \textit{R, I} & 1099, 1099 & Kryoneri & \cite{2018TNSTR.285....1I}  & This work \\ \vspace{0.1cm}
SN 2018hgc & 00:42:04.56 & -02:37:40.80 & Ia & \textit{VR} & 284 & Aristarchos & \cite{2018ATel12104....1B} & \cite{2018ATel12110....1D} \\ \vspace{0.1cm}
SN 2018hhn & 22:52:32.06 & +11:40:26.70 & Ia & \textit{VR} & 296 & Aristarchos & \cite{2018TNSTR1570....1G} & \cite{2018TNSCR1585....1P} \\ \vspace{0.1cm}
SN 2018hna & 12:26:12.05 & +58:18:51.10 & II-pec$^{b}$ & \textit{VR} & 272 & Aristarchos & \cite{2018TNSTR..57....1I} & 
    \cite{2018ATel12258....1P, 2019ATel12897....1R} \\ \bottomrule
\end{tabular}
\end{adjustbox}
\caption*{
$^{a}$Initially, it was spectroscopically classified as a Type IIn by \cite{2018ATel11379....1Z}. \\
$^{b}$Initially, it was classified as a normal Type II supernova. According to \cite{2019ApJ...882L..15S}, this SN is a 1987A-like event object.}
\label{tableobslog}
\end{table*}
 
    The increasing rate of discovery of bright supernovae (SNe) a few days after their explosion and several days before their maximum light has opened a new window into the physics of SNe. Time-domain surveys such as the All-Sky Automated Survey for Supernovae\footnote{\url{http://www.astronomy.ohio-state.edu/assassin/index.shtm}.} \citep[ASAS-SN;][]{2014ApJ...788...48S}, the Asteroid Terrestrial-impact Last Alert System\footnote{\url{http://fallingstar.com/}.} \citep[ATLAS;][]{2018PASP..130f4505T}, the Zwicky Transient Facility\footnote{\url{https://www.ztf.caltech.edu/}.} \citep[ZTF;][]{2019PASP..131a8003M}, the Catalina Real-Time Transient Survey\footnote{\url{http://crts.caltech.edu/}.} \citep{2009ApJ...696..870D}, and the DLT40\footnote{\url{http://dark.physics.ucdavis.edu/dlt40/DLT40}} \citep{2018ApJ...853...62T} enable us to obtain quick follow-up observations of infant SNe. They can be used to investigate the progenitor channels, their circumstellar environment, and the nature of the explosions. Even though several SNe of different types (e.g. Type Ia and IIP) have shown peculiar signs in their early light curves and spectra (early blue or UV excess, carbon absorption lines), most photometric surveys probe SNe on timescales of days and, therefore, their intraday behavior both before and after maximum brightness remains unknown. Rapid, high-cadence photometric follow-up observations of young Type I, II or superluminous SNe are warranted to probe the properties of their progenitors. Early photometric monitoring of Type Ia SNe has revealed crucial characteristics of the explosive stars. Many studies have already found intriguing clues of undulations in very early light curves of SNe. \cite{2016ApJ...820...92M} reported a significant detection of excess luminosity in multiple filters at 16 and 15~days prior to maximum in the light curves of SN Ia 2012cg. They compared the light curves and the \textit{B--V} colors to the \cite{2010ApJ...708.1025K} models and found that the observations are consistent with models for the interaction between a SN Ia and a main-sequence companion star of 6~M$_\odot$. In this single degenerate scenario, the SN Ia will expand until it encounters a companion, when matter is compressed and heated. As the ejecta surrounds the companion star and a shock is formed, a gap opens in the layers of the ejecta material and emission from the shock-heated material can escape through it \citep{2000ApJS..128..615M}. Optical and UV excess can be produced for a few days after the explosion from the radiative diffusion of the deeper layers of the ejecta. A similar case may have been observed for the Type Ia SN iPTF14atg at early times, which was interpreted as the collision of the heated ejecta with a binary companion. Strong but declining UV emission was reported within four days of its explosion \citep{2015Natur.521..328C}, along with another interesting feature, persistent C II 6580~\AA\, absorption. Early observations of the Type Ia SN 2017cbv \citep{2017ApJ...845L..11H} and iPTF16abc \citep{2018ApJ...852..100M} revealed early blue bumps, along with a combination of unburned material in the early optical spectra. These cannot be explained by either SN shock breakout or the SN ejecta colliding with a stellar companion. The features were interpreted as evidence for ejecta interaction with the diffuse circumstellar material or an unusual nickel distribution, which were inferred by comparing them to models. Recently, \cite{2019arXiv190402171F} used early-time light curves of 18 Type Ia SNe observed in the first six sectors of TESS data to constrain the radii of any companion stars, and thus the progenitor channels.
    
    Pre-maximum light curves of Type II SNe have not only revealed properties of the progenitors, but have also similarly reported evidence of bumps. \cite{2016ApJ...820...23G} demonstrated the potential of high-cadence photometry of SNe, using {\it Kepler} photometry of two Type II-P SNe. For the first time they observed the shock breakout in optical light and measured precise rise times, as well as the radius of the progenitor red supergiants. The very early observations of SN 2016X \citep[Type II-P;][]{2018MNRAS.475.3959H}, showed another possible UV peak within 1-2 days of the explosion, which was assumed to be the breakout of a blast shockwave through the outer envelope of the progenitor star after the core-collapse explosion. Early photometric observations of SN 2017gmr in the \textit{U} and \textit{B} band \citep{2019ApJ...885...43A} revealed undulations, which were not further interpreted. These results demonstrate the power of early discovery and high-cadence monitoring of all types of SNe to constrain the progenitor channels and explosion mechanisms of SNe.

\begin{figure*}[h!]
   \centering
   \includegraphics[width=1\textwidth]{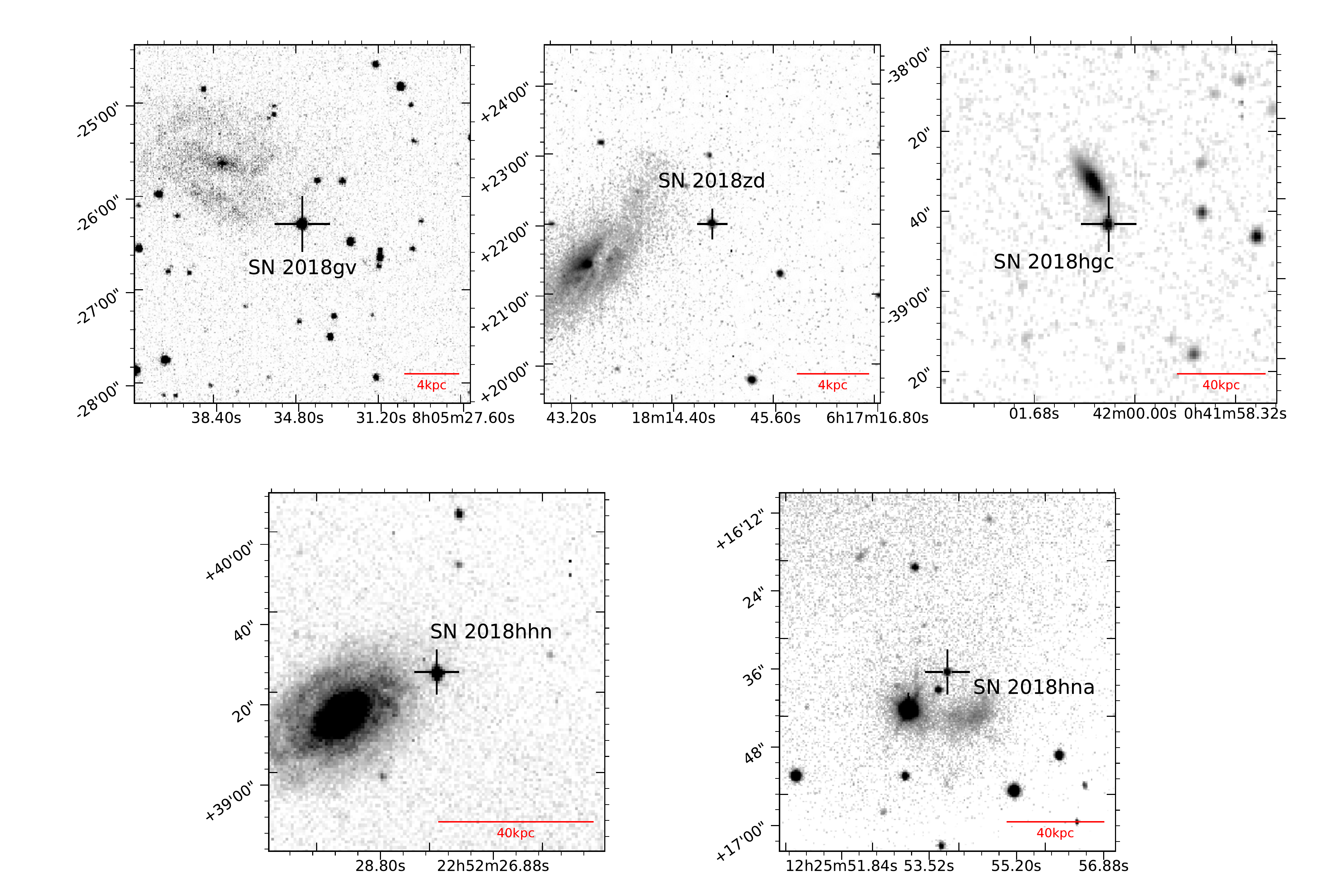}
      \caption{Finder charts of the targets illustrating the position of the SNe in the outskirts of their host galaxies. The bars show the scale, which was converted from apparent to absolute units (kpc).}
         \label{finderchart}
   \end{figure*}

Ground-based high-precision photometric monitoring studies are crucial to provide evidence for such variability in the optical light curves.
The first ground-based high-cadence study of \cite{2016A&A...585A..19B} presented photometry of the nearby Type Ia supernova SN2014J, 15--18~days after maximum in the galaxy M82 \citep{2014CBET.3792....1F}. They reported evidence for rapid variability at a 0.02--0.05~mag level on timescales of 15--60~min on four nights. The following scenarios were suggested for the origin of the variability: (i) clumping of the ejecta \citep{2010ApJ...720.1500H}; (ii) interaction of the ejecta with the circumstellar material \citep{2014MNRAS.443.2887F}; (iii) asymmetry of the ejecta \citep{2008ARA&A..46..433W} and (iv) the onset of the secondary maximum \citep{2000ApJ...530..757P, 2015MNRAS.449.3581J}.

No predictions exist, however, for the intraday behavior of the photometric evolution of infant SNe, the duration of the bumps and the presence of periodicity on short timescales. High-cadence observations of explosive transients, therefore, provide a way to probe such short timescales over which energy might be produced. For these reasons, we observed several bright SNe (Table \ref{tableobslog}) with the aim of exploring their short timescale behavior and searching for evidence of rapid variability, which is not probed by the vast majority of photometric surveys of transients. We employed high-precision photometric techniques widely applied in the exoplanet community to analyze the data, which have not yet been adopted by the SNe community. This work aims to demonstrate how well these methods work for early SNe observations.

The paper is structured as follows: the observations and data reduction are described in Section 2, the photometric methods are presented in Section 3 and the analysis in Section 4. The results are given in Section 5, while the discussion and conclusions follow in Section 6.

\section{Observations and data reduction}

The selection criteria for our target SNe were primarily based on their spectroscopic classification as young and bright SNe (<18~mag at discovery). Tables \ref{tableobslog} and \ref{tablesnsample} include the observations, the sample and its properties. Furthermore, as we aimed to achieve high-precision photometry, we selected targets that were not heavily contaminated by the host galaxy. The 2.3~m Aristarchos telescope and the 1.2~m Kryoneri telescope were employed to observe the targets. These SNe were selected as the best available targets during our allocated telescope time. Finder charts illustrating the locations of the SNe in their host galaxies are shown in Figure~\ref{finderchart}. The targets were monitored for short-term variability with exposure times that yielded a signal-to-noise ratio (S/N) $>70$. Typically, our exposures ranged from 20~s--120~s with a median cadence of 2 min. Sequences of images of at least 1 hour were obtained, yielding 200--300 images in total per target for 1--5 nights, mainly before and near the maximum light (Table \ref{tableobslog}). 
 
 \subsection{Telescopes \& instrumentation}
 
 We used the RISE2 instrument installed on the 2.3~m f/8 Ritchey–Chretien Aristarchos telescope at Helmos Observatory to observe SNe 2018hgc, 2018hhn and 2018hna. RISE2 is equipped with an ANDOR DW485 camera with an E2V CCD47-20 back-illuminated CCD sensor, which has a $1024\times1024$ pixel light sensitive region (pixel size $13.0\times13.0~\mu$m$^2$) and is used for detecting transiting exoplanets. The field of view of the instrument is $10'\times10'$, providing a pixel scale of $~0.60''$. The instrument uses a single broad filter, similar to the \textit{V+R} passband.
 
 We used the wide-field, high-cadence camera system \citep{2018A&A...619A.141X}, installed on the prime focus of the 1.2~m Kryoneri telescope to observe SN 2018gv and 2018zd. The instrument consists of two Andor Zyla 5.5 cameras, each with a $2560\times2160$ front-illuminated scientific CMOS (pixel size $0.06\times0.06~\mu$m$^2$) resulting in a pixel scale of $~0.4''$ and a $17'\times14'$ field of view. It provides the means to record a target simultaneously in two different wavelength bands. A dichroic with a cut-off at 730~nm splits the light beam, sending each resulting beam to a different camera. The cameras are equipped with \textit{R} and \textit{I}-band Johnson-Cousins filters.

It should be noted that at the time of the observations, it was not clear from the literature that bumps are primarily present in the blue bands. The observations only suggested that bumps appear soon after the time of explosion. For these reasons, we opted to use the available cameras with faster readout times and larger field of views, which are ideal for differential photometry and high-cadence observations, without taking the wavelength of the available filters into account.


\begin{table*}[]
\centering
\caption{Properties of the observed SNe sample}
\begin{tabular}{@{}lccccccl@{}}
\toprule
Name & $t_\text{max}$ & phase$^{a}$ & $m_\text{peak}$ & $M_\text{peak}$ & $z$ & $E(B-V)^{b}$ & Host galaxy               \\
 & (days) & (days) & (mag) & (mag) & & (mag) & \\ \midrule \midrule SN 2018gv  & 2458150.7 (\textit{V}) & $-$9 &   12.7    &    $-$18.98                & 0.005   &   0.051     & NGC 2525                \\ SN 2018zd  & 2458188.6 (\textit{V})  & -1/0/+1 &      13.7                &           $-$16.90         & 0.003   &     0.036   & NGC 2146                \\ SN 2018hgc & 2458411.4 (\textit{g}) &  $-$4/$-$3 & 17.3                     &     $-$19.57               & 0.052   &   0.026     & 2MASX J00420581-0237394 \\
SN 2018hhn & 2458417.1 (\textit{o})  &+2/+3/+4/+6   &        16.7            &      $-$19.03              & 0.029   &    0.970    & UGC 12222               \\
SN 2018hna &  2458499.1 (\textit{o})  & $-$80/$-$76  &   14.2                &      $-$15.92              & 0.002   &    0.046    & UGC 7534                 \\ \bottomrule
\end{tabular}
\caption*{
$^{a}$The phase is given with respect to the light curve maximum, according to the light curve fitting.\\
$^{b}$The Galactic reddening towards the SNe targets has been estimated using the NASA/IPAC NED Galactic Extinction Calculator adopting the $R_\text{V}= 3.1$ extinction law of \cite{1989ApJ...345..245C} and the extinction map given by \cite{2011ApJ...737..103S}. The host galaxy extinction was assumed to be negligible.}
\label{tablesnsample}
\end{table*}

     \begin{figure*}[h!]
   \centering
   \includegraphics[width=1\textwidth]{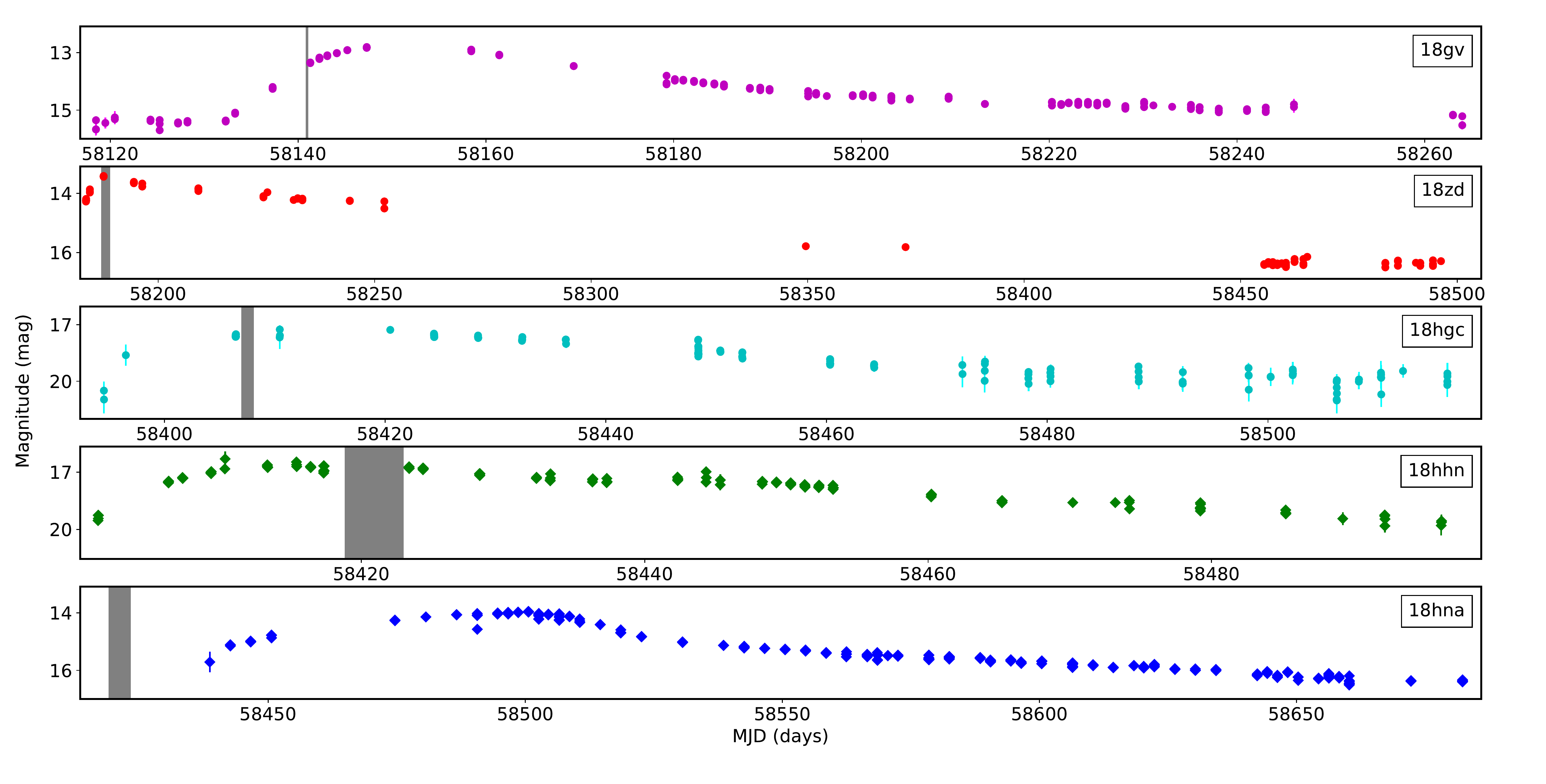}
      \caption{ASAS-SN light curves for SN 2018gv (\textit{V}-band) with the flat beginning corresponding to the host galaxy emission}, SN 2018hgc (\textit{g}-band), SN 2018zd (\textit{V}-band) and ATLAS light curves for SN 2018hhn (\textit{o}-band) and SN 2018hna (\textit{o}-band), respectively. The light curves are presented in apparent magnitudes as a function of time. The gray shaded areas indicate the MJD during which our high-cadence observations were obtained.
         \label{18lc}
   \end{figure*}

    \subsection{Observational methodology \& reduction}
   
   In order to achieve high photometric accuracy and be sensitive to low amplitude undulations, we adopted techniques from the exoplanet community, with the purpose of eliminating the systematic errors. When performing differential photometry (Section 3), accurate bias-subtraction and flat-fielding are of major importance. According to \cite{2007MNRAS.375.1449I}, the Poisson noise is 200~$e^{-}$ for a typical detector with a gain of a few $e^{-}$ ADU$^{-1}$ and the flat illumination level is of 20,000~ADU~pixel$^{-1}$=40,000~$e^{-}$ pixel$^{-1}$. Thus, a typical photometric aperture with a radius of 3 pixels contributes $\sim$1~mmag photon noise. For this reason, we obtained a considerable amount of biases (150-300 frames) and twilight flat-fields (25-100 frames) each night to reduce the Poisson noise to less than 0.2~mmag \citep{2007MNRAS.375.1449I}. The bias frames were averaged together using the \textit{minmax} in the reject option of the \textit{zerocombine} task in IRAF with a view of keeping radiation events out of the master bias frame. The master flat frame was the result of combining all the frames using a median mode. The median value is an excellent way of removing the effects of hot pixels and cosmic rays, so these extreme values do not affect the calculation, as they would, if they would averaged. The reject option was set to \textit{avsigclip}, in which case the "typical" sigma would have been determined from the data itself rather than an a priori knowledge of the noise characteristics of the CCD. Other related issues that can limit the photometric precision are: i) the positioning of the telescope, ii) fringing issues, and iii) the differential variations on the quantum efficiency of the pixels. With the aim of minimizing the contribution of these effects, we repositioned each star almost on the same pixel of the detector using the autoguiding system of each telescope. The read-out-noise of the detectors are insignificant, as it can be as low as a few $e^{-}$ ($<10e^{-}$ for RISE2 and Andor Zyla cameras).
   
   We performed the photometric reduction scheme of bias correction, trimming of the overscan and flat-fielding correction using standard routines in the IRAF\footnote{IRAF is distributed by the National Optical Astronomy Observatories, which are operated by the Association of Universities for Research in Astronomy, Inc., under cooperative agreement with the National Science Foundation (NSF).} \textit{ccdproc} package. Aperture photometry was extracted for each supernova, as the targets were isolated (Figure~\ref{finderchart}). In cases where relatively large undulations were visible in the light curves (e.g., SN 2018gv, Section 5), we also used point spread function (PSF) photometry to assess the signal. 
   
   ASAS-SN light curves, which were publically available\footnote{https://asas-sn.osu.edu/} \citep{2014ApJ...788...48S} and ATLAS light curves \citep{2017ApJ...850..149S,2018AJ....156..241H}, which were provided by the ATLAS team (private communication) of each SN, are shown in Figure~\ref{18lc}. The grey shaded areas represent the MJD during which the high-cadence observations were obtained during our allocated telescope time at the Helmos and Kryoneri observatories.
  
\section {Photometry}

\subsection{Aperture photometry}

All images were initially aligned to avoid small shifts, when the autoguiding system was not functioned properly or stopped unexpectedly during the observational nights with the \textit{imal2}. \textit{Imal2} is a task of \textit{VAPHOT}, which is an improved version of the IRAF task \textit{imalign} and is intended for aligning a large number of frames.  It will measure each frame and then shift it. It does not stop, if a frame cannot be shifted in comparison with \textit{imalign} task.  It keeps track which images have been shifted and it makes an entry named \textit{IMSHIFT}
into the frame-header, and will not measure or shift an image
again. The header entry tells the size of the x, y shifts, so reversing
the shifting is possible. The differential time-series photometry was based on CCD aperture photometry performed using \textit{VAPHOT} \citep{2013ascl.soft09002D}, a package that works within the IRAF environment, similar to the standard aperture photometry task $phot$.  The choice of the optimally-sized apertures was of high importance in obtaining photometric points with the lowest possible noise. We used \textit{apcalc}, a subroutine of the \textit{VAPHOT} package to measure the optimum aperture radius. \textit{apcalc} finds the zero point in the derivative $\partial (S/N)/{\partial r}$, where \textit{S} is the signal, \textit{N} is the noise and \textit{r} is the aperture radii. The algorithm calculates the values $\sigma_\text{psf}, N_\text{ph,tot}$ and $\sigma_\text{1pix}$ using the equations: \begin{align}  S= N_\text{ph}(r)=N_\text{ph,tot}\left(1-\exp(\frac{-r^2}{2\sigma^2_\text{psf}})\right), \end{align} \begin{align}N=\sqrt{\sigma^2_\text{Nph}(r)+\sigma^2_\text{BG}+\sigma^2_\text{scin}}, \end{align}
where $S$ is the signal, $N_\text{ph}$ is the number of photons inside an aperture of a radius \textit{r}, $\sigma_\text{psf}$ is the width of a Gaussian PSF, $N_\text{ph,tot}$ is the total number of photons in the limit of an infinite aperture from the star, $N$ is the total noise inside the radius \textit{r} and $\sigma_\text{scin}$ is the scintillation noise, which is a constant, regardless the magnitude of the star or the aperture used. The photon noise inside a radius \textit{r} from the star, $\sigma_\text{Nph}$ is given by: \begin{align}\sigma_\text{Nph}=\sqrt{N_\text{ph}(r)}. \end{align}
The uncertainty $\sigma_\text{BG}(r)$ of the contribution of the sky-background to the total signal within the aperture is calculated by:
\begin{align}\sigma_\text{BG}(r)=\sigma_\text{1pix}r\sqrt{\pi}, \end{align} where $\sigma_\text{1pix}$ is the average noise of one background pixel.
The ratio of the above equations gives the S/N, the well known "CCD equation" \citep{1989PASP..101..616H}. Aperture photometry was performed with radii selected to yield the maximum S/N. \textit{apcalc} finds the optimum aperture diameter in units of the FWHM of the PSF and \textit{VAPHOT} performs the aperture photometry on a series of frames, with aperture sizes determined individually for each star in a reference frame. We selected the frame with the lowest FWHM as the reference image.

\subsection{PSF photometry}

We proceeded to perform PSF photometry with \textit{DAOPHOT} \citep{2011ascl.soft04011S} on SN 2018gv, which indicated signs of variability or bumps after the analysis (Section 4). We aimed to assess the contamination of the host galaxy and validate the accuracy of the aperture photometry results. We first ran \textit{daofind} in IRAF to find all the stars on each frame above the appropriate brightness threshold, discriminating between extended sources, stars, cosmic rays and CCD defects and performed aperture photometry with \textit{phot}. We then chose a set of bright isolated stars spread across each frame as PSF stars with \textit{pstselect} and we built a semi-empirical PSF model. Using the centroid of a star as the profile center and the sky level as determined for aperture photometry in \textit{allstar}, the PSF model is shifted and scaled to fit the observed frame by non-linear least-squares. The scaling yields the magnitude estimate. We found the instrumental PSF magnitudes to be shifted by 0.08~mag compared to the aperture photometry measurements, while the light curve shape and features remained identical.

\section{Analysis}
Initially, differential light curves were created by extracting the instrumental magnitudes from the aperture photometry and taking the difference between the supernova magnitude and 10--15 comparison star candidates inside the field of view, which previously had been checked for intrinsic variability. We then followed two approaches to subtract systematic effects. One approach was to remove the seeing-correlated effects following \cite{2007MNRAS.375.1449I}. The other approach was the implementation of the signal-reconstruction-mode, the Trend Filtering Algorithm \citep[TFA,][]{2005MNRAS.356..557K} from the VARTOOLS program \citep{2016A&C....17....1H} to filter all sources of systematic noise from the light curves, while preserving real variability. In the final SNe light curves, we applied the least-squares regression line, which is the line that makes the vertical distance from the data points to the regression line as small as possible in order to find the intraday slope. Least squares fitting is a way to find the best fit curve or line for a set of points. In this technique, the sum of the squares of the offsets (residuals) are used to estimate the best fit instead of the absolute values of the offsets, while the errors are measured based on the distances of the data points from the fitting curve.

\subsection{Seeing-correlated variations}

\begin{figure}[!ht]
   \centering
   \includegraphics[width=\hsize]{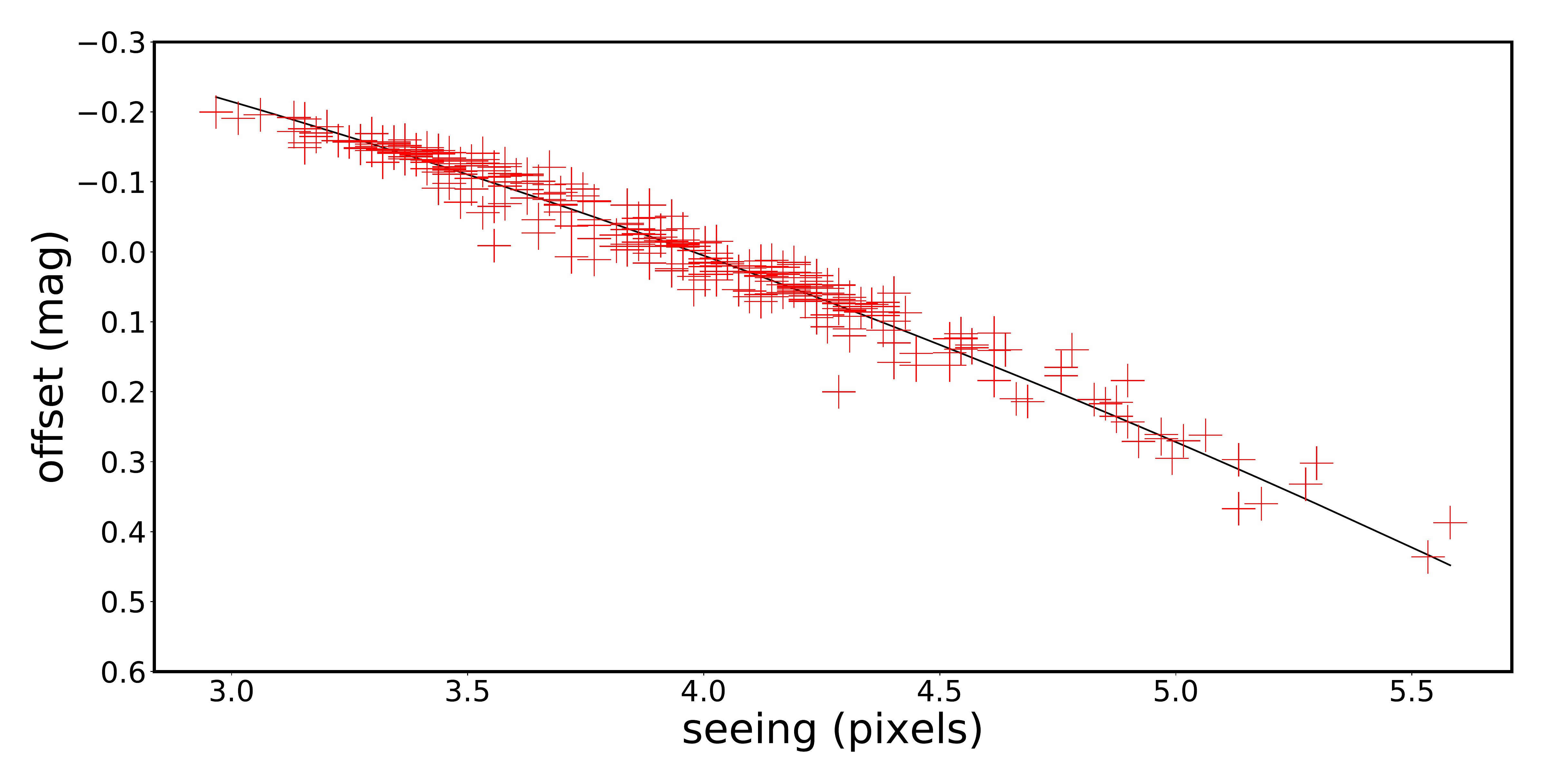}
      \caption{Seeing-correlated effects for the first night of observations for SN 2018zd. The panel shows the correlation between the offset of the magnitude to the median as a function of the measured seeing. The data are plotted as red crosses and the quadratic polynomial fit is illustrated by the solid black line. 
              }
         \label{seeingnoise}
   \end{figure}

\cite{2007MNRAS.375.1449I} demonstrated the strong correlation between the FWHM and the magnitude for blended objects. As the FWHM of the point sources increases due to variations in the seeing, the amount of blended flux increases. This method can be used for finding objects that are blended and subtracting the variations in FWHM. To characterize the level of blending in each light curve, we measured the seeing-correlated shifts of the target from its median magnitude. This was done by calculating the correlation between the offset of the magnitude to the median as a function of the measured seeing and by fitting a quadratic polynomial. Figure~\ref{seeingnoise} demonstrates this methodology for SN 2018zd and Figure~\ref{seeingeffects} for all the objects. We therefore subtracted the fitting function from the light curve of our targets. Figures~\ref{multi2gv} and \ref{multi1zd} illustrate the subtraction of the seeing correlation from the instrumental light curves of our target.

  \subsection{Signal reconstruction}
  
 Image and photometric reduction methods cannot identify weak and short-duration periodic irregular dimmings in the presence of noise and various systematics. To address this deficiency, several methods have been developed such as \cite{2003AcA....53..241K}, SysRem by \cite{2005MNRAS.356.1466T}, TFA by \cite{2005MNRAS.356..557K}, and PDT (Photometric DeTrending Algorithm Using Machine Learning) by \cite{2009MNRAS.397..558K} to remove systematics during the post-processing phase. A detailed comparison of these methods is beyond the scope of this paper. Possible sources of these colored effects are due to uncorrected instrumental effects, the changing of the observing conditions, the imperfections in the data reduction or intrinsic systematic variations. These systematic and random noise sources are responsible for creating trends in time-series data that undermine the signals from the stars. 
 
  \begin{figure*}[h!t]
\centering
\includegraphics[width=1\textwidth]{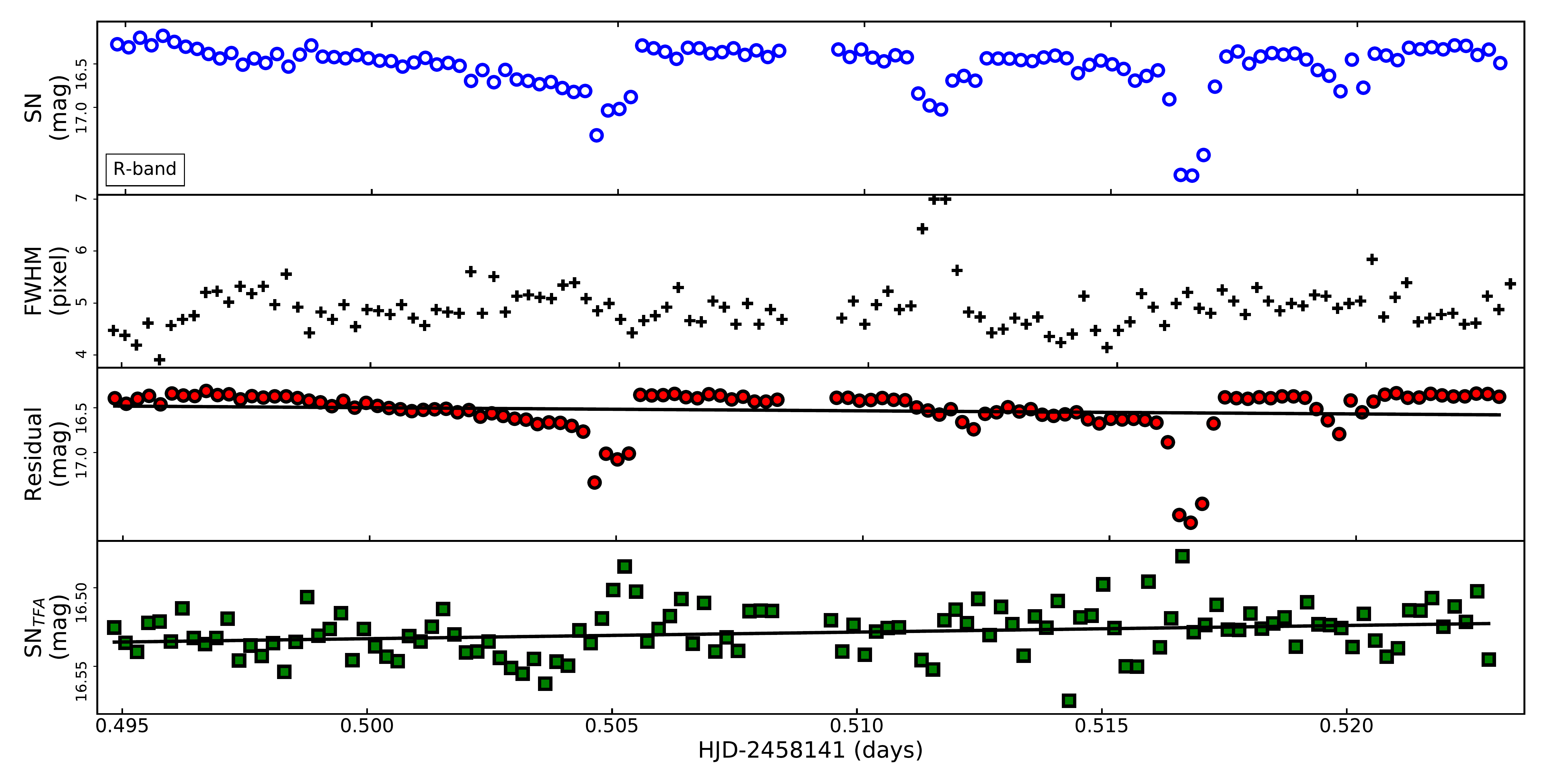}
\bigbreak
\includegraphics[width=1\textwidth]{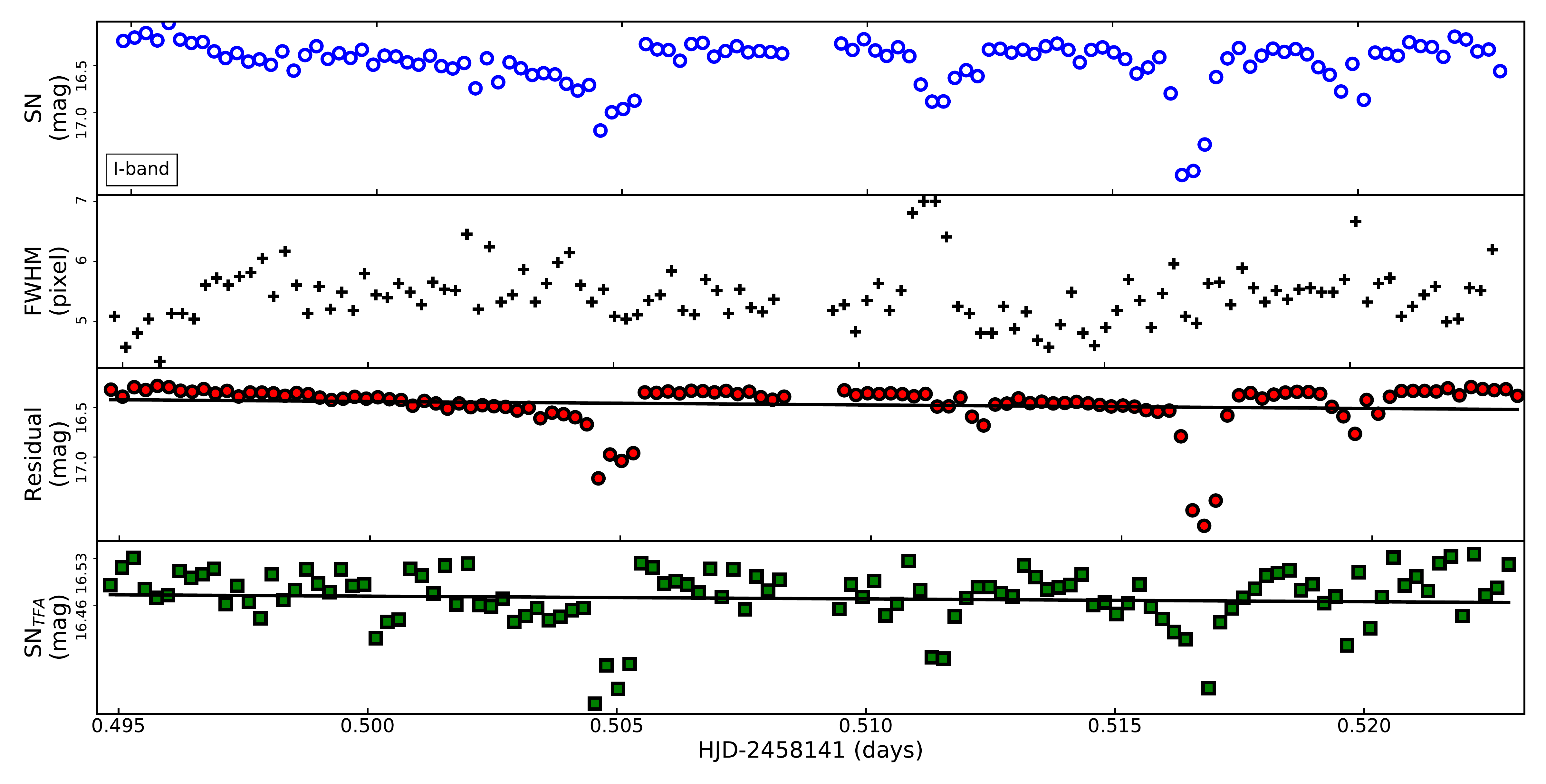}
\caption{Light curves of SN 2018gv in the \textit{R} (upper plot) and \textit{I} bands (lower plot), respectively, obtained with the Kryoneri telescope during the third run. The top panels show the instrumental photometric light curve from VAPHOT. The second panels plot the seeing as a function of time. The third panels present the residual supernova light curve after the subtraction of the FWHM features. The reconstructed light curve after the implementation of TFA for the systematic effects appears on the last panel.}\label{multi2gv}
\end{figure*}
 
 For our analysis we used TFA from the VARTOOLS program \citep{2016A&C....17....1H}, which searches for temporal features observed in the frame and subtracts the part of the target time-series that is produced by the systematics. TFA utilizes the fact that the same types of systematics appear in several stars in the same frame. For this reason, a list of 10 light curve templates for stars, which appeared non-variable through aperture photometry in the same field of view, was created manually to compute the systematic errors in each frame for each target. TFA uses all these available reference stars of the given frame in the template and reproduces each target light curve by a linear combination of these templates, optimized by least-squares. After a signal has been identified in the residuals between the original time-series and the systematics computed by TFA, the shape of the signals is reconstructed. Figure~\ref{multi2gv} demonstrates the implementation of TFA (bottom panel) compared to the initial instrumental light curve (top panel). Photometric accuracy close to the photon noise was achieved and the rms was substantially improved in all cases. Table \ref{measurements} shows the values of light curve rms scatter before and after the implementation of TFA.
 
\begin{figure}[!h]
   \centering
   \includegraphics[width=\hsize]{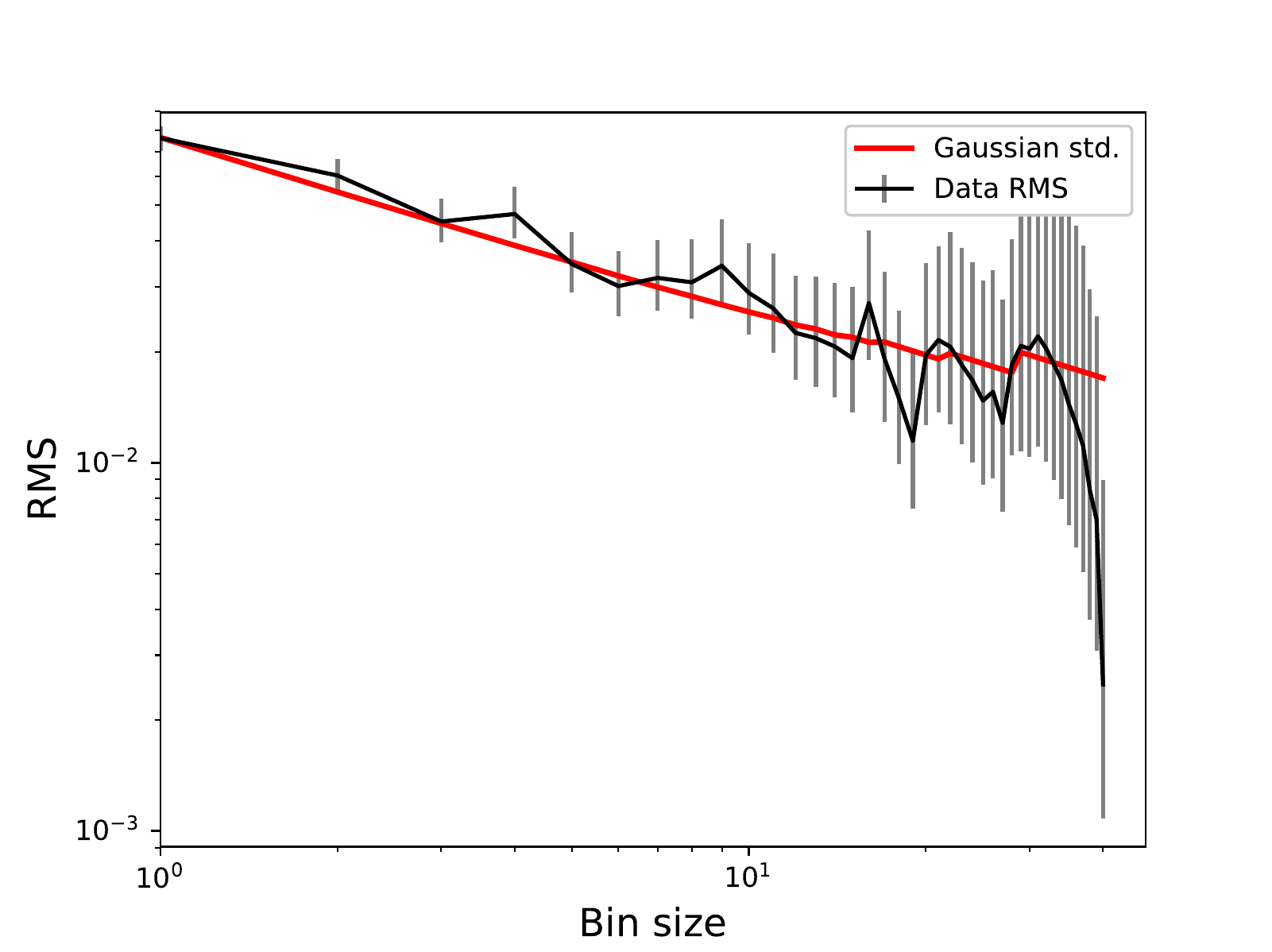}
      \caption{Binned rms residuals vs. bin size plot of the photometry of SN 2018gv on the second night. The rms of the binned residuals, $rms_\text{N}$, is shown as a black curve as a function of bin sizes. The grey error bars represent the uncertainty of $rms_\text{N}$. The red curve shows the expected $rms \sigma_\text{N}$ and the saw-tooth shape of the black curve arises from the change in \textit{M}.
              }
         \label{noise}
   \end{figure}


\subsection{Noise estimation}
We estimated the red and white noise for light curves that showed undulations, which remained after applying corrections for systematics, which are relatively common for ground-based data. We implemented the time-averaging procedure outlined by \cite{2008ApJ...683.1076W} through a Markov Chain Monte Carlo (MCMC) analysis to quantify the significance of the variability or bumps detected. We derived the residuals after subtracting the best-fitting model, in our case the quadratic polynomial fit, from the data points. Then, we measured the mean values of the residuals that have been split into time-ordered bins of N elements, which are not overlapping, and we calculated the root mean square of the binned residuals, $rms_\text{N}$. The process was repeated for a range of bin sizes (number of data points in each bin) from one to half the data size. The uncertainty of $rms_\text{N}$ is computed as $\sigma_\text{rms}=rms_\text{N}/\sqrt{2M}$ \citep{2017AJ....153....3C}, where M is the number of bins. The expected rms $\sigma_\text{N}$ (white noise) in the absence of correlated noise is given by: \begin{align} \sigma_\text{N}= \frac{\sigma_\text{1}}{\sqrt{N}}\sqrt{\frac{M}{M-1}},\end{align} where $\sigma_\text{1}$ is the rms for the unbinned residuals \citep{2008ApJ...683.1076W}. However, we have to note that the correlated noise that possibly exists in our data is accounted for via two correction factors, $\beta_\text{w}$ and  $\beta_\text{r}$. The ratio of the rms of the non-binned residuals to the mean photometric error is $\beta_\text{w}$ and  $\beta_\text{r}$ is measured as: \begin{align} \beta_\text{r}=\frac{rms_\text{N}}{\sigma_\text{N}}=\frac{\sigma_\text{N}}{\sigma_\text{1}}\sqrt{\frac{N(M-1)}{M}}. \end{align}

The resulting values of white and red noise measurements are reported in Section 5.1 for SN 2018gv. Figure~\ref{noise} shows the binned rms residuals as a function of the bin size plot for the photometric data of SN 2018gv on the second night. The rms of the binned residuals, $rms_\text{N}$, is shown as a black curve. The grey error bars represent the uncertainty of $rms_\text{N}$. The red curve shows the expected $rms \sigma_\text{N}$ in the absence of correlated noise. The saw-tooth shape of the black curve arises from the change in \textit{M}, which becomes more significant as \textit{N} increases.

\subsection{Light curve fitting}
We finally proceeded with a basic analysis of the SN light curves in Figure~\ref{18lc} by fitting Type I and Type II templates through the SNcosmo framework \citep{2016ascl.soft11017B}. For Type Ia light curves we used the most recent version of the SALT2 light curve fitter \citep[SALT2.4,][]{2014A&A...568A..22B, 2010yCat..35230007G}. SALT2 characterizes the flux density of a given SN as function of time $t$ and wavelength $\lambda$ as: \begin{align}f(t,\lambda)=x_\text{0}[(M_\text{0}(t,\lambda)+x_\text{1}M_\text{1}(t,\lambda)]e^{cC_\text{L}(\lambda)},\end{align} where $x_\text{0}, x_\text{1}$ and $c$ are the normalization, shape and color parameters, respectively. $M_\text{0}, M_\text{1}$ are the zeroth and first components of the model, and $CL$ is the color law, which gives the extinction in magnitudes for $c=1$. 

For Type II light curves, the spectral model for each SN that we used is given by:
\begin{align}f(p,\lambda)=A\times{M(p,\lambda)},\end{align}
where $M$ is the flux defined on a grid in phase $p$ and wavelength $\lambda$ and $A$ is the amplitude, a single free parameter of the model. The resulting fitting parameters for each SN are presented in Section 5.

\section{Results}

  \subsection{SN 2018gv}

\begin{figure}[h!]
   \centering
   \includegraphics[width=\hsize]{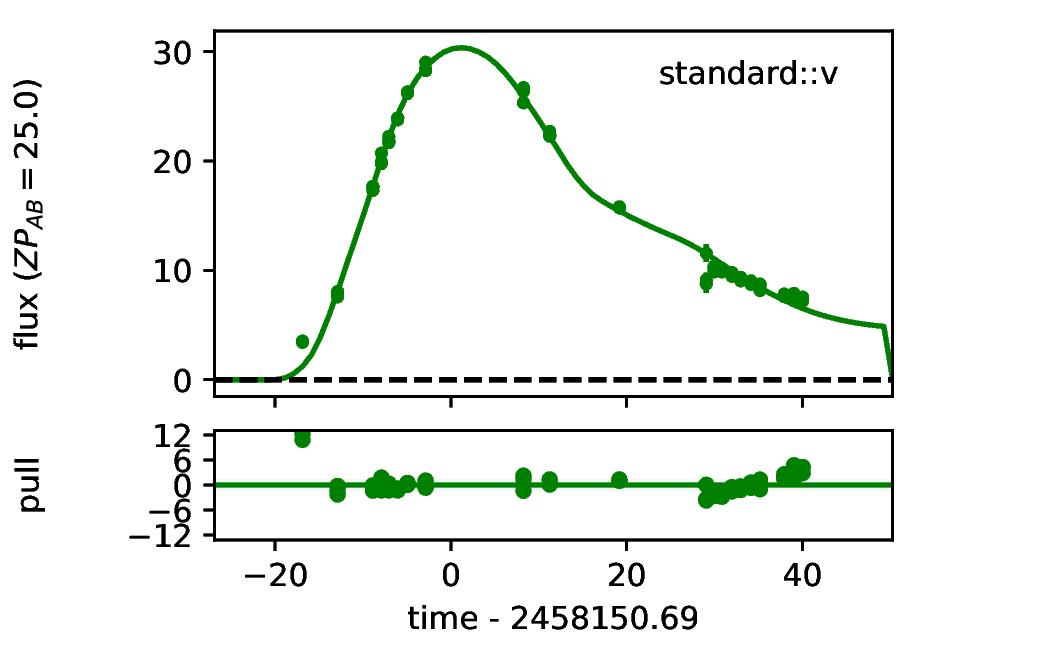}
      \caption{SN 2018gv ASAS-SN \textit{V}-band light curve. A SALT2 fitting model was applied to estimate the $t_\text{max}$.
              }
         \label{multi2gvsalt}
   \end{figure}

 SN 2018gv was discovered by \cite{2018ATel12063....1B} on UT 2018-01-15.68 at 16.5~mag in NGC 2525, a galaxy found in the NED Galaxy Catalogue, with a distance modulus ($m-M$) of 31.1~mag, 16.8~Mpc away \citep{2013AJ....146...86T}. Spectroscopic classifications with the LRIS spectrograph on the 10-m Keck I telescope and with EFOSC on the ESO New Technology Telescope at La Silla in the framework of ePESSTO\footnote{\url{http://www.pessto.org}} using SNID \citep{2007ApJ...666.1024B}, indicated that SN 2018gv was a very young, normal Type Ia supernova, 10-15 days before the maximum brightness. We used the ASAS-SN \textit{V}-band light curves from \cite{2014ApJ...788...48S} and \cite{2017PASP..129j4502K} to fit the parameters of a SALT2 model to the photometric light curve data (Figure~\ref{multi2gvsalt}). We measured a SALT2 shape parameter of $x_\text{1}=1.14 \pm 0.05$, which is consistent with the properties of Type Ia SNe \citep{2019ApJ...886..152Y}, a color parameter $c=2.51 \pm 0.03$ and we determined that SN 2018gv peaked on HJD 2458150.69 $\pm$ 0.03 in the \textit{V}-band, which is in strong agreement with the measurements of \cite{2019arXiv190310820Y}.\\
  
 Optical photometry of SN 2018gv was obtained using the 1.2~m Kryoneri telescope on HJD 2458141.4, nine days before the maximum light according to the SALT2 fit. The observations were conducted during three runs on the same night  because of interruptions due to the weather conditions (clouds). Images in the \textit{R} and \textit{I} bands were obtained simultaneously, with an exposure time of 30~s and a total time coverage of 118~minutes in each band. We kept the target near the same pixel of the detector using the telescope guiding system. The light curve from the first second seems to have some variations of 0.15~mag (Figure~\ref{multi3gv}) even after the subtraction of the seeing-correlated noise. We therefore used nine non-variable comparison stars to create a trendlist template to subtract the systematic effects. After the implementation of TFA, the rms was improved from 0.27 to 0.02~mag and the variations decreased to an amplitude of 0.04~mag in the \textit{R} band. The rms was also improved significantly in the \textit{I} band (Table \ref{seeingnoise}). On the third observing run the light curve shows a 0.05~mag bump (Figure~\ref{multi2gv}), which does not appear to be related to seeing or to systematic effects. We continued with PSF photometry, with the aim of validating our photometric results. The same trend with 0.05~mag amplitude was also apparent in the PSF photometry. We proceeded with the MCMC analysis to estimate the time-correlated noise. The rms of the binned residuals was computed to be 0.02~mag with a number of bins > 20. Therefore the 0.05~mag bump cannot be considered as a significant undulation after the red noise estimation due to its small amplitude. Most of the trends on the first run were due to seeing-correlated effects and were removed. The final light curve has an improved rms of 0.01~mag (Figure~\ref{multi1gv}). The mean intraday decline slope was measured to be 0~mag day$^{-1}$ in \textit{R} band and --0.13~mag day$^{-1}$ in \textit{I} band (Table \ref{measurements}).
  
\subsection{SN 2018zd}
SN 2018zd (otherwise known as Gaia18anr  or ATLAS18mix) was discovered by \cite{2018TNSTR.285....1I} on UT 2018-03-02.49 at 17.8~mag in NGC 2146, a galaxy found in the Ned Galaxy Catalogue with $m - M = 31.28$~mag, 18~Mpc away \citep{2012MNRAS.426.1185A}. It was classified by \cite{2018ATel11379....1Z} on UT 2018-03-05.7 as a Type IIn supernova, thirteen days before maximum light. We used the ASAS-SN \textit{V}-band light curves \citep{2014ApJ...788...48S, 2017PASP..129j4502K} to fit Nugent's templates \citep{2002PASP..114..803N}\footnote{\url{https://c3.lbl.gov/nugent/nugent_templates.html}.} and applied SNANA fitting code \citep{2009PASP..121.1028K} for Type IIn SNe. As SNe Type IIn show intrisic variety in their properties and the templates are based on only a few events (SN 1999el, SN 2006ez and SN 2006ix), the analysis resulted in a poor fit. Therefore, we applied a low-order polynomial fit and we estimated that the SN peaked on HJD 2458188.6, which is the first determination of the SN maximum. Observations were obtained at Kryoneri Observatory, initially based on the spectroscopic classifications as a young Type II SN during three photometric nights. According to the \textit{V}-band fit, it was observed during the maximum light in the \textit{R} and \textit{I} bands. Figure~\ref{multi1zd} shows the light curves for the first observing night.
  
The photometric analysis yielded a precision below 0.05~mag level and did not indicate any undulations in the light curves. We applied the TFA with a template of ten comparison stars, whose light constancy was checked using differential photometry. The estimated values for the intranight decline rate and the rms before and after the implementation of TFA are shown in the last two columns of Table \ref{measurements} for the \textit{R} and \textit{I} bands, respectively. Most of them were caused by seeing transitions as the second panel of each figure shows. On the second night there was a drop of $\sim0.2$~mag, which was subtracted with the TFA (first panel of Figure~\ref{multi2zd}). The drop was due to an image shift that was not perfectly corrected with the image alignment step. The first part of the light curve coverage of the second night of SN 2018zd is excluded from the analysis and Figure~\ref{multi2zd} (104 data points) due to the consecutive large variations of seeing. Most of the trends on the third run (Figure~\ref{multi3zd}) were due to variable seeing. The final light curve had an improved rms of 0.04~mag.
  
\begin{table*}[h]
\centering
\caption{Light curve scatter and slope measurements of the observed SNe}
\begin{tabular}{@{}lcccc@{}}
\toprule
Name  & 
\begin{tabular}[c]{@{}c@{}}Observation epoch$^a$\\ (HJD)\end{tabular} &
\begin{tabular}[c]{@{}c@{}}rms$_\text{0}$$^b$\\ (mag)\end{tabular} & \begin{tabular}[c]{@{}c@{}}rms$_\text{TFA}$$^c$\\ (mag)\end{tabular}  & \begin{tabular}[c]{@{}c@{}}$\alpha_\text{SN,TFA}$ $^d$\\(mag day$^{-1}$) \end{tabular}\vspace{0.1cm} \\ \midrule \midrule
SN 2018gv & 2458141.44 & 0.11, 0.27, 0.11 (\textit{R})  & 0.01, 0.02, 0.01 (\textit{R}) & 0.65 $\pm$ 0.36, $-$0.42 $\pm$ 0.21, $-$0.23 $\pm$ 0.10  (\textit{R})  \vspace{0.05cm}  \\
 & & 0.11, 0.26, 0.09 (\textit{I})  & 0.02, 0.05, 0.02 (\textit{I}) & 0.19 $\pm$ 0.23, $-$0.41 $\pm$ 0.20, $-$0.17 $\pm$ 0.05 (\textit{I}) \vspace{0.05cm}  \\\midrule
SN 2018zd & \begin{tabular}[c]{@{}c@{}} 2458187.43\\ 2458188.27 \\ 2458189.35\end{tabular} & \begin{tabular}[c]{@{}c@{}}0.13 (\textit{R}), 0.12 (\textit{I})\\ 0.54 (\textit{R}), 0.4 (\textit{I})\\ 0.11(R), 0.1 (\textit{I})\end{tabular}  & \begin{tabular}[c]{@{}c@{}}0.04 (\textit{R}), 0.04 (\textit{I})\\ 0.17 (R), 0.01 (\textit{I})\\ 0.04 (\textit{R}), 0.04 (\textit{I})\end{tabular}   & \begin{tabular}[c]{@{}c@{}}0.03 $\pm$ 0.01 (\textit{R}), 0.10 $\pm$ 0.06(\textit{I})\\ $-$0.19 $\pm$ 0.06(\textit{R}), 0.03 $\pm$ 0.01 (\textit{I})\\ $-$0.06 $\pm$ 0.02(\textit{R}), $-$0.03 $\pm$ 0.02 (\textit{I})\end{tabular}   \vspace{0.05cm}      \\\midrule
SN 2018hgc & \begin{tabular}[c]{@{}c@{}}2458407.51\\ 2458408.29\end{tabular}& \begin{tabular}[c]{@{}c@{}}0.72\\ 0.04\end{tabular}           & \begin{tabular}[c]{@{}c@{}}0.49\\ 0.01\end{tabular}               & \begin{tabular}[c]{@{}c@{}}0.31 $\pm$ 0.40\\ -0.18 $\pm$ 0.01 \end{tabular}          \vspace{0.05cm}         \\\midrule
SN 2018hhn &\begin{tabular}[c]{@{}c@{}}2458419.36\\ 2458420.26\\ 2458421.33\\ 2458423.41\end{tabular}  & \begin{tabular}[c]{@{}c@{}}0.03\\ 0.75\\ 1.4\\ 1\end{tabular} & \begin{tabular}[c]{@{}c@{}}0.02\\ 0.01\\ 0.07\\ 0.05\end{tabular} & \begin{tabular}[c]{@{}c@{}}0.36 $\pm$ 0.24\\ $-$0.10 $\pm$ 0.05\\ $-$0.37 $\pm$ 0.05\\ 0.11 $\pm$ 0.19\end{tabular} \vspace{0.05cm} \\\midrule
SN 2018hna & \begin{tabular}[c]{@{}c@{}}2458419.58\\ 2458423.58\end{tabular}& \begin{tabular}[c]{@{}c@{}}0.07\\ 0.22\end{tabular}           & \begin{tabular}[c]{@{}c@{}}0.01\\ 0.01\end{tabular}               & \begin{tabular}[c]{@{}c@{}}0.29 $\pm$ 0.14\\ 0.01 $\pm$ 0.01\end{tabular}         \vspace{0.05cm}          \\ \bottomrule
\end{tabular}
\caption*{$^a$The start of the time series.\\
$^b$The initial rms. \\
$^c$The rms after the implementation of TFA.\\
$^d$The intraday decline rate with respect to the linear fitting.}
\label{measurements}
\end{table*}
  
\subsection{SN 2018hgc}
SN 2018hgc (otherwise ASASSN-18xr, MLS181006:004205-023744 or ATLAS18wrj) was discovered at 17.9~mag on UT 2018-10-10 by the ASAS-SN group \citep{2018ATel12104....1B} in 2MASX J00420581-0237394 (otherwise WISEA J004205.96-023739.1), a galaxy found in the NED Galaxy Catalogue at a luminosity distance of 216~Mpc, which implies $m - M = 36.68$~mag. The luminosity distance was derived using the redshift and the Cosmology-Corrected Quantities $H_\text{o}$ =  73.00~km~s$^{-1}$~Mpc$^{-1}$, $\Omega_\text{matter}$ =   0.27 and $\Omega_\text{vacuum}$ =   0.73. On UT 2018-10-14, it was classified as a Type Ia supernova \citep{2018ATel12110....1D}. We used the ASAS-SN \textit{g}-band light curve to fit the parameters of a SALT2 model to photometric light curve data and to calculate the maximum date for the first time. We measured a SALT2 shape parameter of $x_\text{1}=4.8$ $\pm$ 1.1, which is consistent with the properties of Type Ia SNe \citep{2019ApJ...886..152Y}, a color parameter $c=25.19$ $\pm$ 0.61 and we determined that SN 2018hgc peaked on HJD 2458411.44 $\pm$ 0.32 in the \textit{g}-band (Figure~ \ref{salt2}). 
  
The photometric data were obtained with the Aristarchos telescope approximately four days before the maximum light. The target was observed for two consecutive nights with an exposure time of 60--90s in the \textit{VR} broadband filter. The reconstructed light curves for both nights seem smooth below 0.05~mag, after the systematic errors were subtracted. The signs of bumps with an amplitude of $\sim$ 0.04~mag cannot be considered as rapid variability due to the large scatter. The rms improved on both nights after the implementation of TFA (Table \ref{measurements}) and the intraday decline rate was $\sim0.31$ $\pm$ 0.40  and --0.18 $\pm$ 0.01 ~mag day$^{-1}$. Figure~\ref{multi1hgc} illustrates the results of the analysis of the photometric data.
 
\subsection{SN 2018hhn}
SN 2018hhn (otherwise known as ATLAS18wrh or Gaia18del) was firstly discovered by \cite{2018TNSTR1570....1G} in UGC 12222 on UT 2018-10-14, at 17.1~mag and was classified by \cite{2018TNSCR1585....1P} as a Type Ia supernova approximately 11 days before maximum light. The distance of its host is about $z=0.029$, which implies a $m - M = 35.11$~mag, 105~Mpc away \citep{2009ApJS..182..474S}. We use the ATLAS \textit{o}-band light curves \citep{2014ApJ...788...48S, 2017PASP..129j4502K} to fit the parameters of a SALT2 model to the photometric light curve data. We measured a SALT2 shape parameter of $x_\text{1}=0.51$ $\pm$ 0.69, which is consistent with the properties of Type Ia SNe \citep{2019ApJ...886..152Y}, a color parameter $c=-1.51$ $\pm$ 0.48 and we measured that SN 2018hhn peaked on HJD 2458417.10 $\pm$ 0.62 in \textit{o}-band, which is the first determination of the SN maximum (Figure~ \ref{salt2}).
  
We obtained observations between +2 to +6~days after maximum, using the 2.3~m Aristarchos telescope for four nights (Figures~\ref{multihhn27} and \ref{multihhn29} present the light curves). The photometric results (bottom panel of Figure~\ref{multihhn29}) indicate stable light curves with no undulations above 0.05~mag after the subtraction of the systematic errors with TFA. The rms was improved to 0.01--0.07~mag and the intraday decline rate was measured to range between $-$0.37 $\pm$ 0.05 to 0.36 $\pm$ 0.24~mag day$^{-1}$ over the four observing nights.
  
\subsection{SN 2018hna}
SN 2018hna (otherwise known as Gaia18dff or ZTF18acbwaxk) was discovered by \cite{2018TNSTR..57....1I} in UCG 7534 on 2018-10-23, at 16.3~mag and was classified by  \cite{2018ATel12258....1P} and \cite{2019ATel12897....1R}, 39 days later, as a normal rising Type II supernova. It appeared to have a low absolute magnitude and long rise time. It was located at a distance of 10.5~Mpc \citep{2013AJ....145..101K}, which implies a $m - M = 30.11$~mag, hosted by a possible metal poor faint dwarf galaxy found in the Ned Galaxy Catalogue. We used the ATLAS \textit{o}-band light curve and attempted to fit Nugent's templates \citep{2002PASP..114..803N} for Type II and IIn (based on blackbody templates from \citet{2002ApJ...573..144D}) as well as Type IIP/IIL templates from \citet{2011ApJ...738..162S} and Type IIP, IIn and II-pec from SNANA software \citep{2009PASP..121.1028K}. None of the templates provided a good fit covering the early and late evolution of its light curve. According to \cite{2019ApJ...882L..15S}, SN 2018hna is a 1987A-like event object, which is relatively rare, and not well represented by normal Type IIP SN templates. For this reason, we derived the peak epoch after fitting a low-order polynomial curve to the ATLAS data and HJD 2458499.1 is adopted as the epoch of maximum.
  
Two nights of photometric \textit{VR} observations were obtained with the 2.3-m Aristarchos telescope at Helmos Observatory, 80 and 76~days before the peak brightness according to the polynomial fit. The light curves of the SN (Figure~\ref{multihna}), after the implementation of TFA for both nights, do not show any significant undulation, as the rms was improved to 0.01~mag. The intraday slope ranged from 0.01 $\pm$ 0.01 to 0.29 $\pm$ 0.14~mag day$^{-1}$(Table \ref{measurements}).
    
\section{Discussion and conclusions}

The exact shape of the light curves of SNe in the early stages contains critical information about the properties of the progenitors (initial radii, explosion mechanisms), the existence and the composition of the circumstellar environment and the characteristics of any companion stars \citep{2010ApJ...708..598P, 2011ApJ...728...63R, 2016ApJ...826...96P, 2019MNRAS.483.3762K}. Theoretical models of SNe do not predict the behavior of SNe light curves on short timescales (minutes or hours), which might be the key observation for studying the progenitors of SNe and their circumstellar material. 

Theoretically, the light emanating from SNe originates from the following: a) radioactive decay of nickel and its associated daughter products in the ejecta, b) energy that comes from the collision with dense circumstellar material, which converts the kinetic energy of the ejecta into radiative energy \citep[e.g. SN 2010jl,][]{2012AJ....144..131Z, 2013A&A...555A..10T, 2013ApJ...769...67P, 2016ApJ...826...96P}, c) shock heating caused by the interaction of the ejecta with the companion star, in the case of binary systems and d) shock-deposited energy from the initial explosion. The key feature of these events is that the energy deposition and subsequent radiation do not need to follow a smooth, continuous form but may exhibit discrete, brief events as the 1--2 day bumps that have already been reported in a number of young SNe. Recently, models produced by \cite{2018MNRAS.476.2840M} predict \textit{u} and \textit{g} band bumps on timescales of days in the early stages of SNe due to high-density structures in the circumstellar medium caused by the wind acceleration close to the progenitor. 

Supernova explosions have mean expansion velocities on the order of 2,000-10,000~km s$^{-1}$. If we assume that the outer edge of the explosion moves at $10^{4}$~km s$^{-1}$, it will cover a distance of $2\times10^{15}$~cm over several weeks. Therefore, it is more likely for supernovae to exhibit features of variability during the first week after the explosion. Moreover, SNe light curves with blue bumps in the early evolutionary stages are now observed and reported more often, due to the increased observing cadence of many time-domain surveys; however, the origin of these events remains controversial.  A recent example is the Type II SN 2017gmr \citep{2019ApJ...885...43A}: SWIFT and ground based observations soon after the explosion revealed a short timescale (below one day) bump at a scale of 0.1-0.2~mag in \textit{U} and \textit{B} bands, which has not been further investigated. The observation of the intraday behavior of SNe is one of the keys for opening a new window to the study of SNe light curves and their properties, which at the moment lack predictions from theoretical models.

We have presented the first systematic, ground-based, high-cadence photometric study of early-time light curves of five bright SNe. We suggested methods adopted from the exoplanet community that can be applied to SNe light curve analysis to improve photometric precision and to validate the bumps that have been recently reported and we have demonstrated that these techniques do work. Even though we did not find any significant bumps in our observed light curves, undulations in bluer bands cannot be ruled out. The observation and analysis of the SNe light curves included four main steps, which we propose as a methodology for identifying early bumps: 

\begin{enumerate}
\item The acquisition of a large number of calibration flats and biases and the utilization of the guiding system to keep the target on the same pixel. 

\item The implementation of both optimal aperture photometry and PSF photometry to create the light curves. Optimal aperture photometry seemed to be sufficient for SNe well separated from the center of the host galaxy. 

\item The subtraction of seeing variations from the initial light curves, which cause variability, following the procedure described in \cite{2007MNRAS.375.1449I}. 

\item The subtraction of systematic effects and the improvement of the photometric accuracy below 0.05~mag using the Trend Filtering algorithm \citep{2016A&C....17....1H}.

\end{enumerate}

Our main results are the following:

\begin{enumerate}

      \item SALT2, Type II templates, and low-order polynomial fits were applied to the ASAS-SN and ATLAS light curves in order to calculate the time of peak. A photometric classification of the Type of SN 2018gv, SN 2018hgc and SN 2018hhn was derived from the template light curve fitting. The dates of light curve maximum for SNe 2018zd and 2018hhn have been reported for the first time. SN 2018hna showed a poor fit due to its long rise time, which cannot be predicted by the present templates.
      \item We measured the intraday slope of the early light curves by applying a linear function to each reconstructed light curve. The estimation of the intraday slope ranged from -0.37 $\pm$ 0.05 to 0.36~mag $\pm$ 0.24 day$^{-1}$.
      \item For SN 2018gv, which indicated undulations at a level of 0.05~mag, a MCMC analysis was performed to account for both white and correlated noise (red) in order to quantify and constrain the accuracy of the photometric results. The bump did not appear to be significant after the noise estimation.
      \item We achieved a photometric precision of 1--4~mmag per target on all nights. The sources did not show any sign of undulation above 0.05~mag after the implementation of the above methods during our observations in the \textit{R}, \textit{I}, and \textit{VR} bands. This does not rule out the existence of possible undulations in bluer bands. 
   \end{enumerate}
   
   A key component of future work will be to investigate undulations in the bluest bands with the tools that have been presented in this work and confidently identify structure in the early time light curves. We encourage the SN community to undertake ground-based, high-cadence photometric surveys and obtain blue photometric observations of different types of SNe as close as possible to the explosion time in order to understand whether bumps are ubiquitous. Wide-field transient surveys will discover and monitor nearby infant SNe over the next few years and will provide us with more instances of early SN discoveries, which are needed to investigate intraday undulations. A large sample of young SNe will enable stringent constraints on the behavior of SNe light curves and will reveal properties of the explosion mechanisms and the circumstellar environment of every type of SNe.

\begin{acknowledgements}
We acknowledge support of this work by the project "PROTEAS II" (MIS 5002515), which is implemented under the Action "Reinforcement of the Research and Innovation Infrastructure", funded by the Operational Programme "Competitiveness, Entrepreneurship and Innovation" (NSRF 2014-2020) and co-financed by Greece and the European Union (European Regional Development Fund). AZB acknowledges funding from the European Research Council (ERC) under the European Union’s Horizon 2020 research and innovation programme (grant agreement number 772086). This research has made use of NASA’s Astrophysics Data System Bibliographic Services. This research has made use of the SIMBAD database, operated at CDS, Strasbourg, France. This research has made use of the NASA/IPAC Extragalactic Database (NED) which is operated by the Jet Propulsion Laboratory, California Institute of Technology, under contract with the National Aeronautics and Space Administration. We acknowledge ESA Gaia, DPAC and the Photometric Science Alerts Team (\url{http://gsaweb.ast.cam.ac.uk/alerts}). We thank the ATLAS science team for providing their light curve data. The Asteroid Terrestrial-impact Last Alert System (ATLAS) project is primarily funded to search for near earth asteroids through NASA grants NN12AR55G, 80NSSC18K0284, and 80NSSC18K1575; byproducts of the NEO search include images and catalogs from the survey area. The ATLAS science products have been made possible through the contributions of the University of Hawaii Institute for Astronomy, the Queen's University Belfast, the Space Telescope Science Institute, and the South African Astronomical Observatory.
\end{acknowledgements}

\bibliographystyle{aa}
\bibliography{references}

\begin{appendix}

\section{SNe light curves for all observing runs}

\begin{figure*}[h]
\centering
\includegraphics[width=1\textwidth]{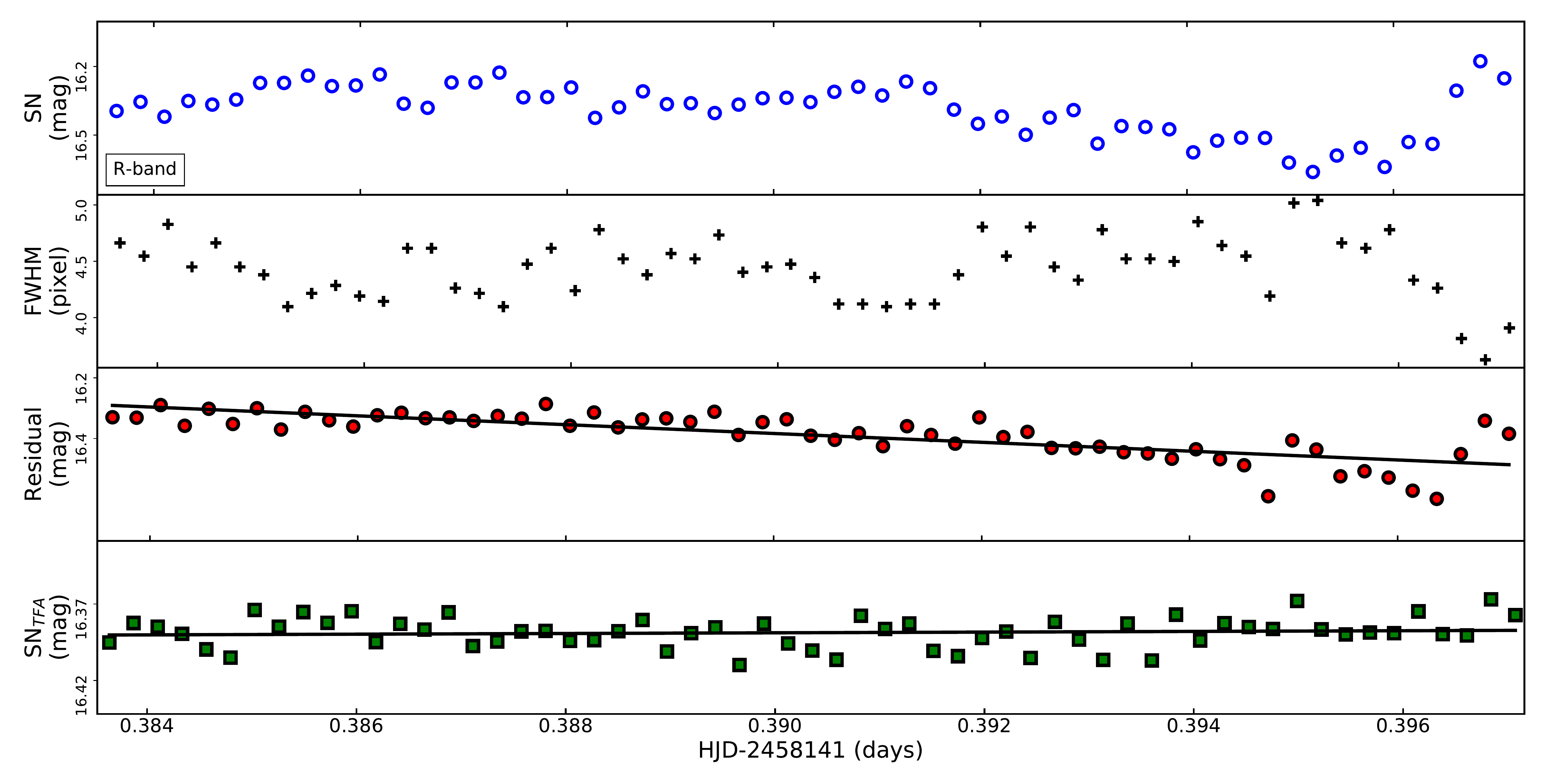}
\bigbreak
\includegraphics[width=1\textwidth]{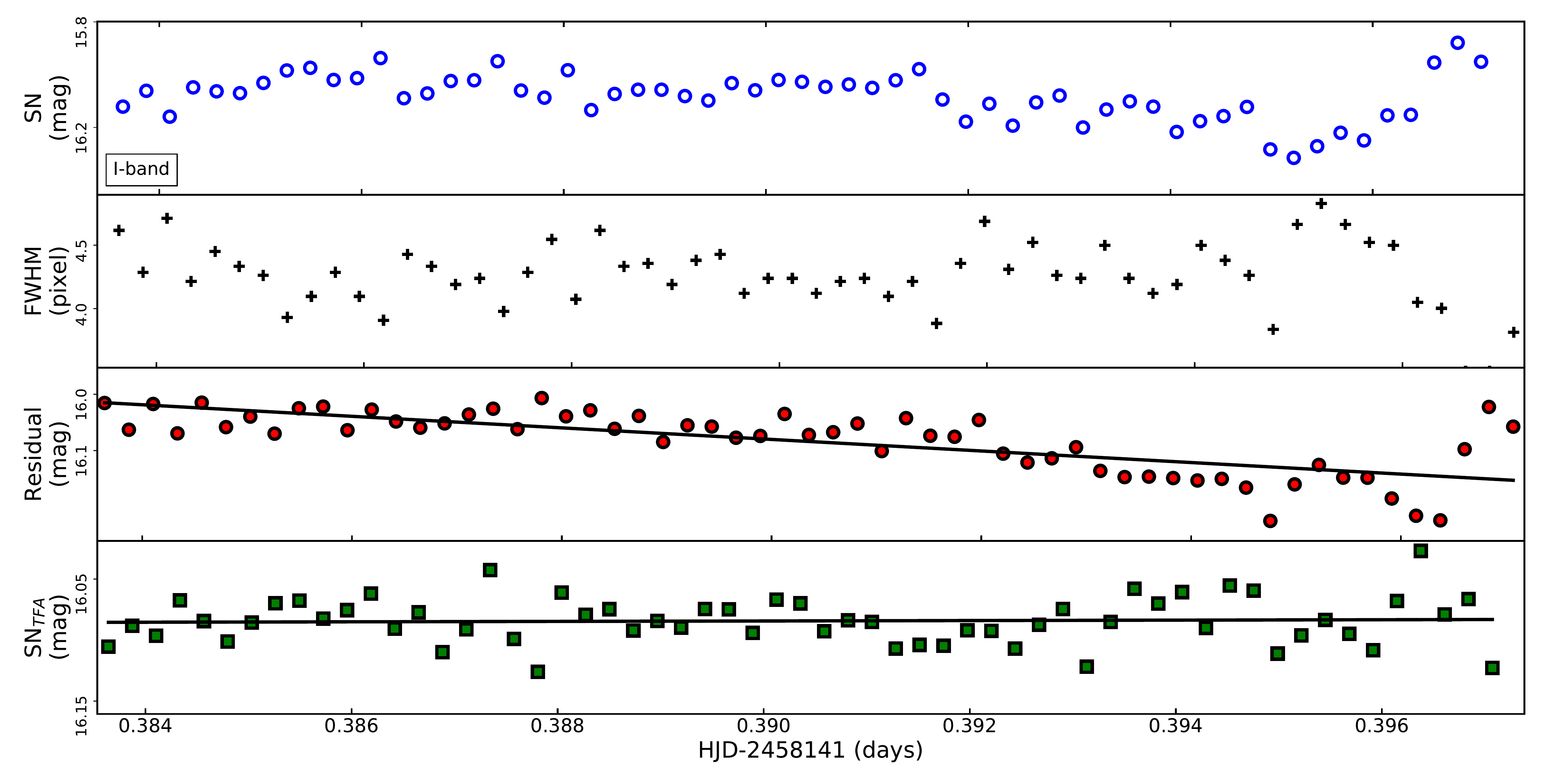}
\caption{Light curves of SN 2018gv in \textit{R} and \textit{I} bands during the first part of the first night of observations.}\label{multi1gv}
\end{figure*}

\begin{figure*}[h]
\centering
\includegraphics[width=1\textwidth]{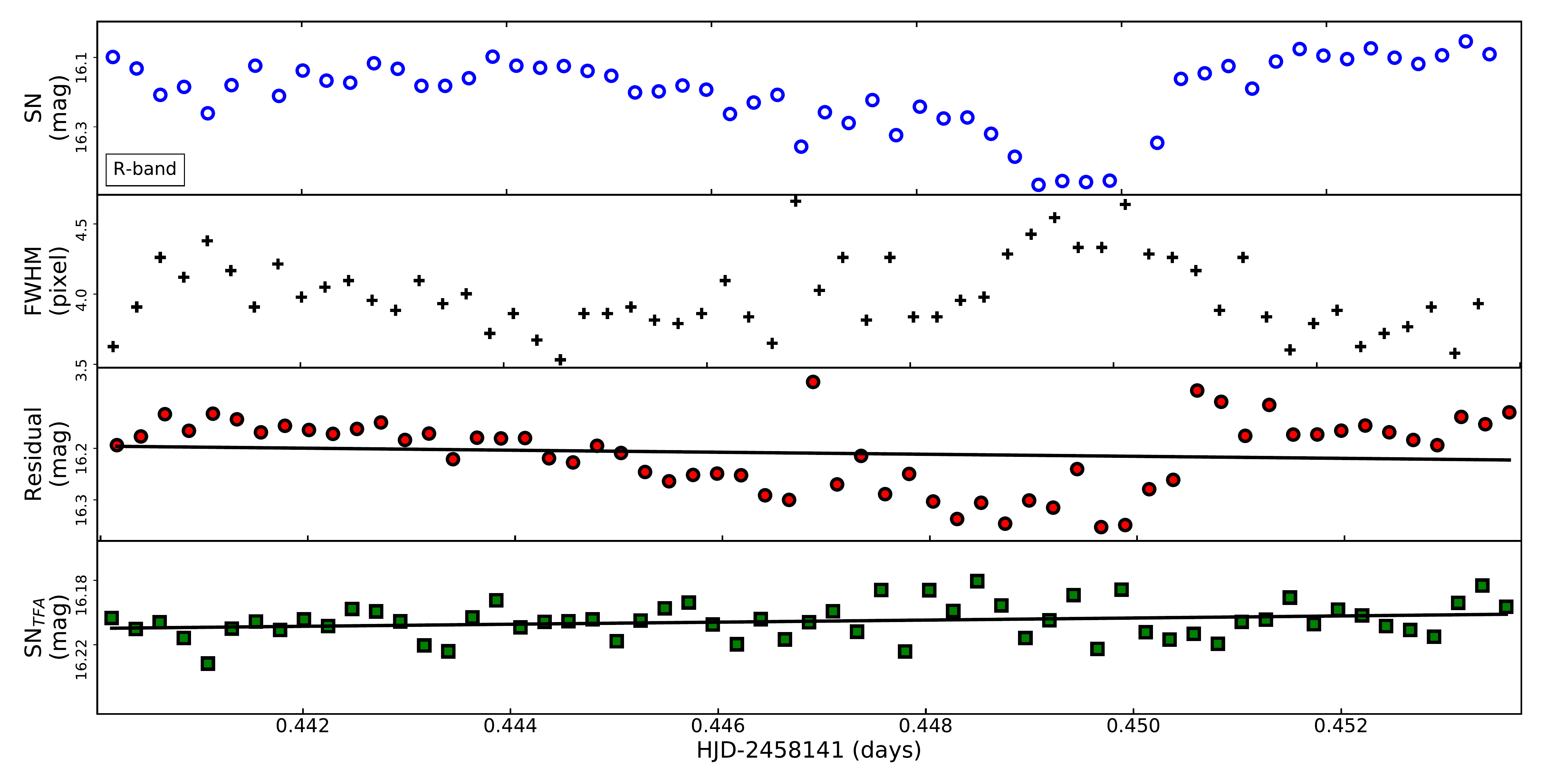}
\bigbreak
\includegraphics[width=1\textwidth]{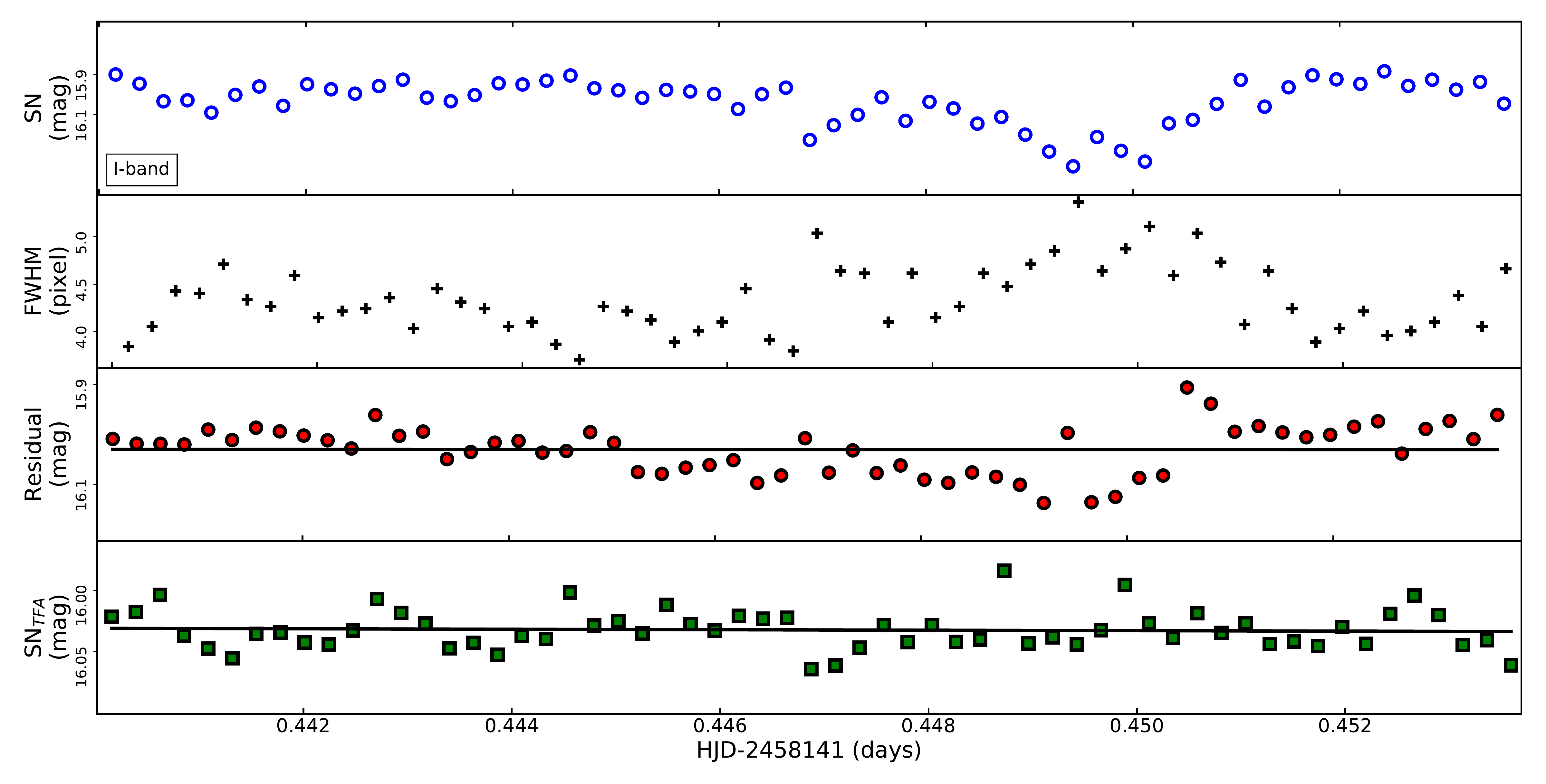}
\caption{Same as Figure~\ref{multi1gv}, but for the second part of the first night of observations.}\label{multi3gv}
\end{figure*}

\begin{figure*}[h]
\centering
\includegraphics[width=1\textwidth]{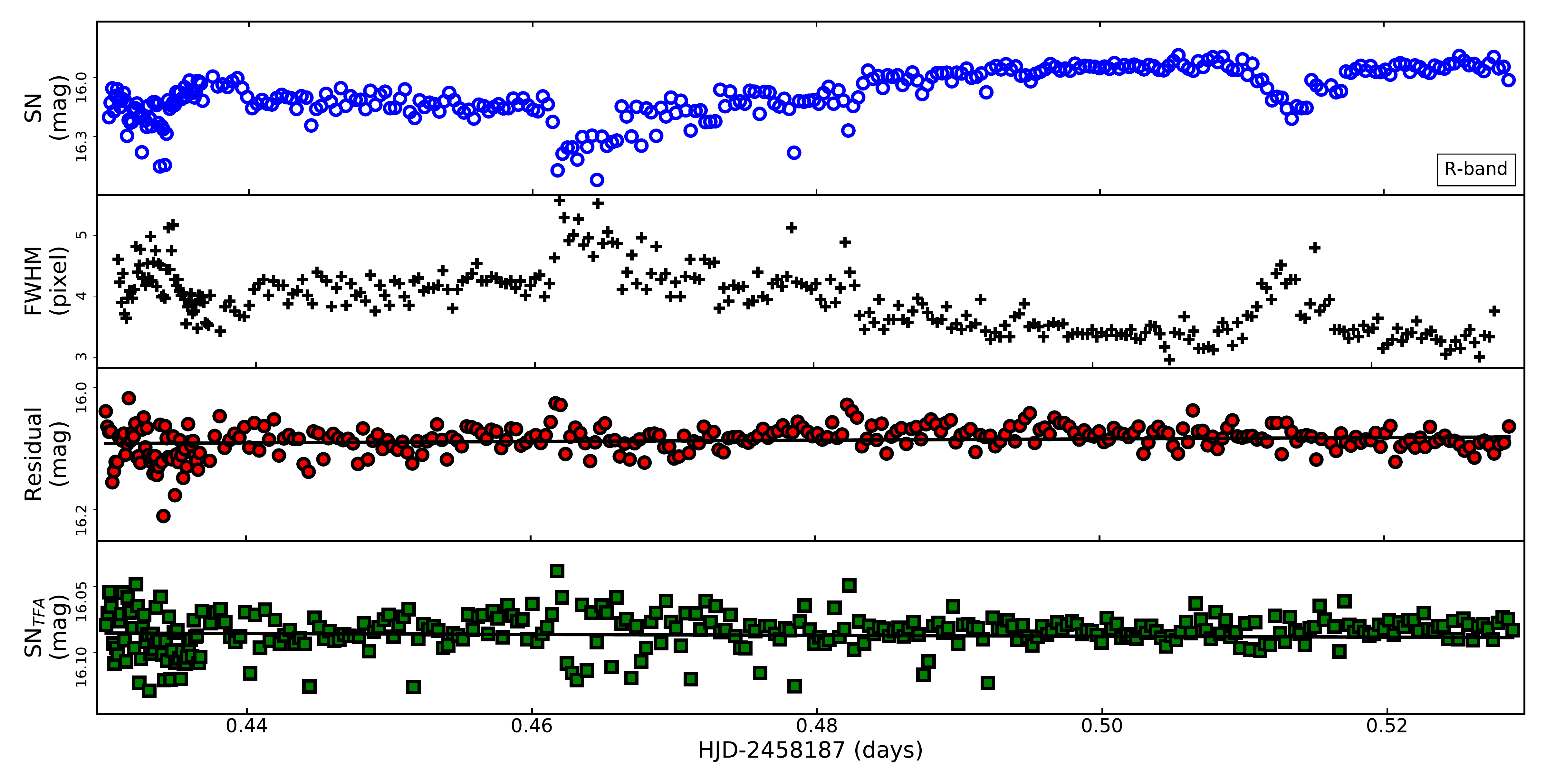}
\bigbreak
\includegraphics[width=1\textwidth]{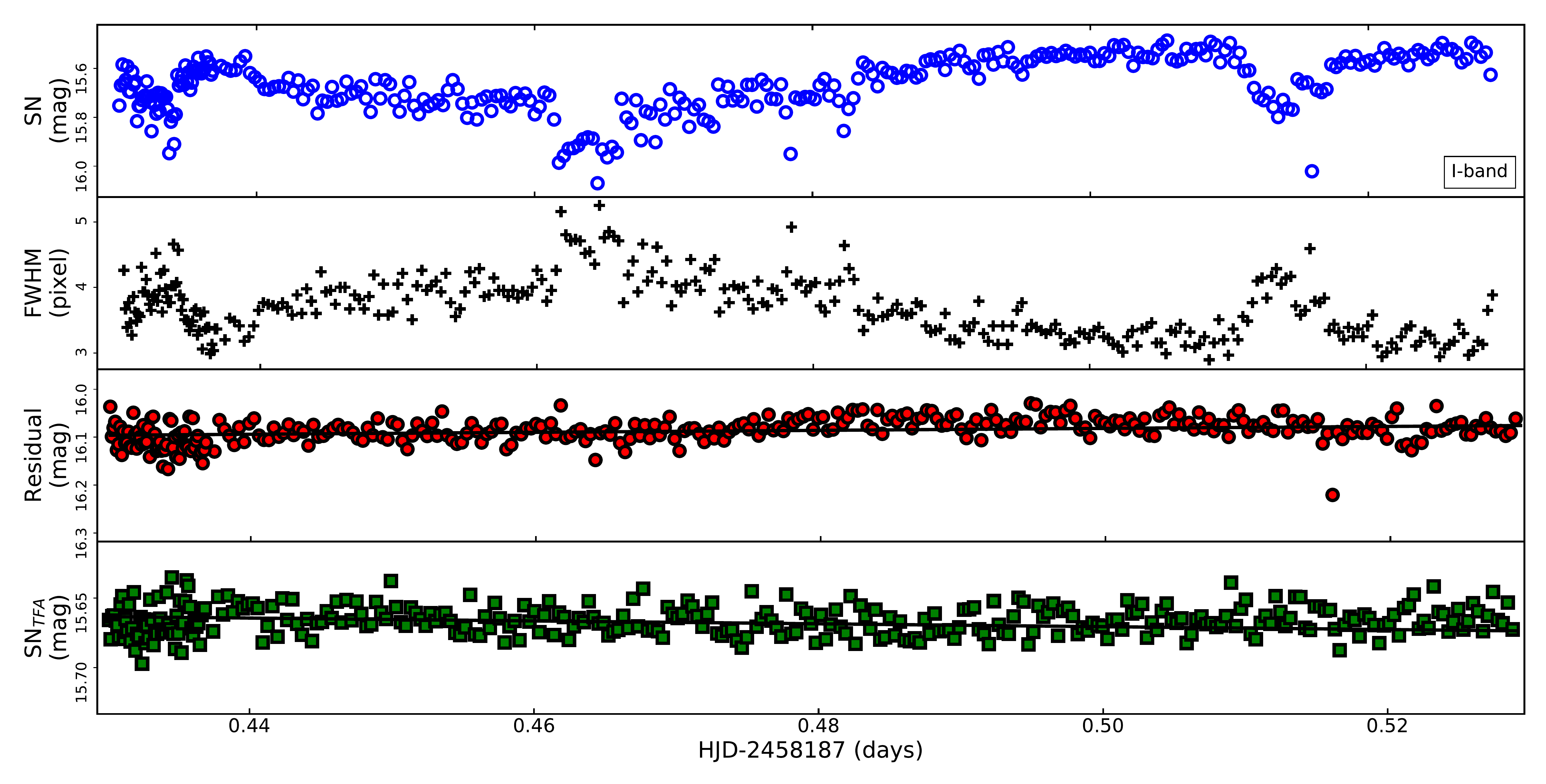}
\caption{Same as Figure~\ref{multi1gv}, but for SN 2018zd for the first night of observations.}\label{multi1zd}
\end{figure*}

\begin{figure*}[h]
\centering
\includegraphics[width=1\textwidth]{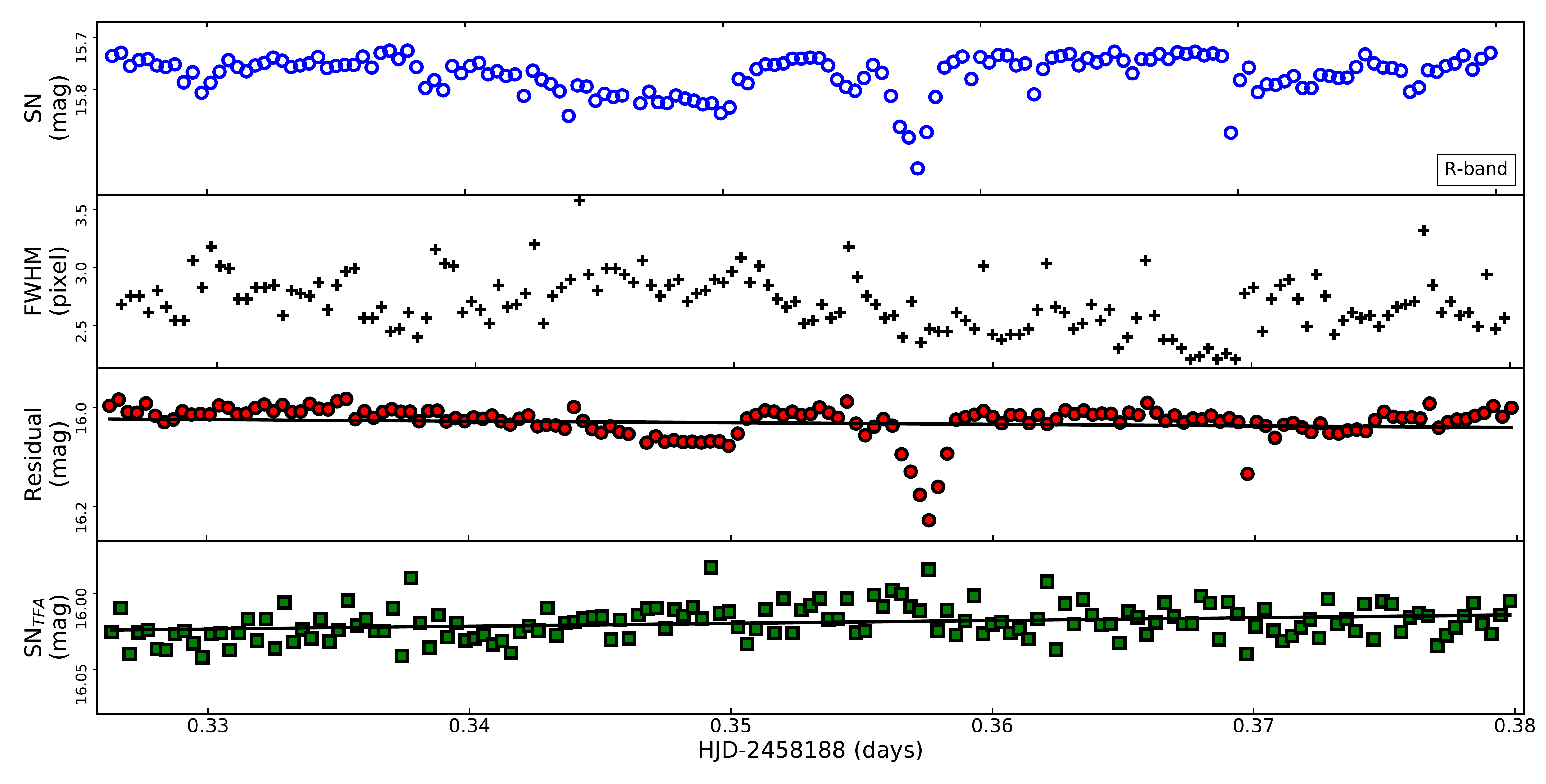}
\bigbreak
\includegraphics[width=1\textwidth]{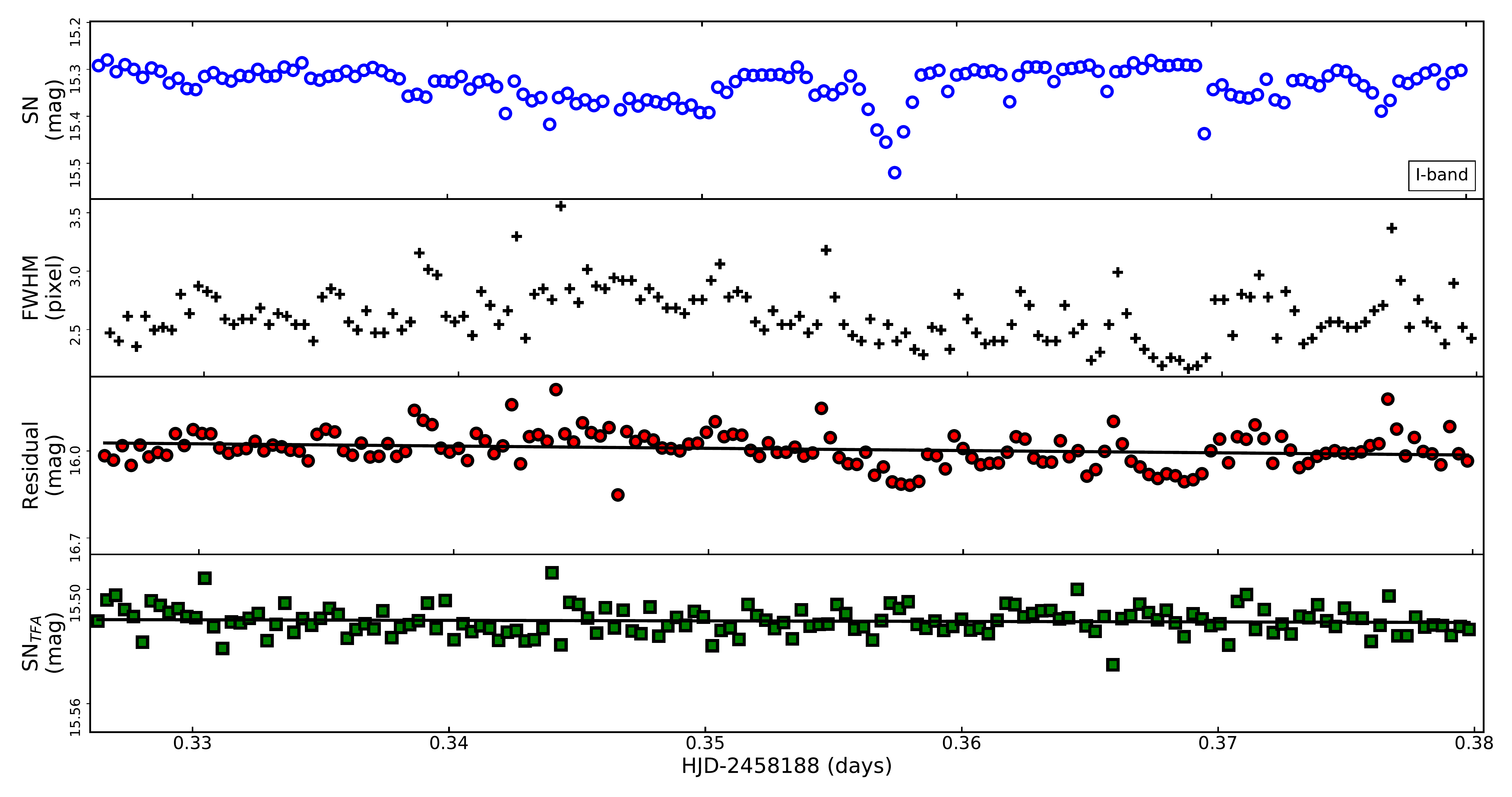}
\caption{Same as Figure~\ref{multi1gv}, but for SN 2018zd for the second night of observations.}\label{multi2zd}
\end{figure*}

\begin{figure*}[h]
\centering
\includegraphics[width=1\textwidth]{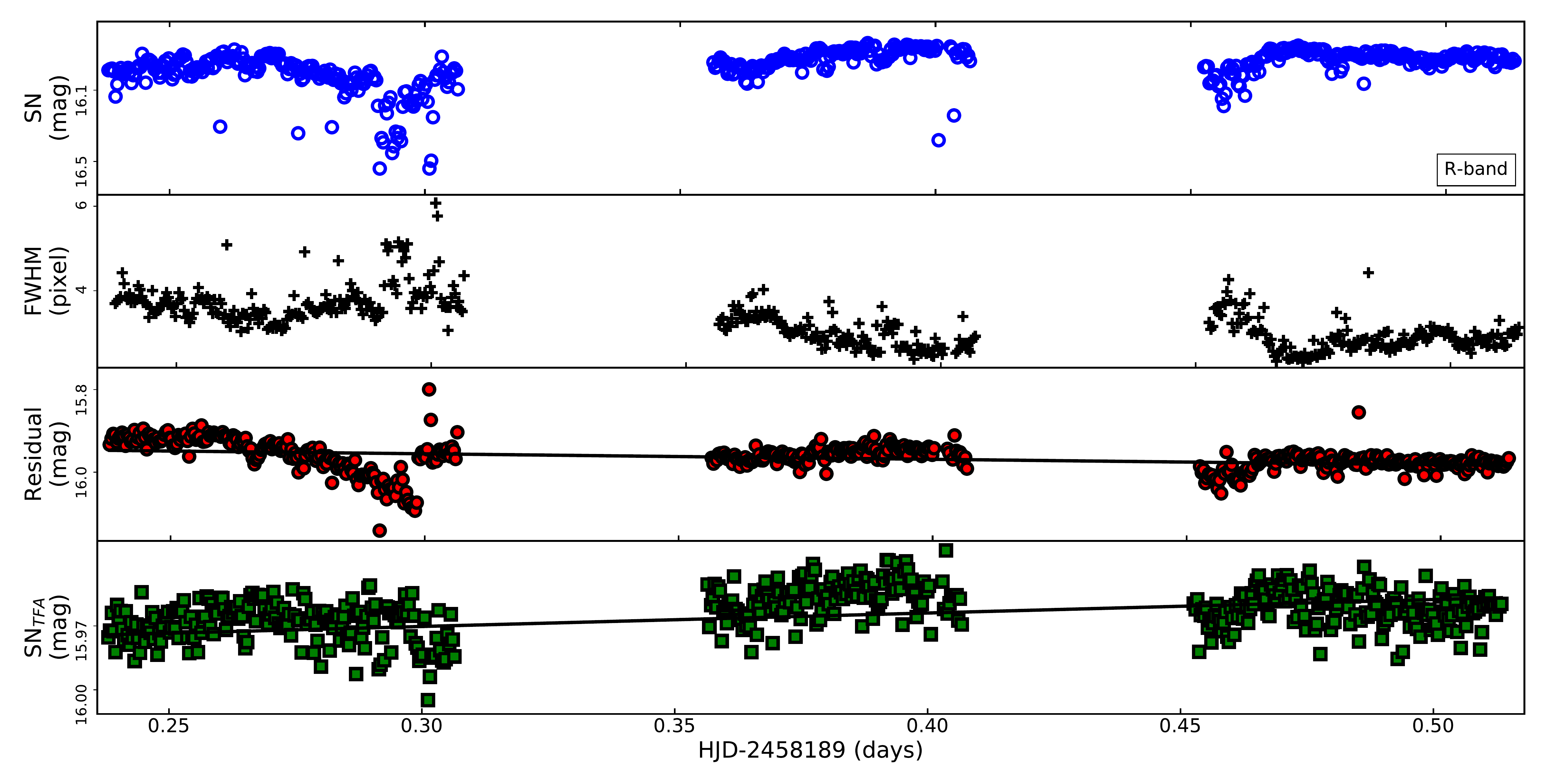}
\bigbreak
\includegraphics[width=1\textwidth]{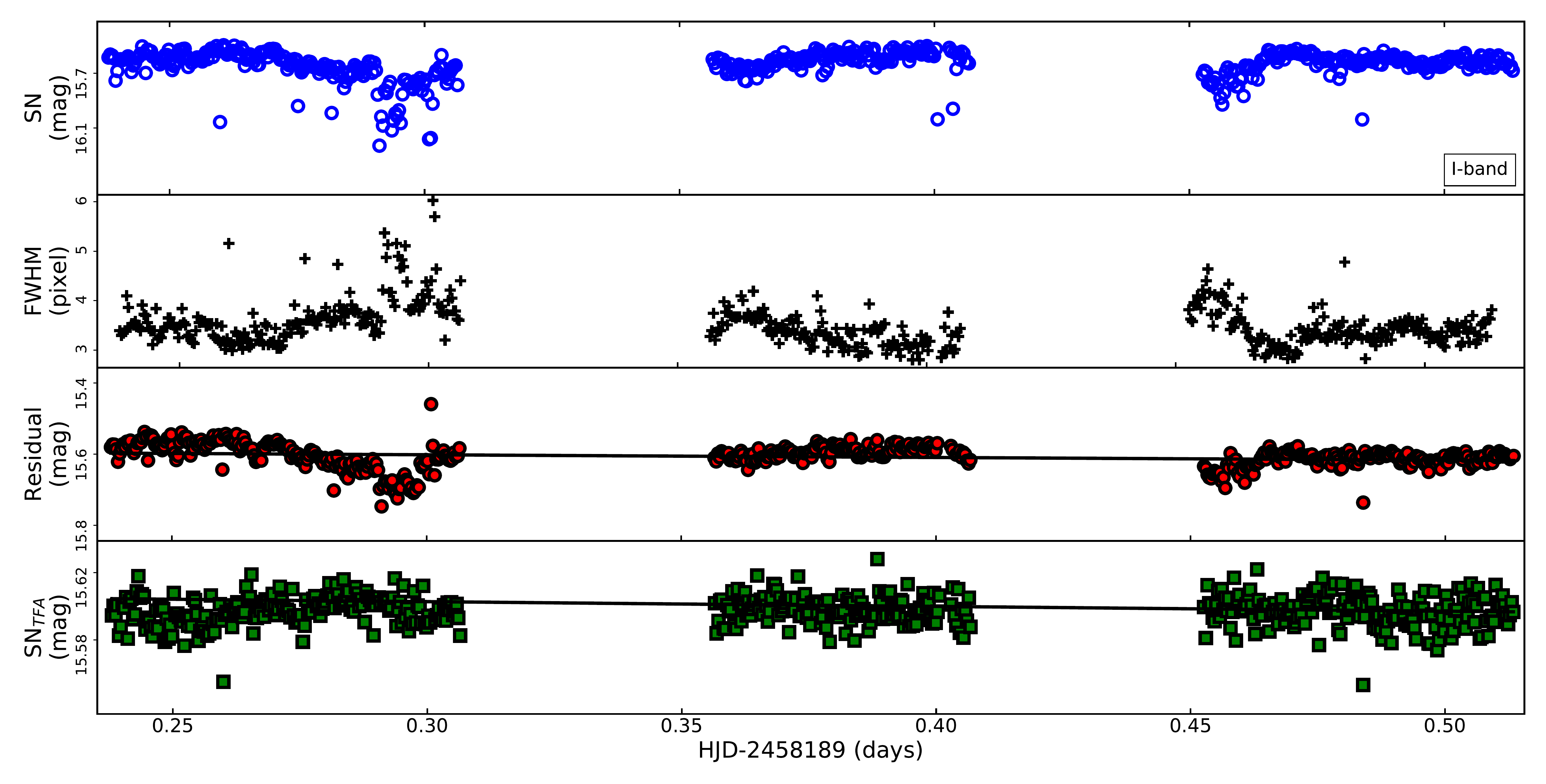}
\caption{Same as Figure~\ref{multi1gv}, but for SN 2018zd for the third night of observations.}\label{multi3zd}
\end{figure*}

\begin{figure*}[h]
\centering
\includegraphics[width=1\textwidth]{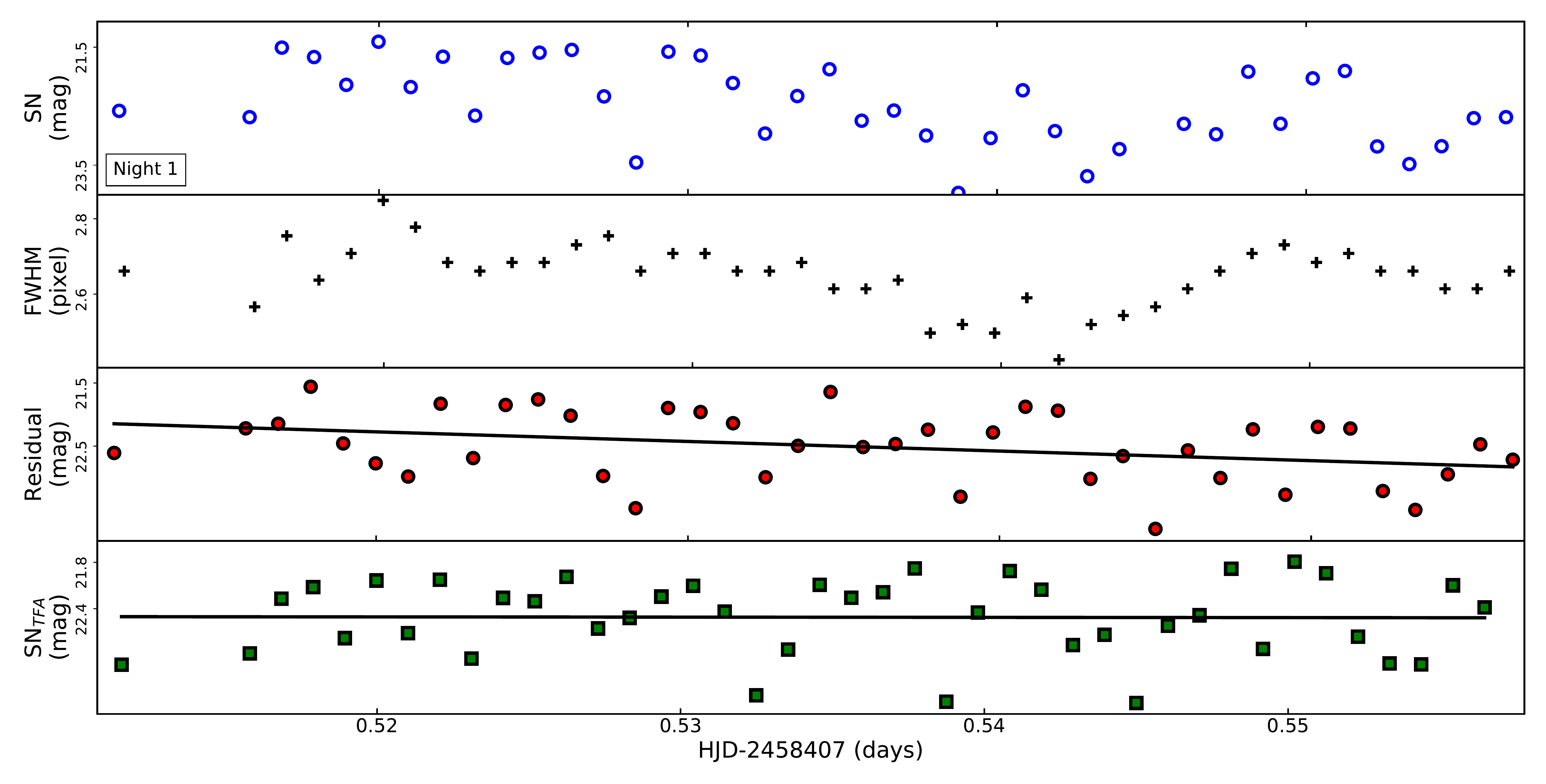}
\bigbreak
\includegraphics[width=1\textwidth]{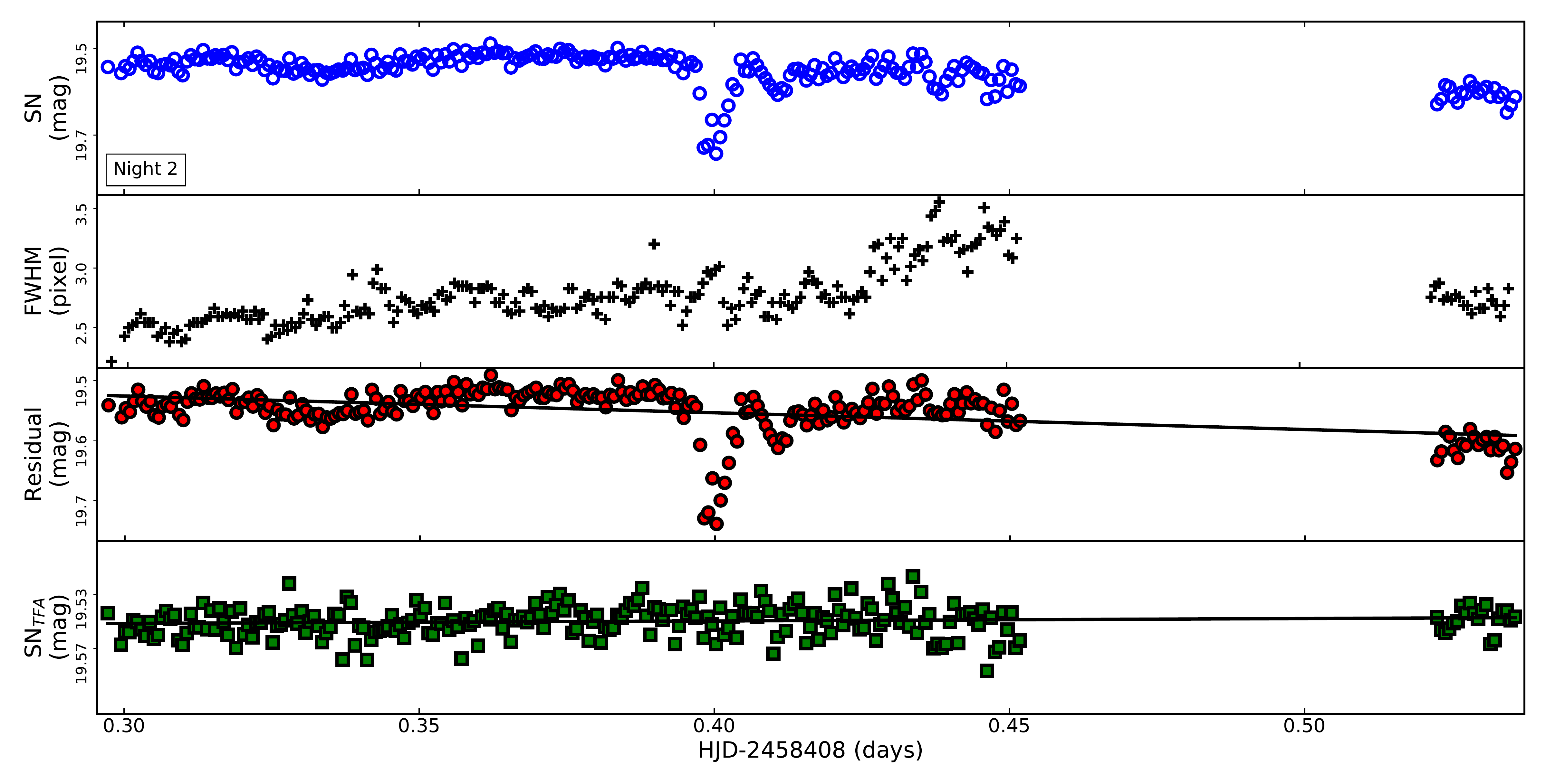}
\caption{Light curves of SN 2018hgc in the \textit{VR} band during two consecutive nights, respectively.}\label{multi1hgc}
\end{figure*}

\begin{figure*}[h]
\centering
\includegraphics[width=1\textwidth]{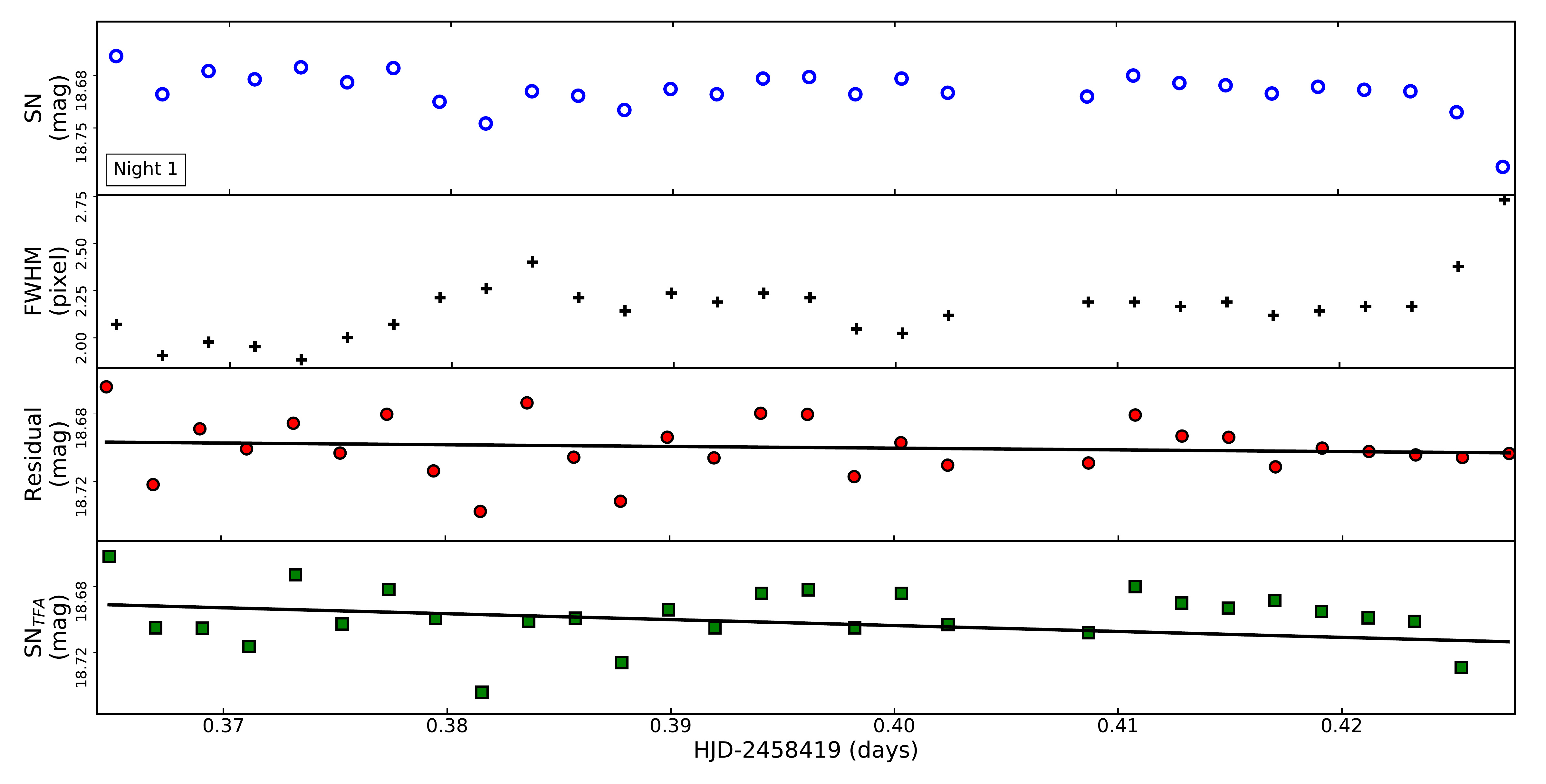}
\bigbreak
\includegraphics[width=1\textwidth]{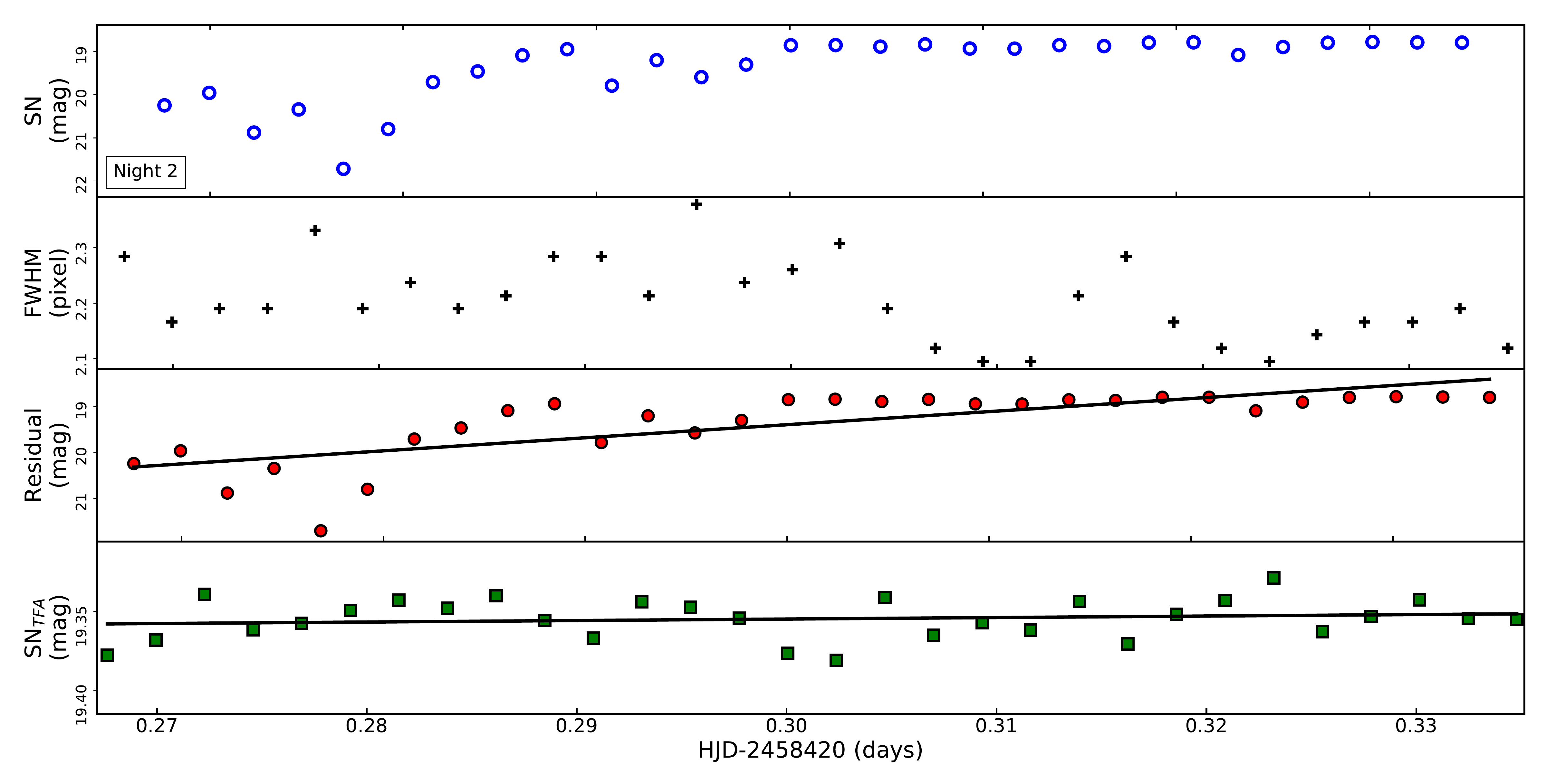}
\caption{Same as Figure A.6, but for SN 2018hhn for the first two consecutive nights, respectively.}\label{multihhn27}
\end{figure*}

\begin{figure*}[h]
\centering
\includegraphics[width=1\textwidth]{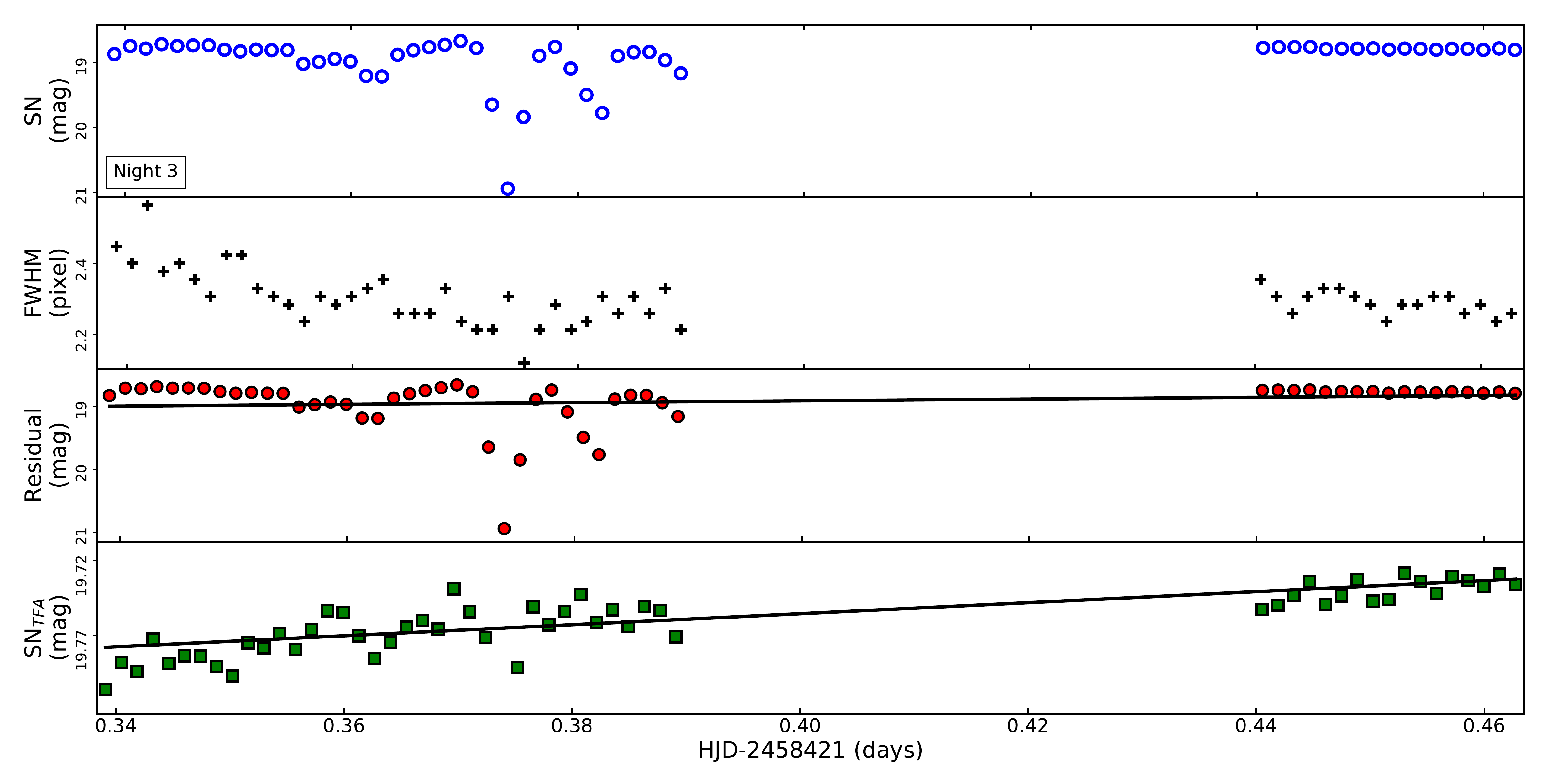}
\bigbreak
\includegraphics[width=1\textwidth]{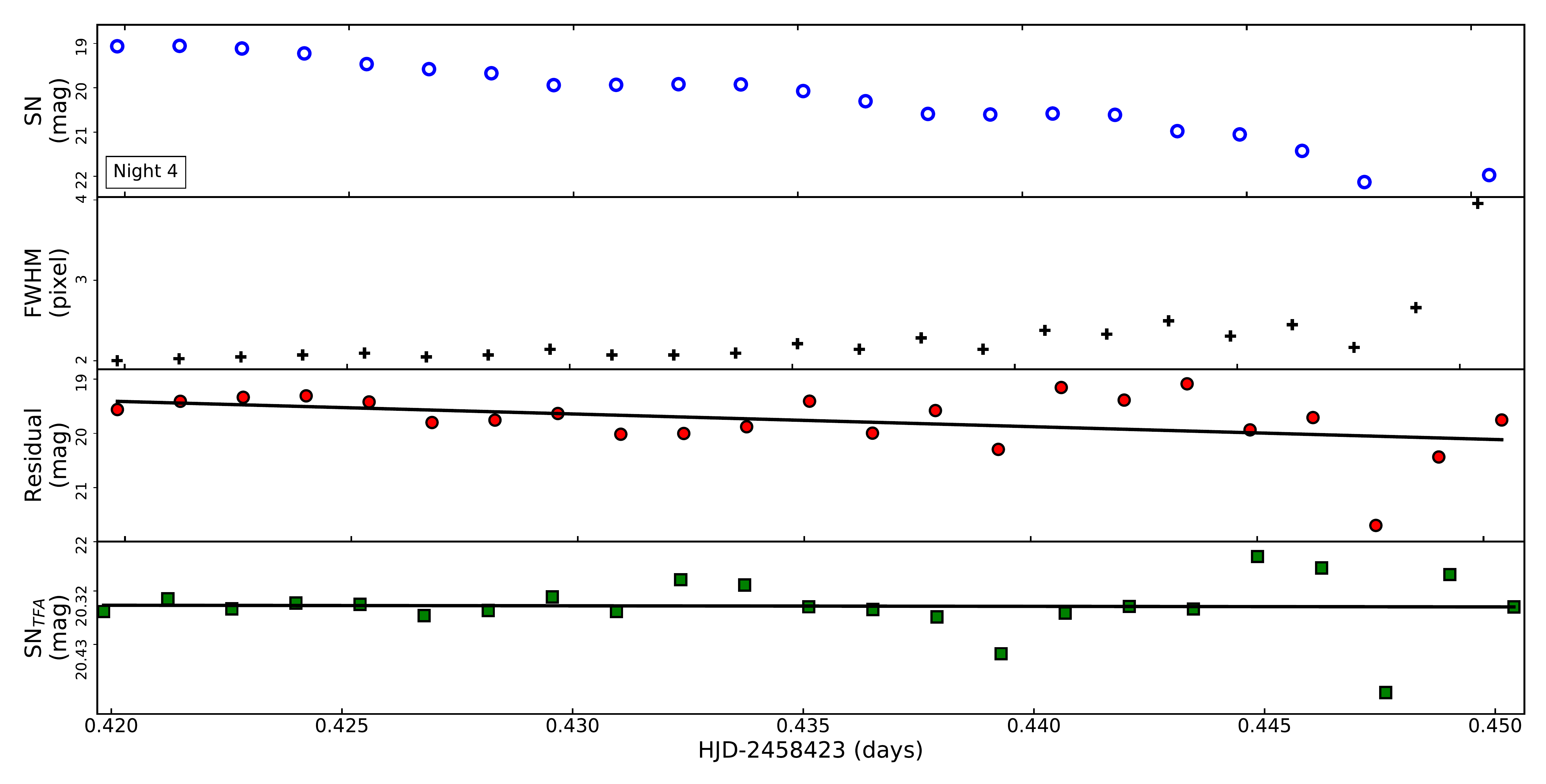}
\caption{Same as Figure~\ref{multi1hgc}, but for SN 2018hhn for the third and fourth night, respectively.}\label{multihhn29}
\end{figure*}

\begin{figure*}[h]
\centering
\includegraphics[width=1\textwidth]{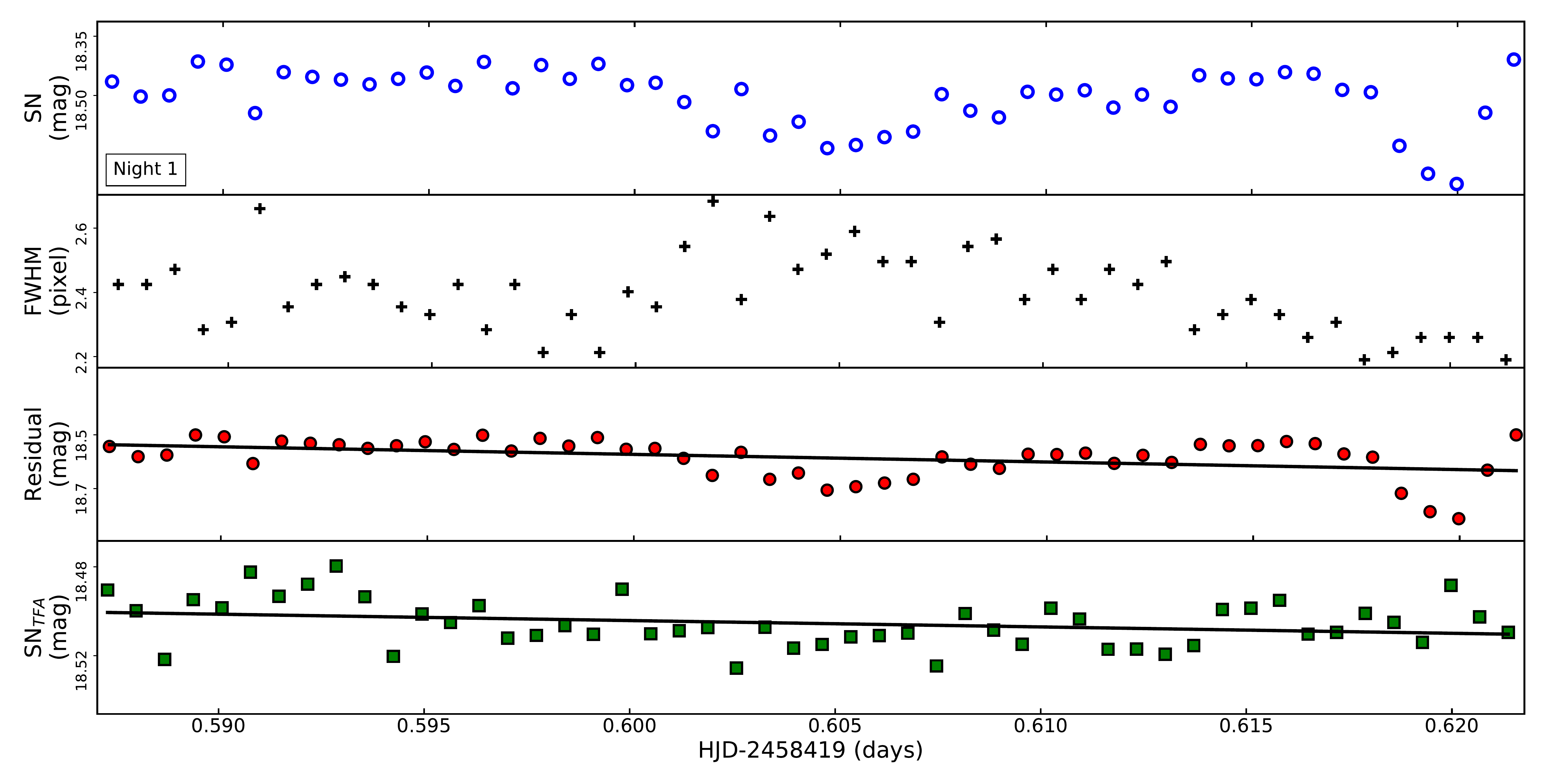}
\bigbreak
\includegraphics[width=1\textwidth]{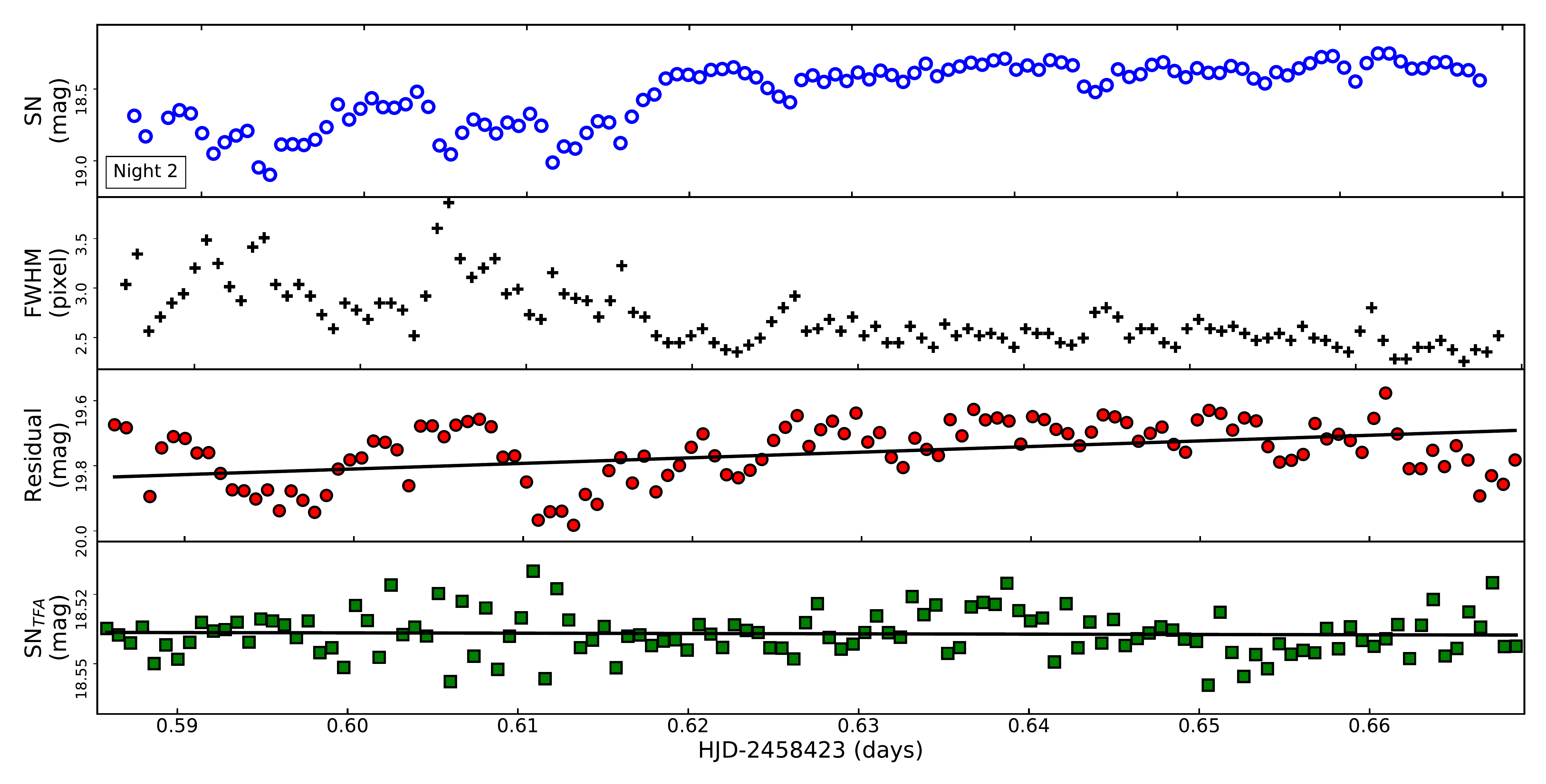}
\caption{Same as Figure~\ref{multi1hgc}, but for SN 2018hna for the first two nights.}\label{multihna}
\end{figure*}

\begin{figure*}
\centering
\subfloat[Run 1]{
  \includegraphics[width=57mm]{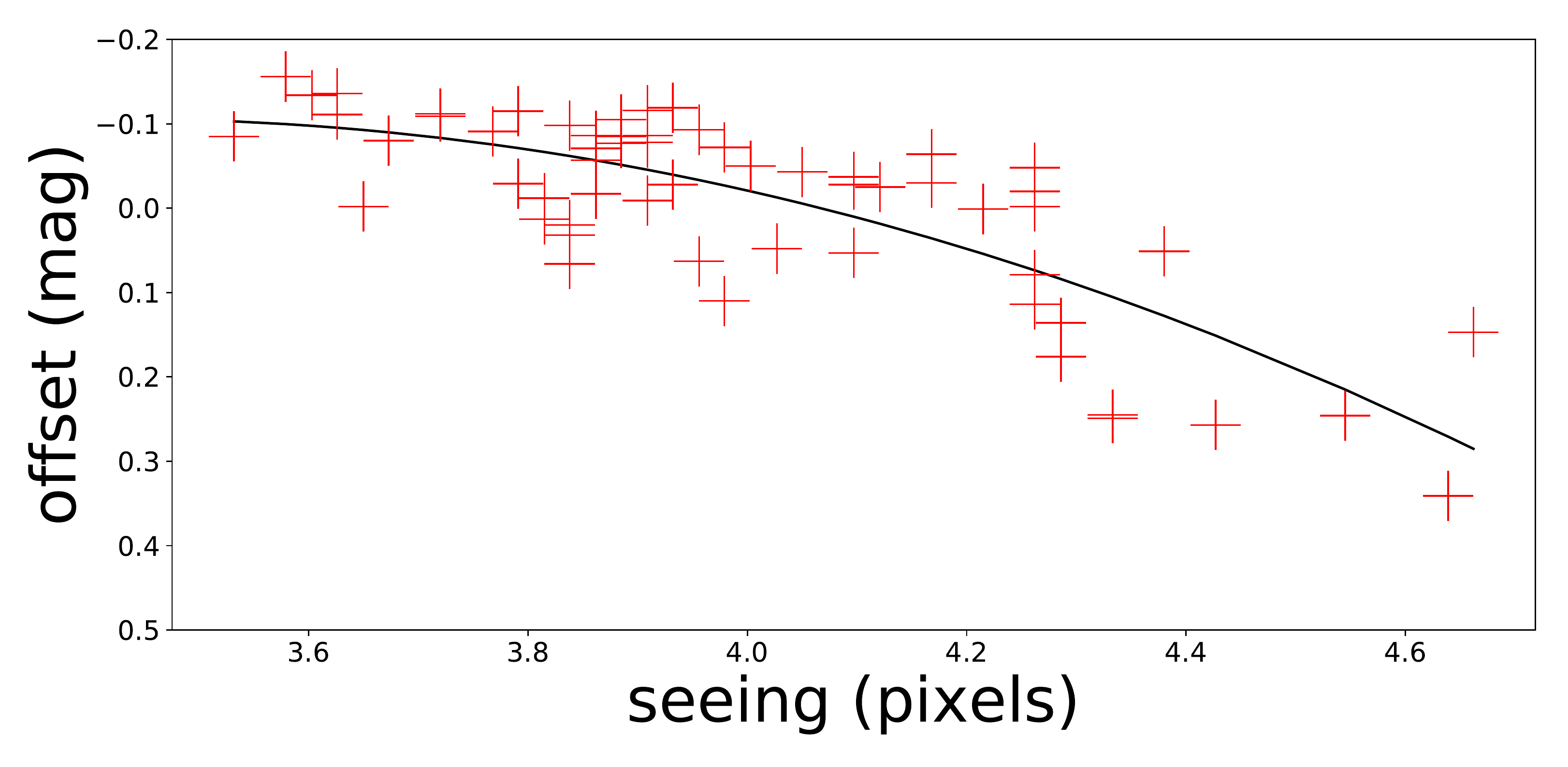}
}
\subfloat[Run 2]{
  \includegraphics[width=57mm]{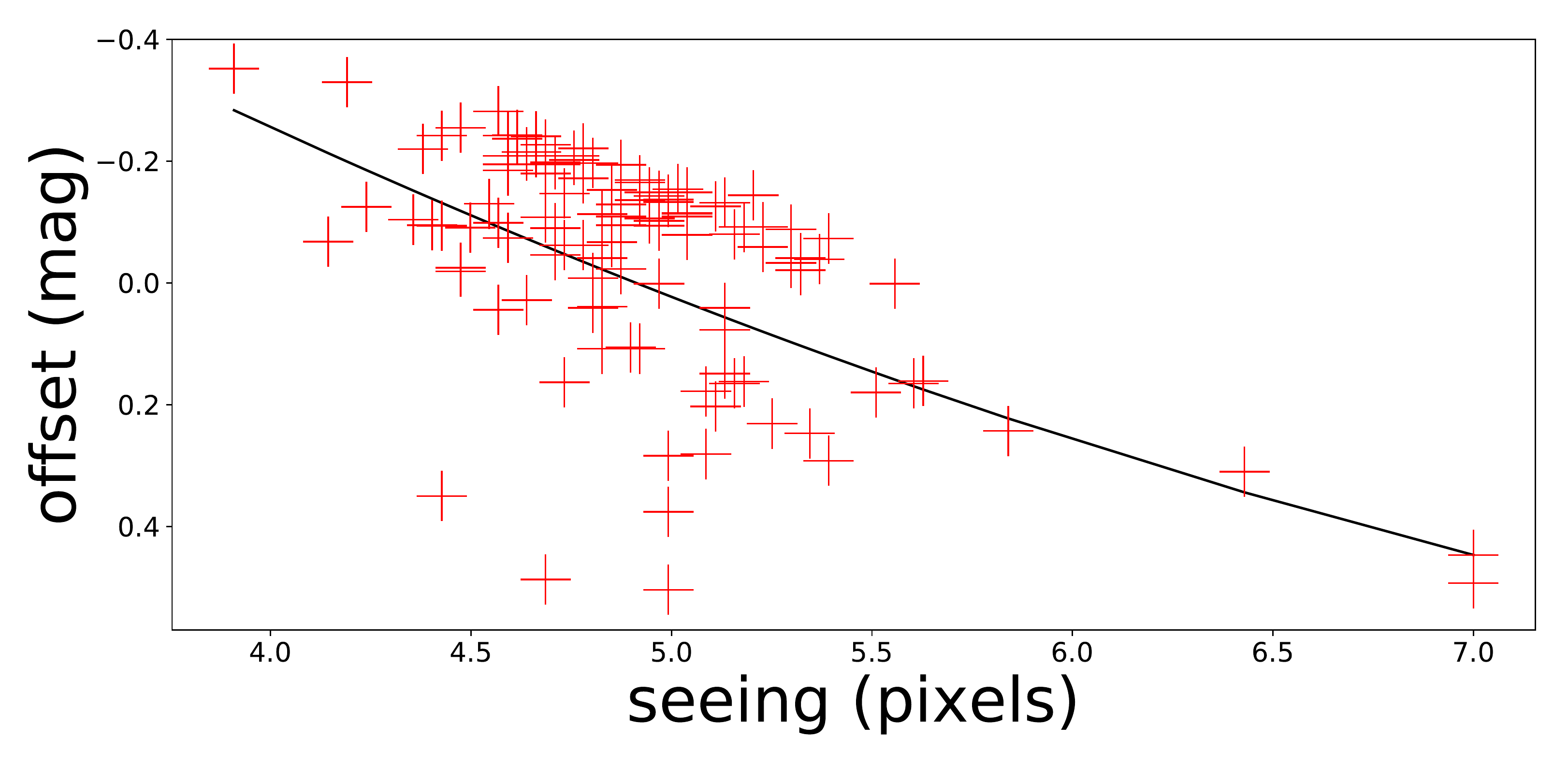}
}
\subfloat[Run 3]{   
  \includegraphics[width=57mm]{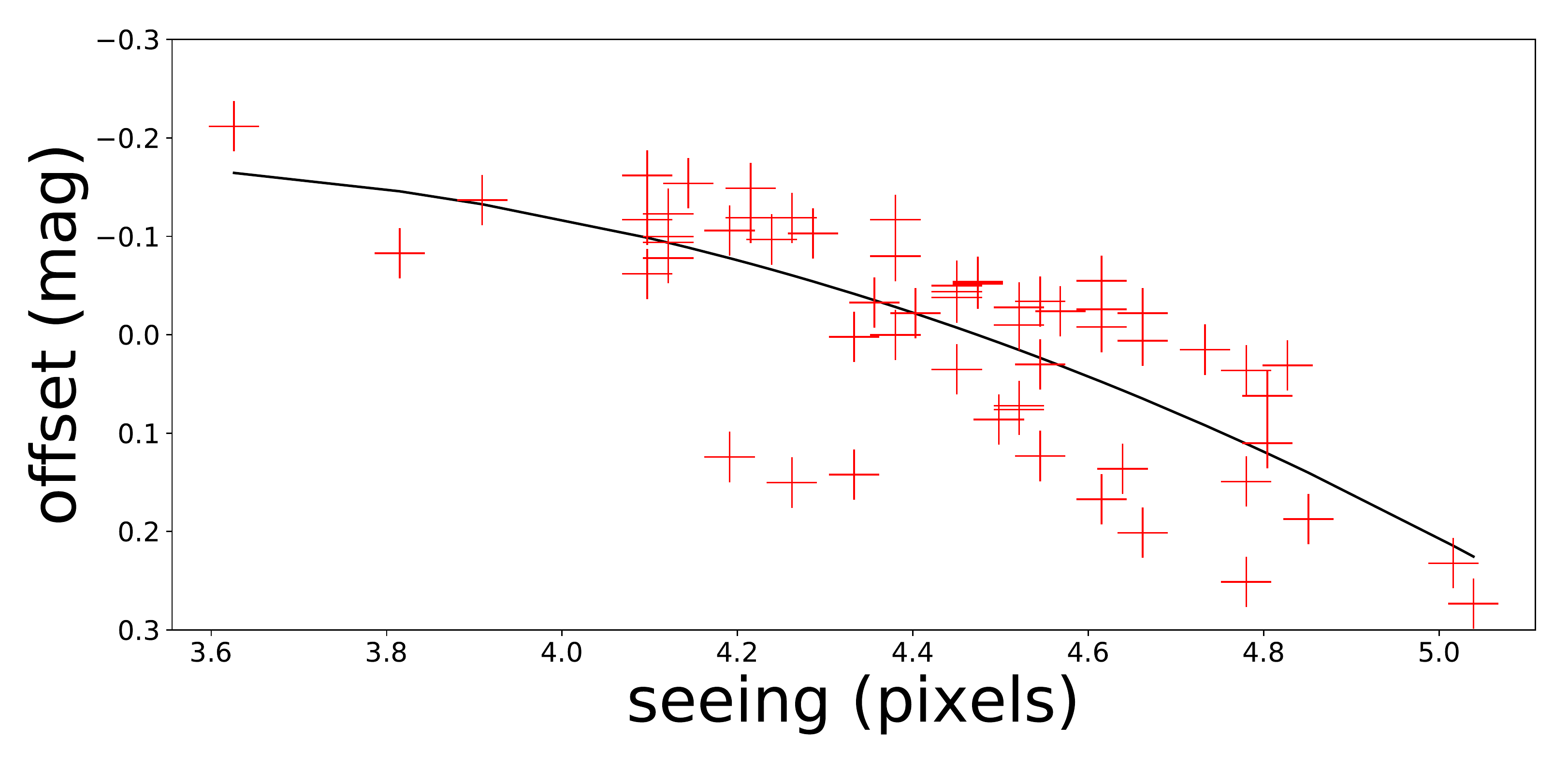}
  }
\hspace{0mm}
\subfloat[Run 1]{
  \includegraphics[width=57mm]{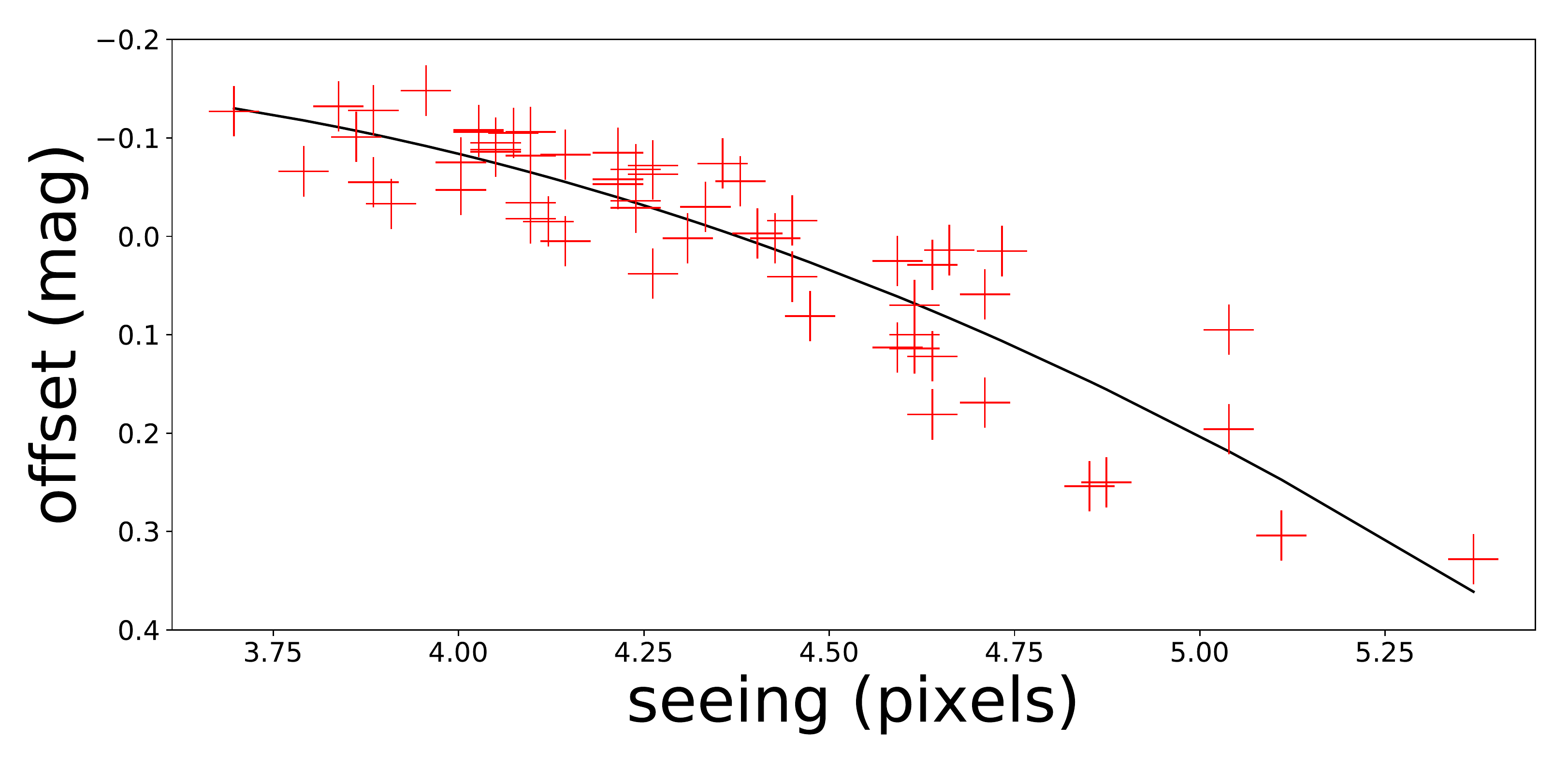}
}
\subfloat[Run 2]{
  \includegraphics[width=57mm]{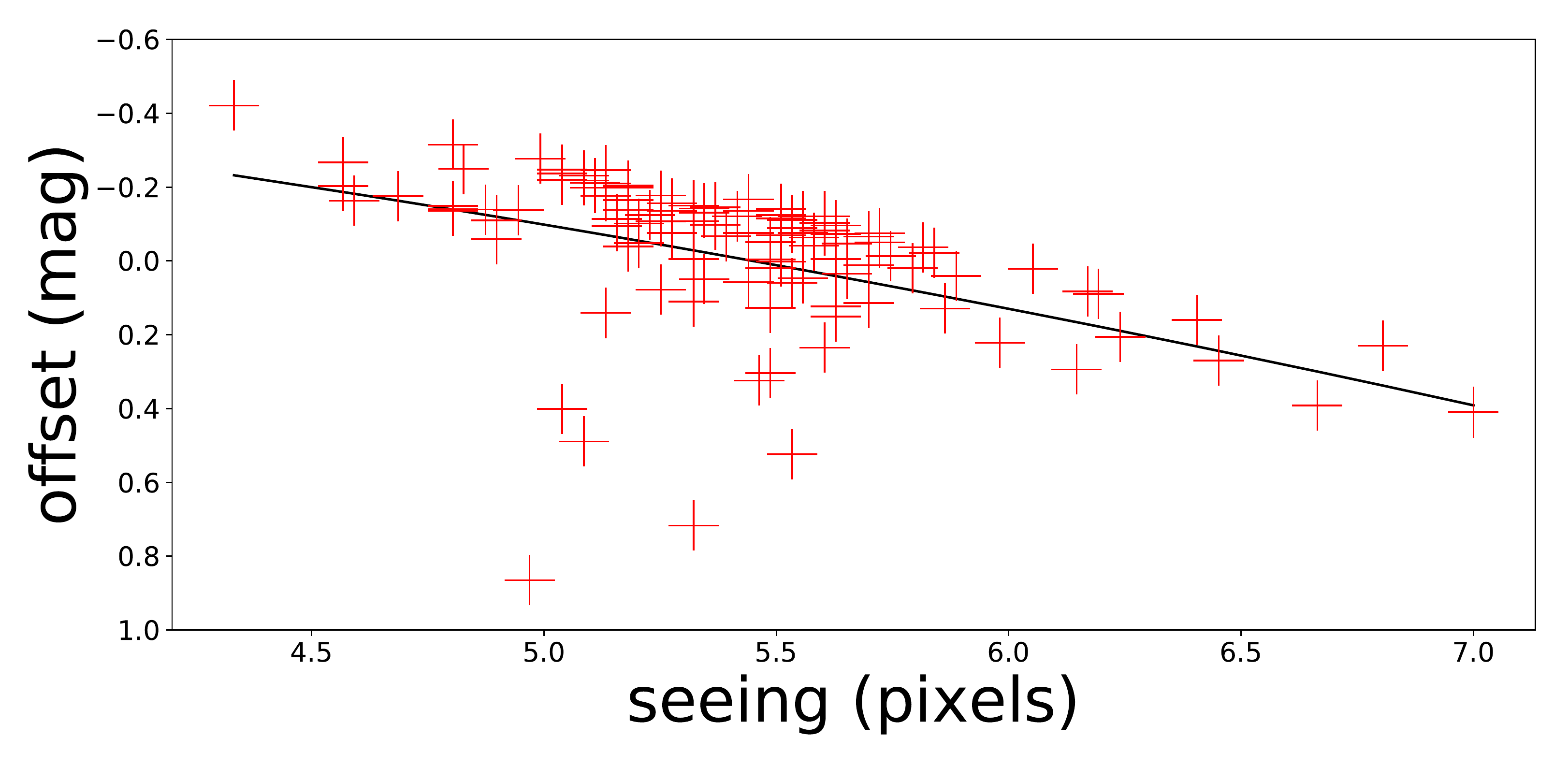}
}
\subfloat[Run 3]{
  \includegraphics[width=57mm]{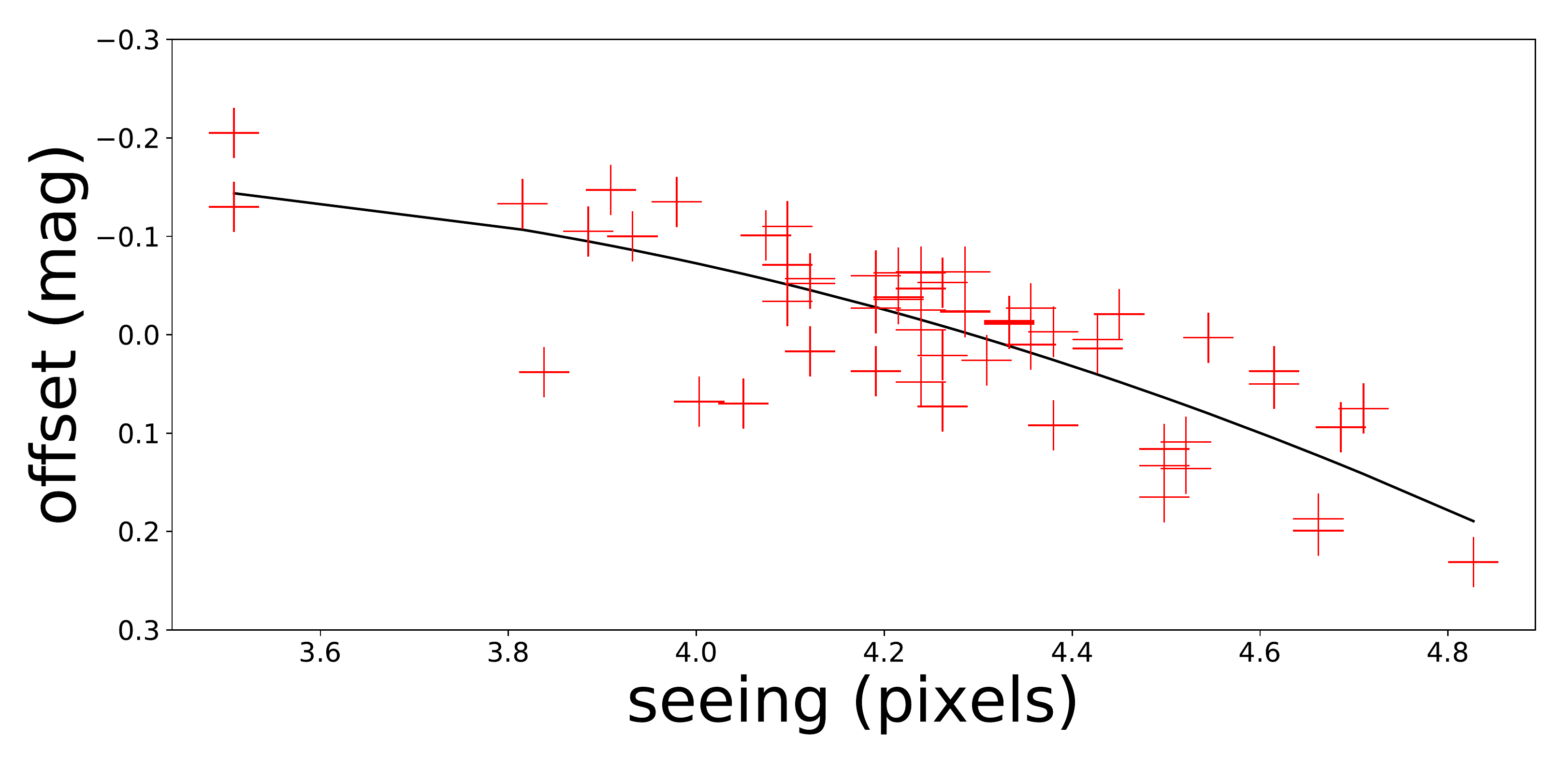}
  }
\hspace{0mm}
\subfloat[Night 1]{
  \includegraphics[width=57mm]{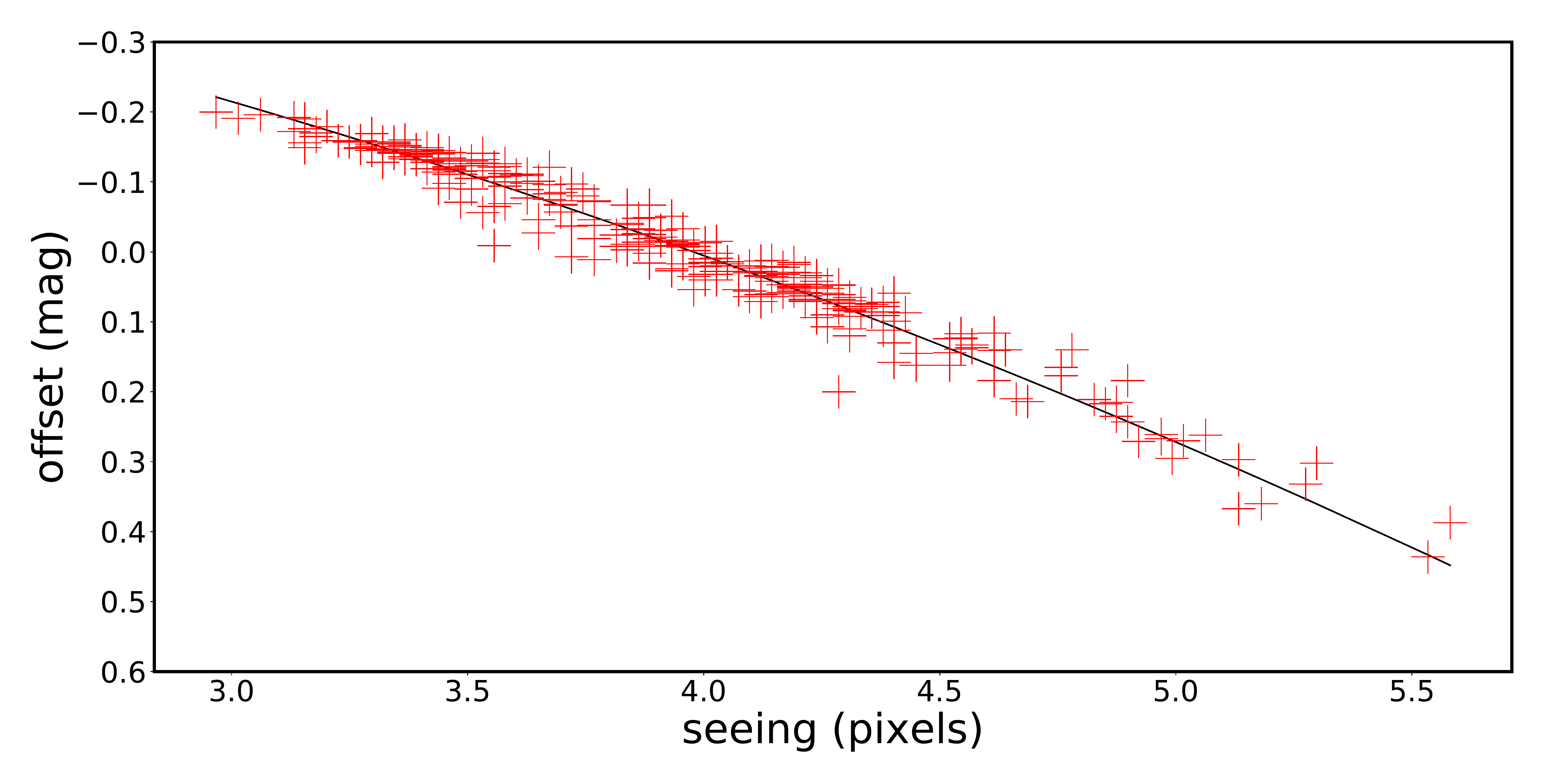}
}
\subfloat[Night 2]{
  \includegraphics[width=57mm]{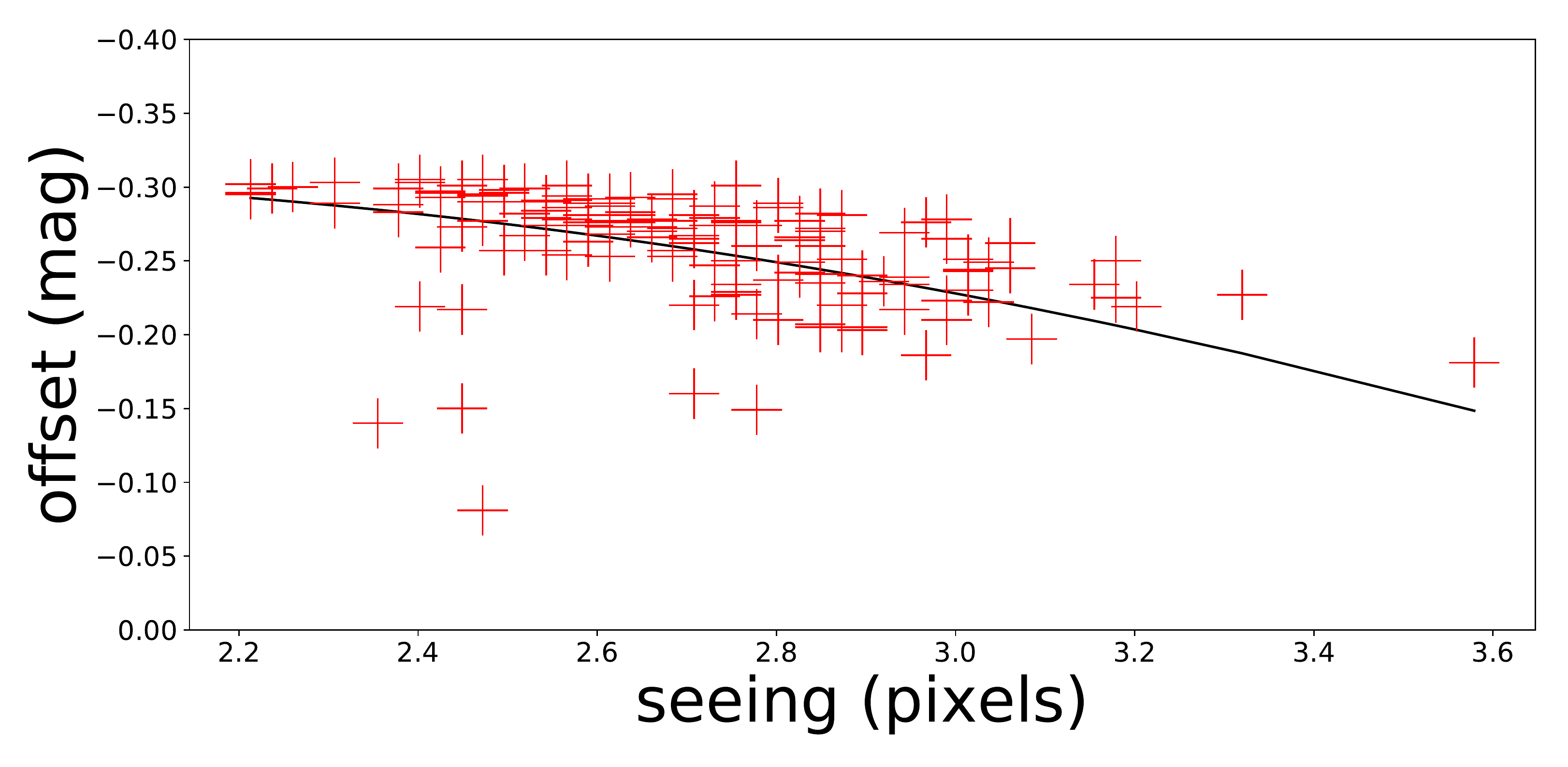}
}
\subfloat[Night 3]{   
  \includegraphics[width=57mm]{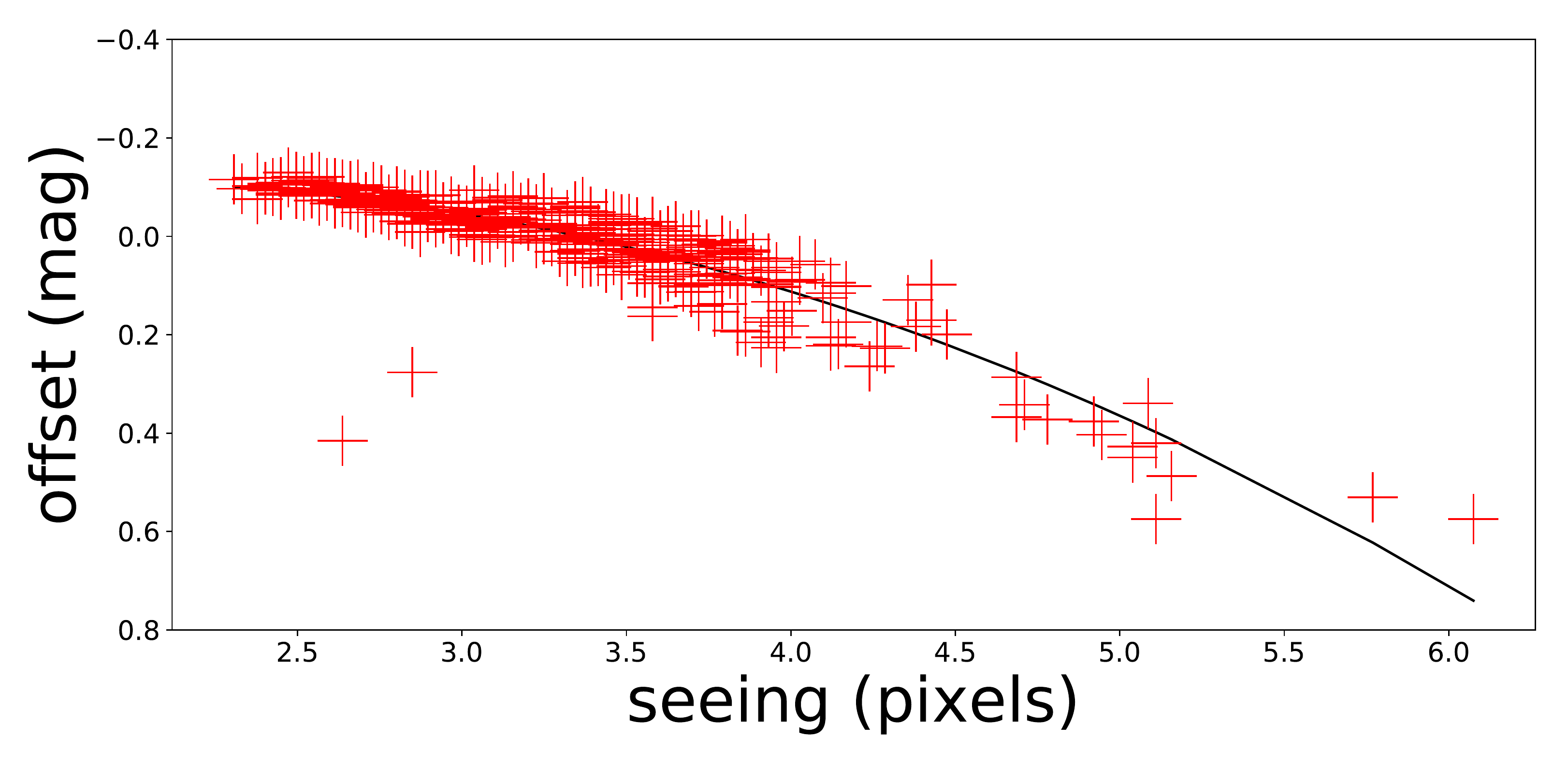}
  }
\hspace{0mm}
\subfloat[Night 1]{
  \includegraphics[width=57mm]{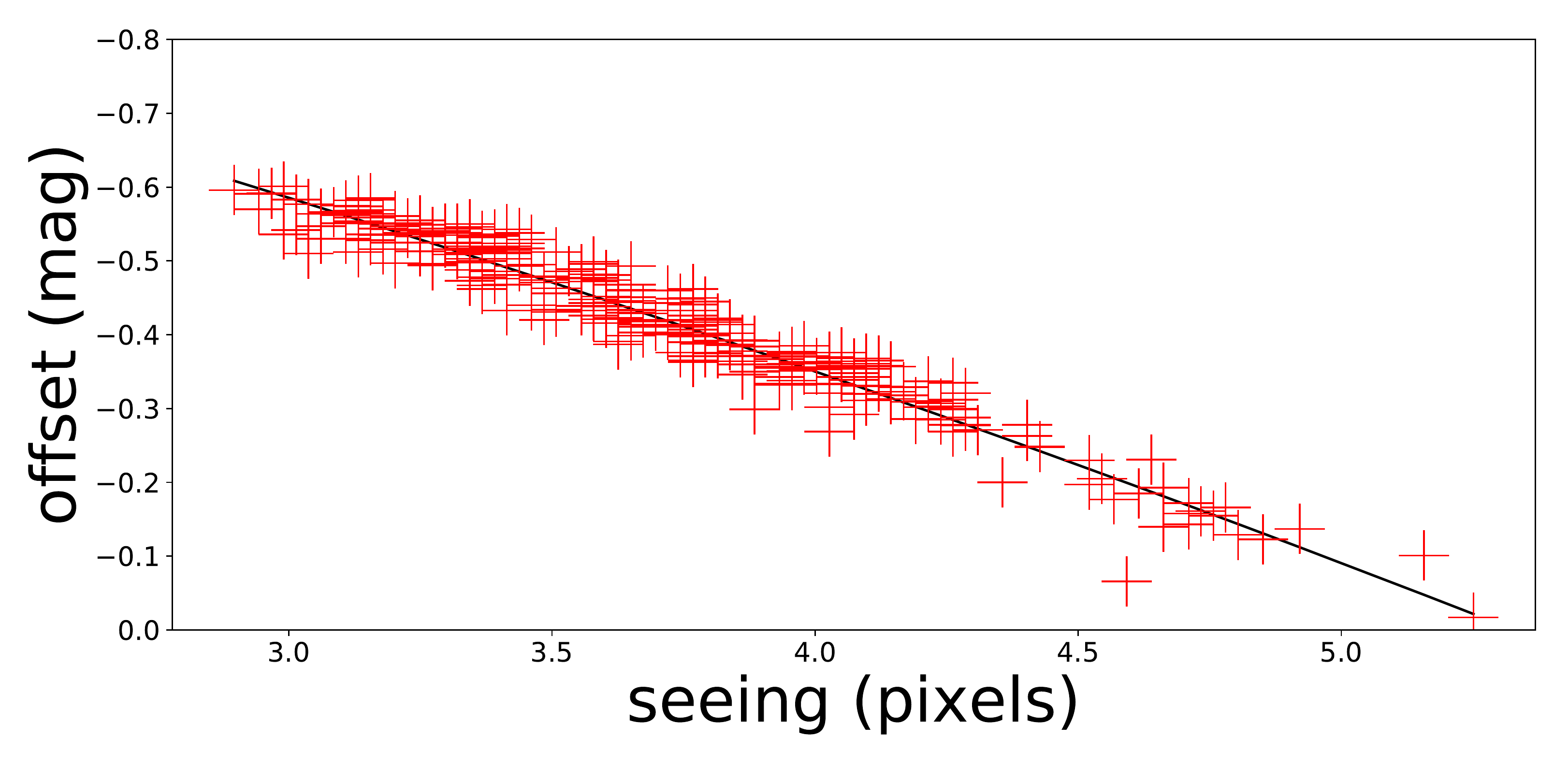}
}
\subfloat[Night 2]{
  \includegraphics[width=57mm]{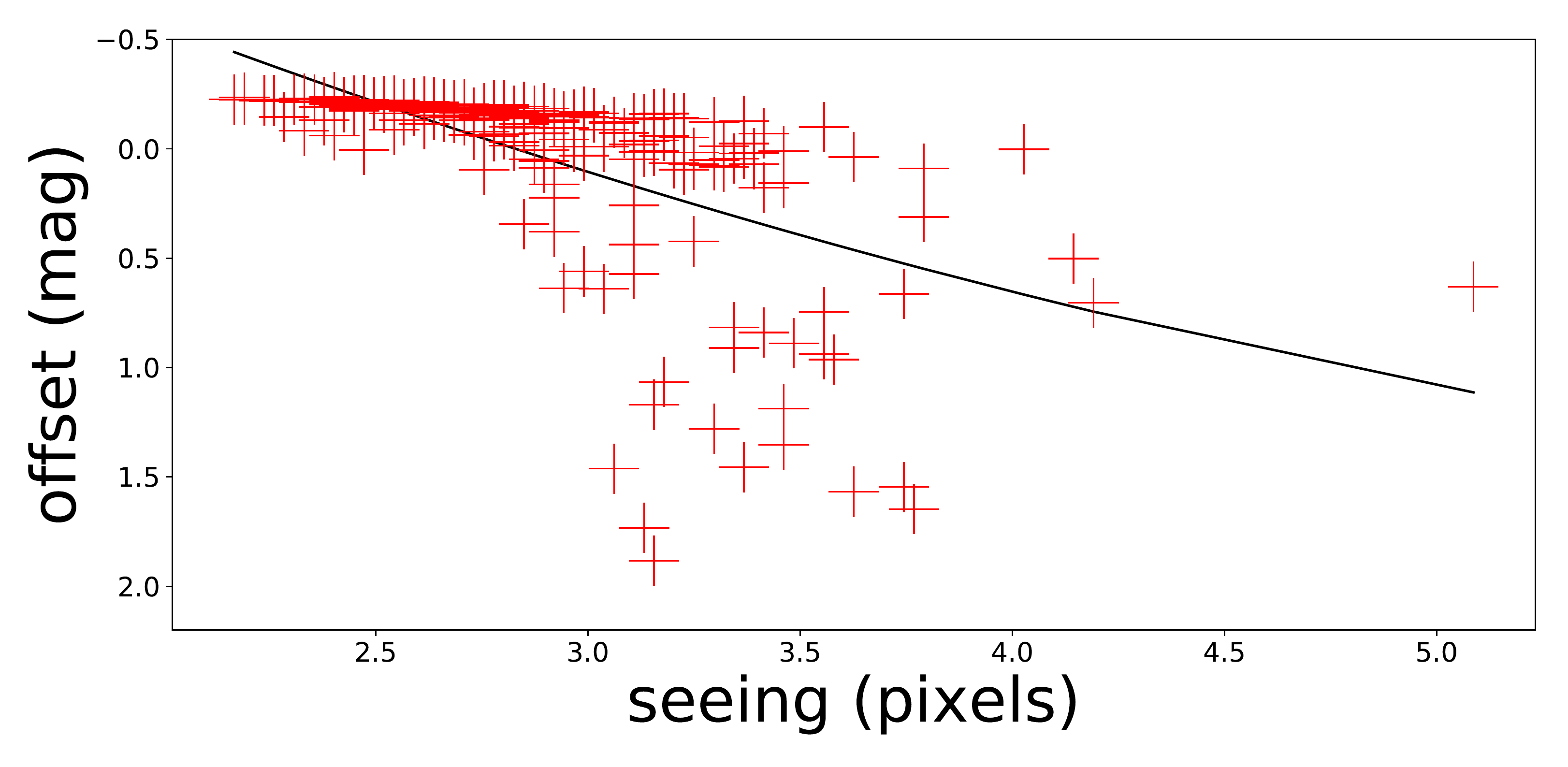}
}
\subfloat[Night 3]{
  \includegraphics[width=57mm]{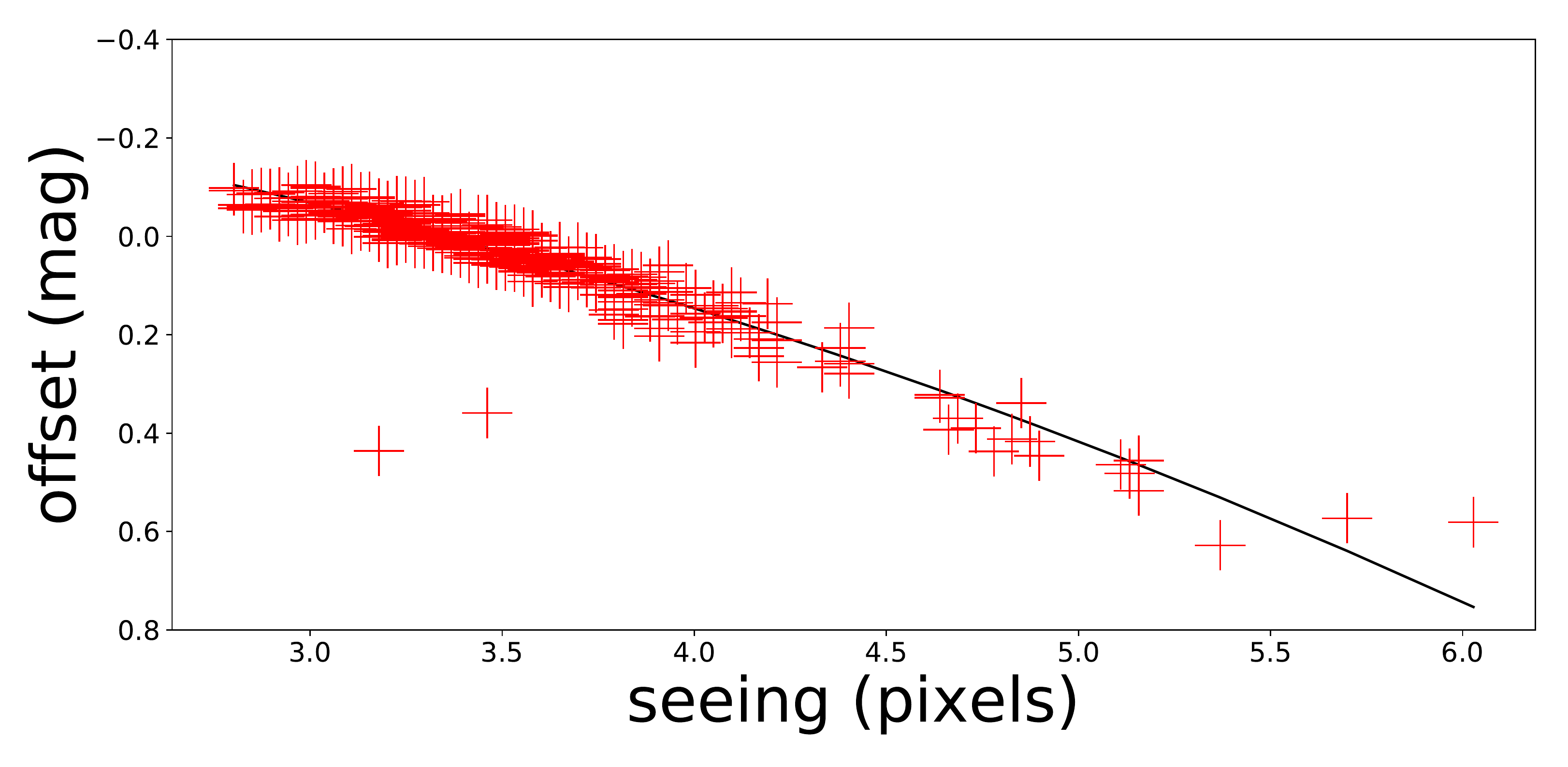}
  }
\hspace{0mm}
  \subfloat[Night 1]{
  \includegraphics[width=57mm]{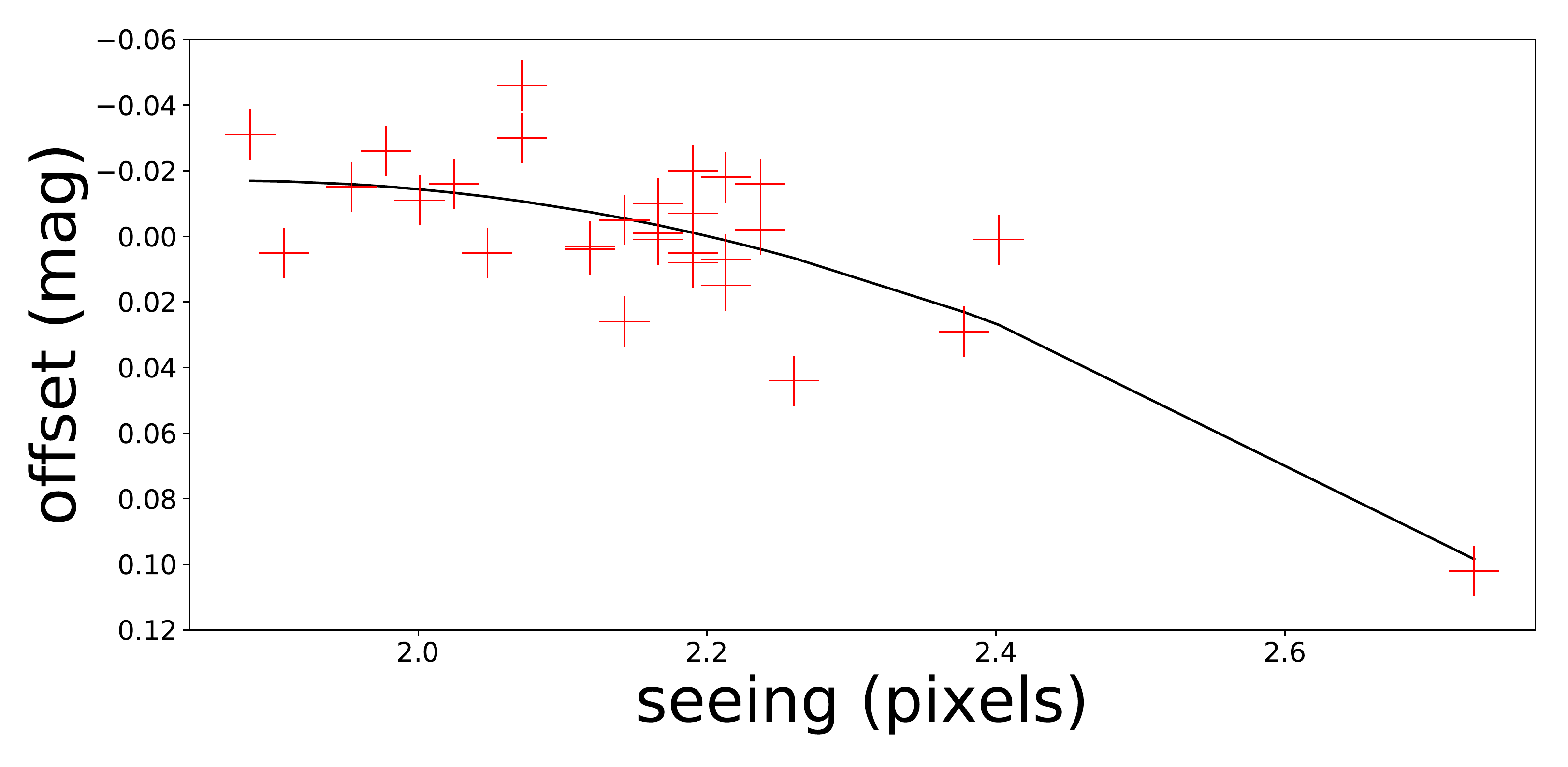}
}
\subfloat[Night 2]{
  \includegraphics[width=57mm]{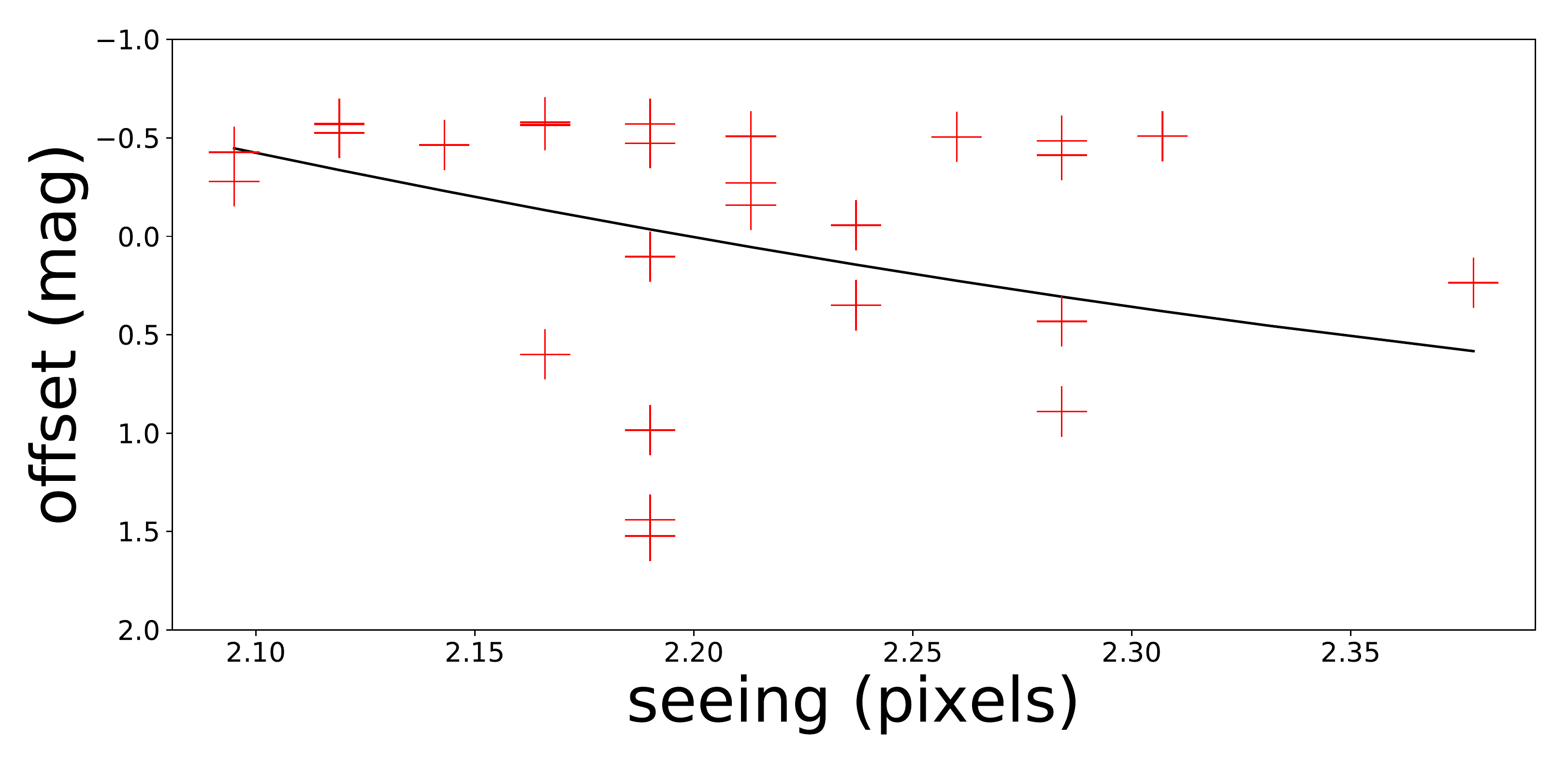}
}
\subfloat[Night 3]{   
  \includegraphics[width=57mm]{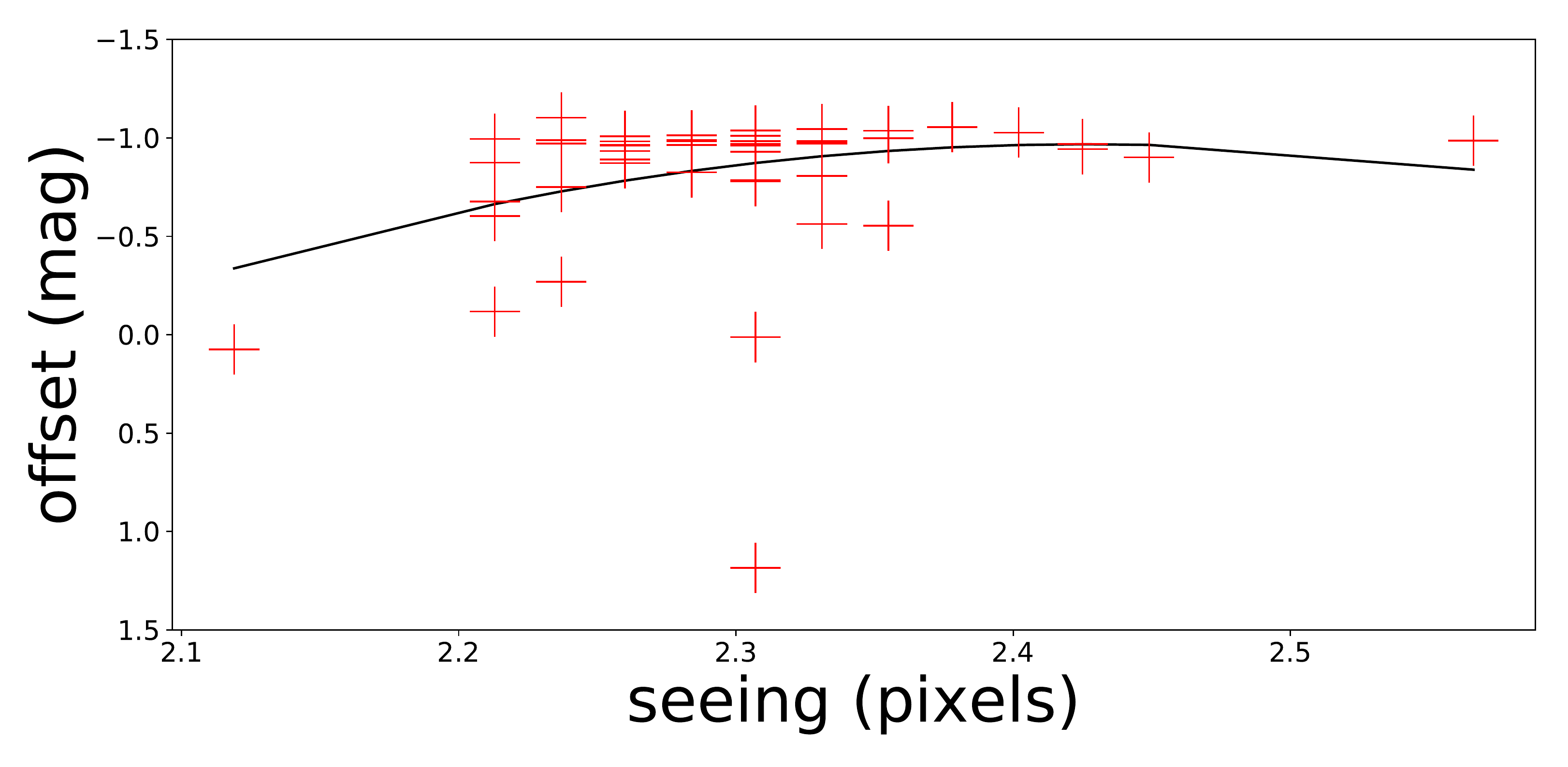}
  }
\hspace{0mm}
\subfloat[Night 4]{
  \includegraphics[width=57mm]{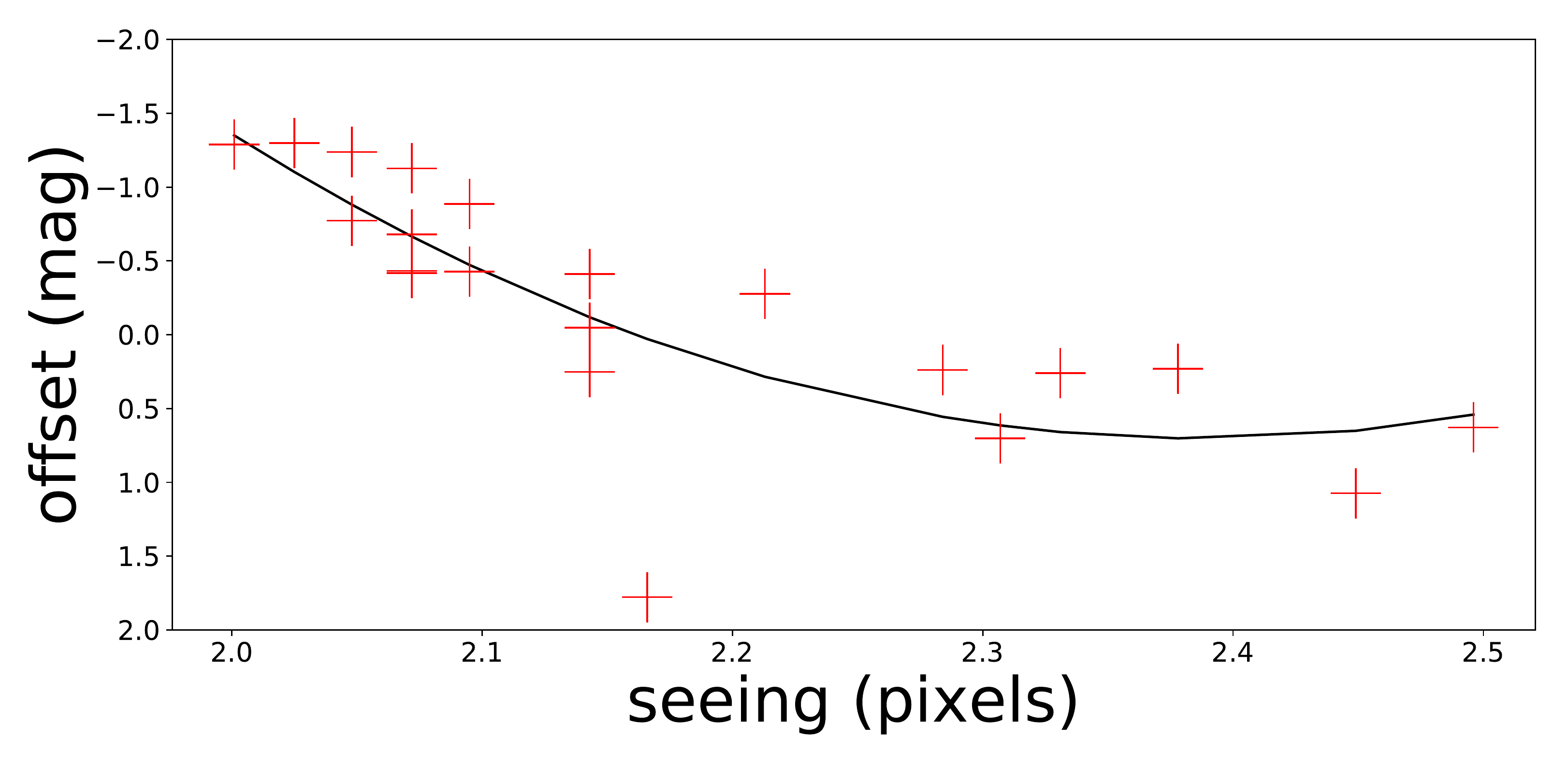}
}
\subfloat[Night 1]{
  \includegraphics[width=57mm]{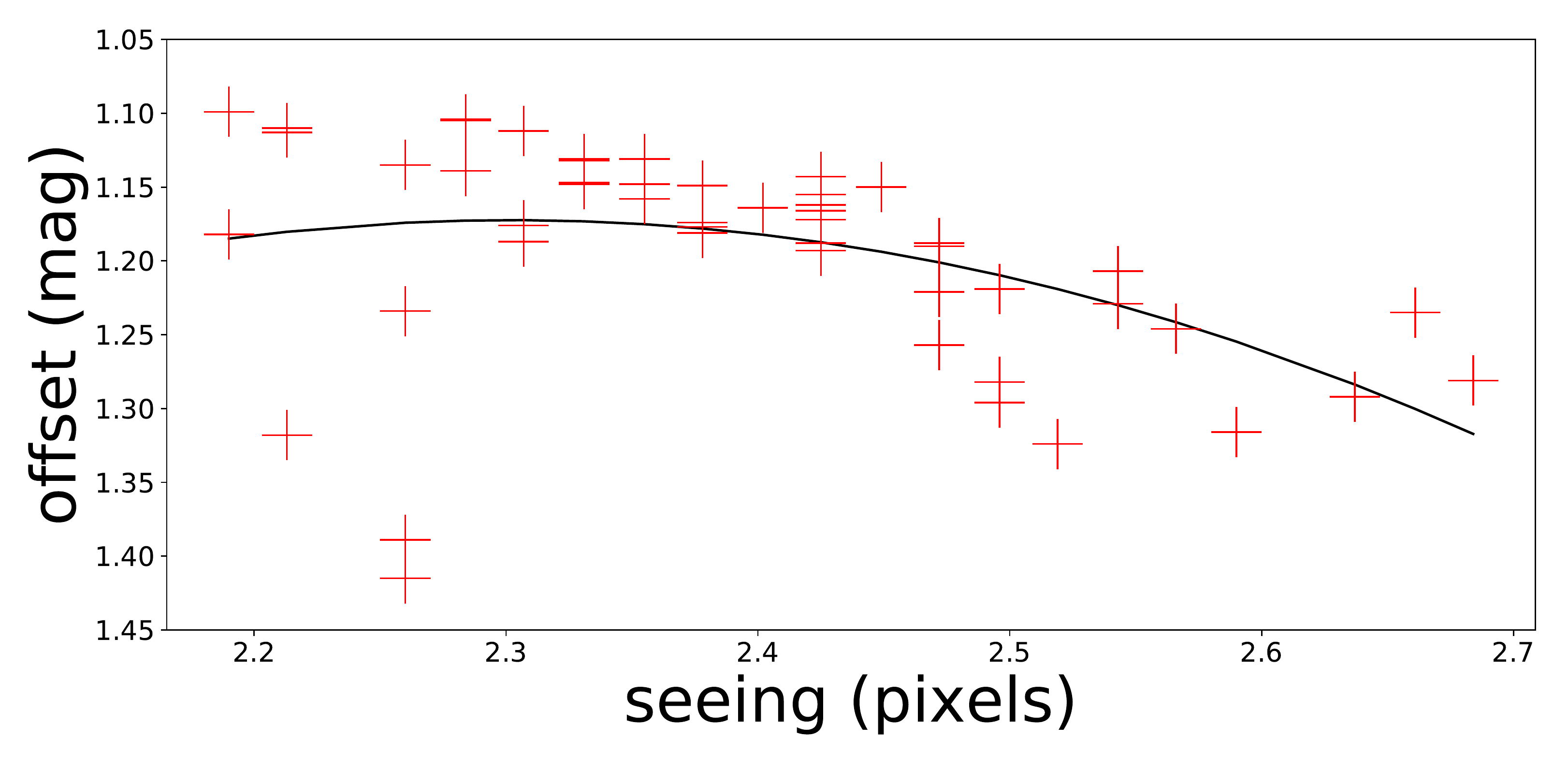}
}
\subfloat[Night 2]{
  \includegraphics[width=57mm]{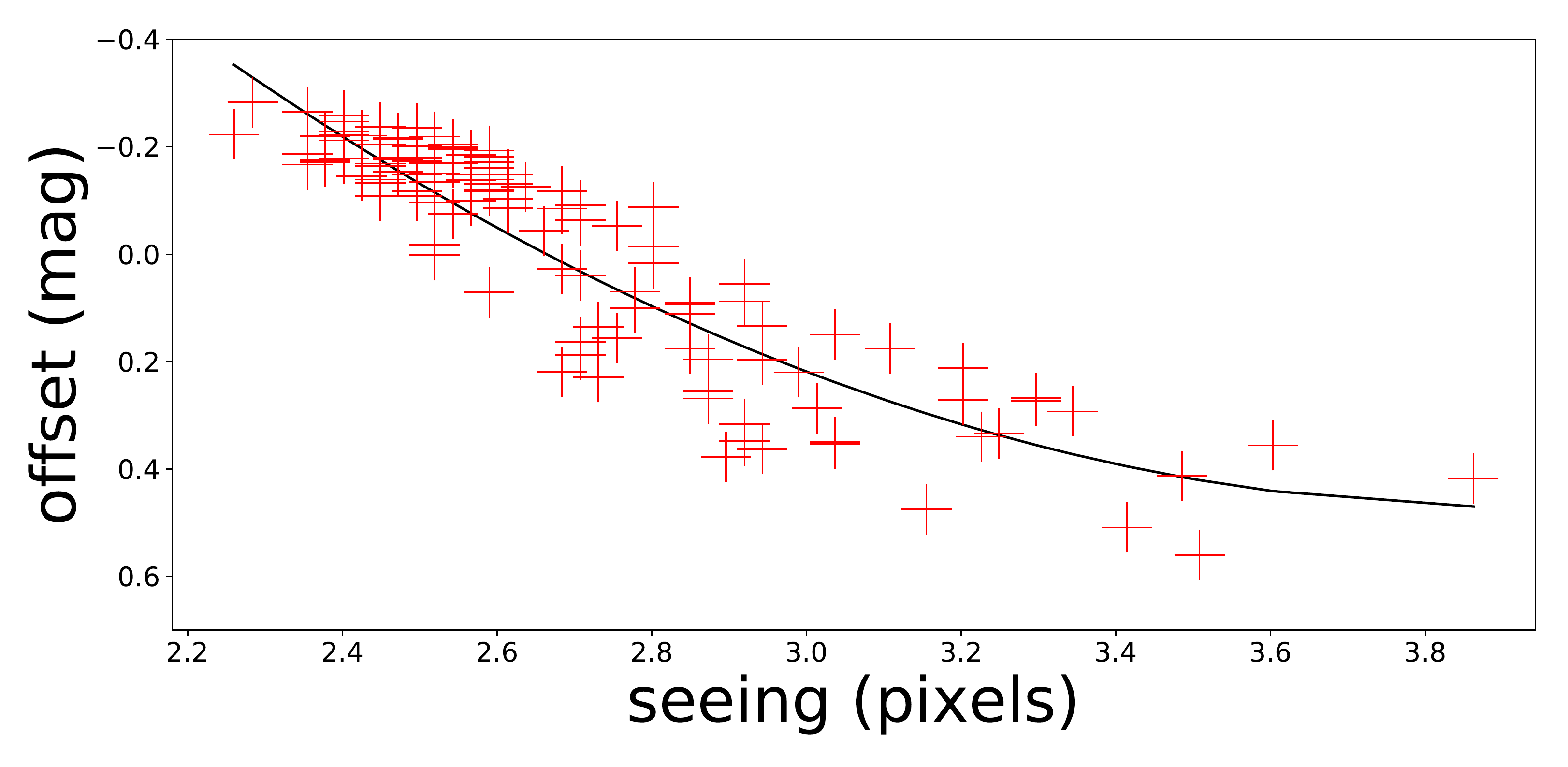}
  }
\caption{Seeing-correlated effects for supernovae.\\
(a), (b), (c), and (d), (e), (f) for SN 2018gv in \textit{R} and \textit{I} bands,\\
(g), (h), (i), and (j), (k), (l) for SN 2018zd in \textit{R} and \textit{I} bands,\\
(m), (n), (o), (p) for SN 2018hhn in \textit{VR} band,\\
(q), (r) for SN 2018hna in \textit{VR} band.}\label{seeingeffects}
\end{figure*}

\begin{figure*}[h]
\centering
\includegraphics[width=1\textwidth]{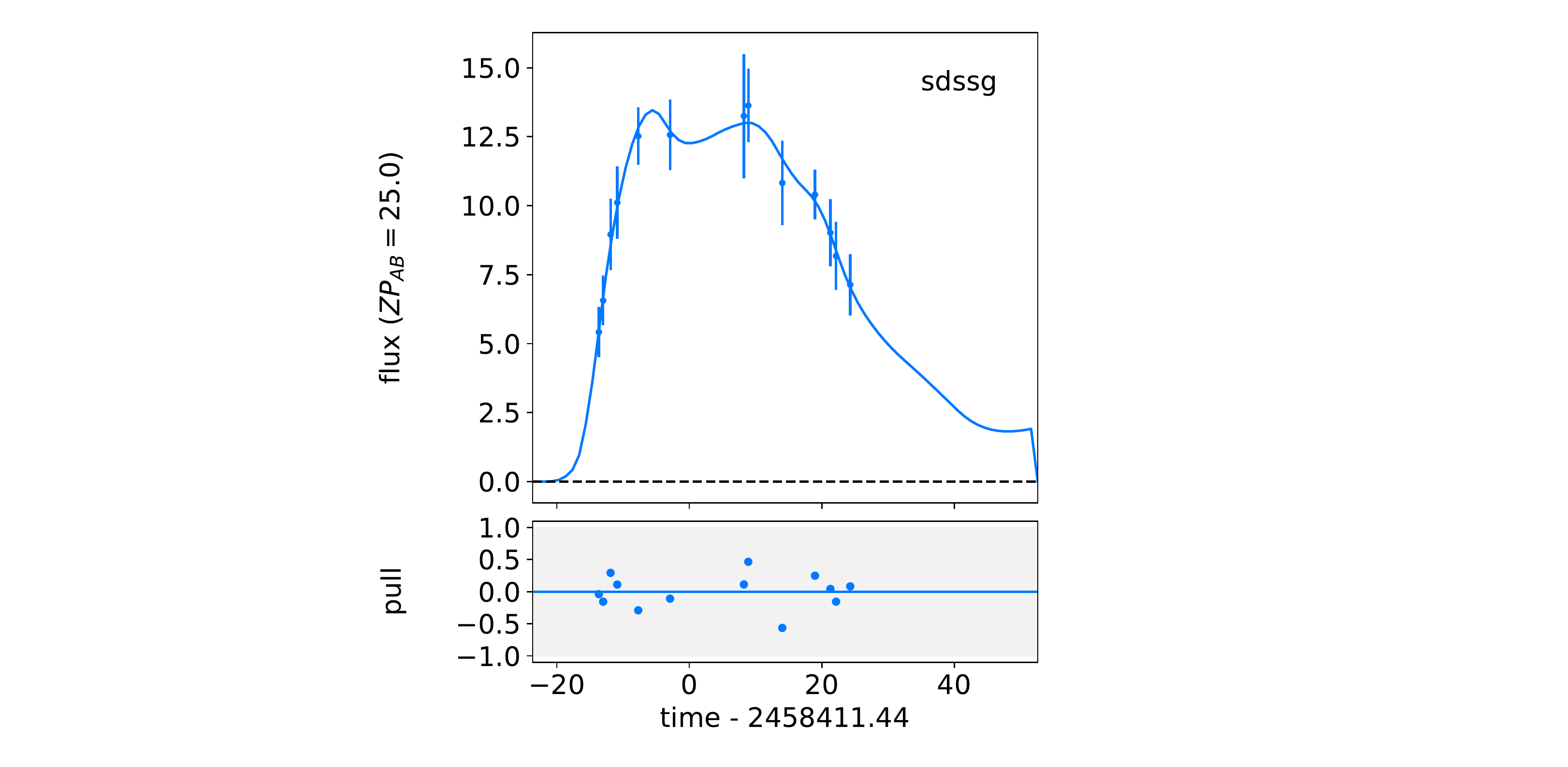}
\bigbreak
\includegraphics[width=1\textwidth]{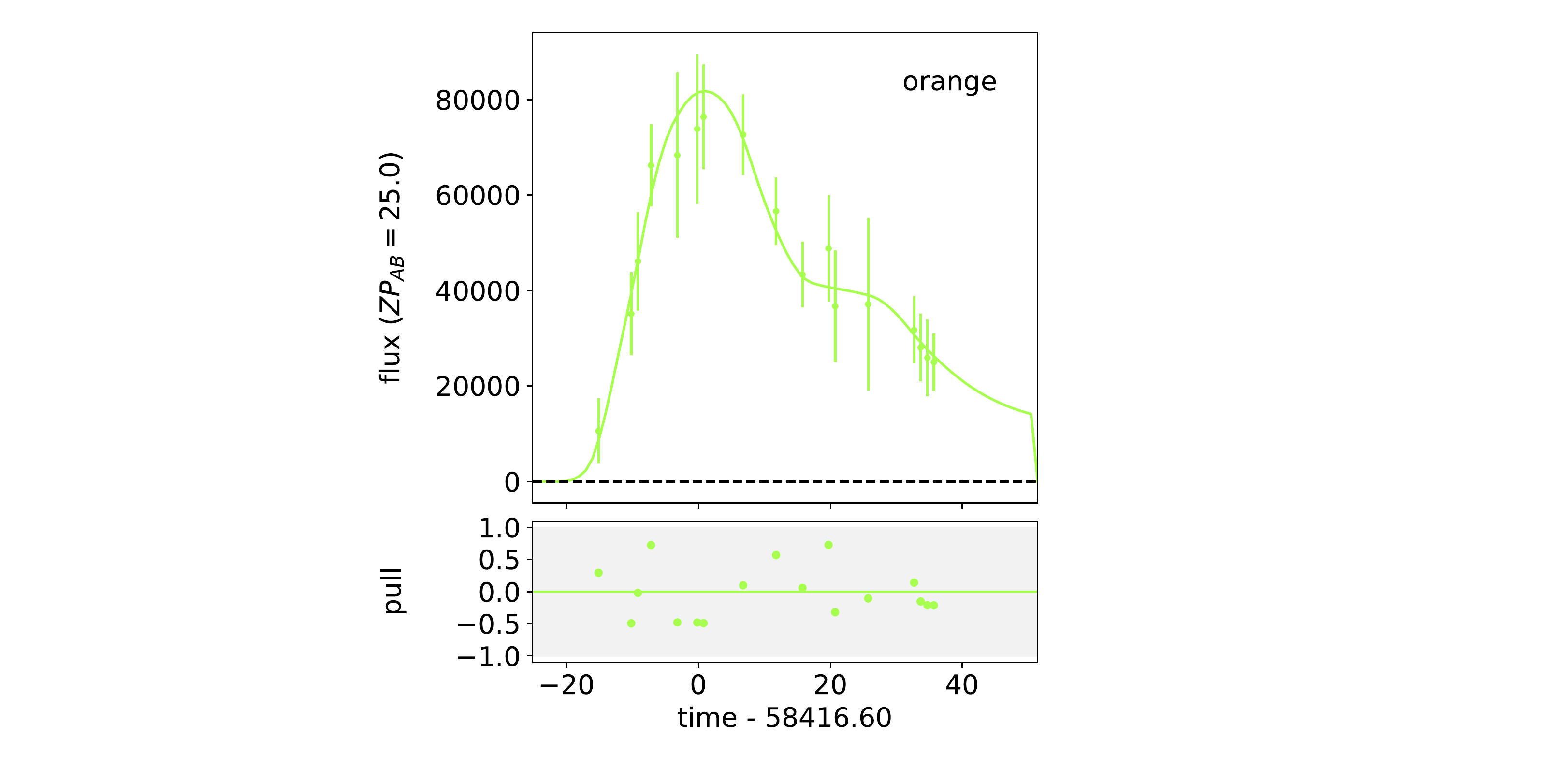}
\caption{SN 2018hgc ASAS-SN \textit{V}-band and SN 2018hhn ATLAS \textit{o}-band light curves. A SALT2 fitting model was applied to estimate the time of maximum light.}\label{salt2}
\end{figure*}

\end{appendix}

\end{document}